\documentclass[twocolumn,aps,prb,10pt,superscriptaddress,longbibliography]{revtex4-2}
\usepackage{amsmath}
\usepackage{glossaries}
\usepackage{graphicx}
\usepackage[version=4]{mhchem}
\usepackage[
    separate-uncertainty=true,
]{siunitx}
\usepackage[dvipsnames,svgnames]{xcolor}
\usepackage{hyperref}
\hypersetup{
pdfnewwindow=true,      
colorlinks=true,        
linkcolor=Blue,         
citecolor=Blue,         
filecolor=Blue,         
urlcolor=Blue           
}

\newacronym{aimd}{AIMD}{\textit{ab initio} molecular dynamics}
\newacronym{dft}{DFT}{density functional theory}
\newacronym{md}{MD}{molecular dynamics}
\newacronym{nep}{NEP}{neuroevolution potential}

\DeclareSIUnit\angstrom{\text{Å}}
\DeclareSIUnit\atom{\text{atom}}

\newcommand{\sautoref}[1]{\hyperref[#1]{Supplementary Figure~\ref*{#1}}}
\newcommand{\sautorefRange}[2]{%
  \hyperref[#1]{Supplementary Figures~\ref*{#1}--\ref*{#2}}
}
\newcommand{\snautorefRange}[2]{%
  \hyperref[#1]{Supplementary Sections~\ref*{#1}--\ref*{#2}}
}

\makeatletter
\def\frontmatter@RRAP@format{%
  \small
  \centering
}
\makeatother

\begin{document}

\title{NEP89: Universal neuroevolution potential for inorganic and organic materials across 89 elements}

\author{Ting~Liang}
\affiliation{College of Physical Science and Technology, Bohai University, Jinzhou 121013, P. R. China}
\affiliation{Department of Electronic Engineering and Materials Science and Technology Research Center, The Chinese University of Hong Kong, Shatin, N.T., Hong Kong SAR, 999077, P. R. China}

\author{Ke~Xu}
\affiliation{College of Physical Science and Technology, Bohai University, Jinzhou 121013, P. R. China}
 \affiliation{Department of Electronic Engineering and Materials Science and Technology Research Center, The Chinese University of Hong Kong, Shatin, N.T., Hong Kong SAR, 999077, P. R. China}

\author{Eric~Lindgren}
\affiliation{
  Department of Physics and Astronomy,
  Chalmers University of Technology,
  41926 Gothenburg, Sweden
}

\author{Zherui~Chen}
\affiliation{Future Technology School, Shenzhen Technology University, Shenzhen 518118, P. R. China}
\affiliation{College of Applied Sciences, Shenzhen University, Shenzhen 518060, P. R. China}

 \author{Rui~Zhao}
\affiliation{School of Mechanical and Electrical Engineering, Xinyu University, Xinyu 338004, P. R. China}

\author{Jiahui~Liu}
\affiliation{Beijing Advanced Innovation Center for Materials Genome Engineering,  University of Science and Technology Beijing, Beijing 100083, P. R. China}

\author{Esm{\'e}e~Berger}
\affiliation{
  Department of Physics and Astronomy,
  Chalmers University of Technology,
  41926 Gothenburg, Sweden
}

\author{Benrui~Tang}
\affiliation{College of Physical Science and Technology, Bohai University, Jinzhou 121013, P. R. China}

\author{Bohan~Zhang}
\affiliation{College of Physical Science and Technology, Bohai University, Jinzhou 121013, P. R. China}

\author{Yanzhou~Wang}
\affiliation{MSP group, QTF Centre of Excellence, Department of Applied Physics, P.O. Box 15600, Aalto University, FI-00076 Aalto, Espoo, Finland}

\author{Keke~Song}
\affiliation{College of Physics and Information Engineering, Fuzhou University, Fuzhou 350108, China}

\author{Penghua~Ying}
\affiliation{Department of Physical Chemistry, School of Chemistry, Tel Aviv University, Tel Aviv, 6997801, Israel}

\author{Nan Xu}
\affiliation{College of Chemical and Biological Engineering, Zhejiang University, Hangzhou 310058, P. R. China}

\author{Haikuan~Dong}
\affiliation{College of Physical Science and Technology, Bohai University, Jinzhou 121013, P. R. China}

\author{Shunda~Chen}
\affiliation{Department of Civil and Environmental Engineering, George Washington University,
Washington, DC 20052, USA}

\author{Paul~Erhart}
\affiliation{
  Department of Physics and Astronomy,
  Chalmers University of Technology,
  41926 Gothenburg, Sweden
}
\affiliation{
    Wallenberg Initiative Materials Science for Sustainability, Chalmers University of Technology, 41926 Gothenburg, Sweden
}

\author{Zheyong~Fan}
\affiliation{College of Physical Science and Technology, Bohai University, Jinzhou 121013, P. R. China}

\author{Tapio~Ala-Nissila}
\affiliation{MSP group, QTF Centre of Excellence, Department of Applied Physics, P.O. Box 15600, Aalto University, FI-00076 Aalto, Espoo, Finland}
\affiliation{Interdisciplinary Centre for Mathematical Modelling, Department of Mathematical Sciences, Loughborough University, Loughborough, Leicestershire LE11 3TU, UK}

\author{Jianbin~Xu}
\affiliation{Department of Electronic Engineering and Materials Science and Technology Research Center, The Chinese University of Hong Kong, Shatin, N.T., Hong Kong SAR, 999077, P. R. China}

\date[]{%
\parbox{0.95\linewidth}{\centering\small
\vspace*{0.5em}
These authors contributed equally: Ting Liang and Ke Xu\\[0.2em]
Corresponding authors: Shunda Chen (\href{mailto:phychensd@gmail.com}{phychensd@gmail.com}),
Paul Erhart (\href{mailto:erhart@chalmers.se}{erhart@chalmers.se}),\\
Zheyong Fan (\href{mailto:brucenju@gmail.com}{brucenju@gmail.com}), and
Jianbin Xu (\href{mailto:jbxu@ee.cuhk.edu.hk}{jbxu@ee.cuhk.edu.hk})
}%
}


\begin{abstract}
\vspace*{0.8em}
While machine-learned interatomic potentials offer near-quantum-mechanical accuracy for atomistic simulations, many are material-specific or computationally intensive, limiting their broader use. 
Here, we introduce NEP89, a foundation model based on neuroevolution potential architecture, delivering near-empirical-potential speed and high accuracy across 89 elements.
A compact yet comprehensive training dataset covering inorganic and organic materials was curated through descriptor-space subsampling and iterative refinement across multiple datasets.
NEP89 achieves competitive accuracy compared to representative foundation models while being three to four orders of magnitude more computationally efficient, enabling previously impractical large-scale atomistic simulations of inorganic and organic systems. 
In addition to its out-of-the-box applicability to diverse scenarios, including million-atom-scale compression of compositionally complex alloys, ion diffusion in solid-state electrolytes and water, rocksalt dissolution, methane combustion, and protein-ligand dynamics, NEP89 also supports fine-tuning for rapid adaptation to user-specific applications, such as mechanical, thermal, structural, and spectral properties of two-dimensional materials, metallic glasses, and organic crystals.
\end{abstract}
\maketitle

\section{Introduction}

Large-scale atomistic \gls{md} simulations are invaluable tools for elucidating the intricate properties of complex materials. 
However, the fidelity of these simulations critically relies on the accuracy of the underlying interatomic potentials (or force fields). 
\Gls{aimd} simulations offer the desired level of fidelity, but their computational cost is prohibitive in large-scale \gls{md} simulations of millions of atoms and more.

In recent years, artificial intelligence has been successfully applied to construct accurate yet efficient machine-learned interatomic potential models \cite{behler2007generalized, bartok2010gaussian}, substantially advancing the field of atomistic simulations \cite{Behler2016perspective, Frank2020Machine, unke2021machine, Deringer2021Machine, Mishin2021Machine}. 
Early models were developed for specific materials and even specific properties of a given material.
More recent developments have revealed a trend toward building more general-purpose potentials, from a single element \cite{bartok2018machine} to a set of metals \cite{song2024general}, to tens of elements \cite{takamoto2022towards, zhang2024dpa2}, and to almost the entire periodic table \cite{chen2022nuniversal, dengchgnet2023, merchant2023scaling, Xie2024GPTFF, batatia2023foundation, yang2024MatterSim}.
The last category of potential models, although encompassing many species, remains incomplete, as they were mostly constructed based on inorganic materials. 
Moreover, their high computational cost severely limits their practical applicability, particularly for large-scale or long-timescale simulations.

Here, to overcome these limitations, we introduce NEP89, a foundation model that enables high-fidelity simulations of both inorganic and organic materials across 89 elements, while achieving a computational speed several orders of magnitude faster than previous universal interatomic potential models.
Our model is based on the state-of-the-art \gls{nep} approach \cite{fan2021neuroevolution, song2024general}, which combines high accuracy with efficiency.
This approach has recently been demonstrated to achieve a general-purpose interatomic potential for 16 metals and their alloys \cite{song2024general}.

One major obstacle is the lack of a readily usable training dataset that comprehensively incorporates both inorganic and organic materials.
The lack of such a dataset is presumably the reason why previous works have only trained universal potential models either for inorganic \cite{chen2022nuniversal, dengchgnet2023, merchant2023scaling, Xie2024GPTFF, batatia2023foundation, yang2024MatterSim} or organic materials \cite{unke2024biomolecular, eastman2023spice, zhang2024exploring, kovacs2025maceoff}, or both, but with multiple training tasks instead of a single unified one \cite{zhang2024dpa2}.
A crucial contribution in the present work is the curation of existing databases of separate sets of materials to ensure they have consistent target data for training a unified interatomic potential for both inorganic and organic materials. 
Starting from a sub-sampled dataset of OMAT24 \cite{barrosoluque2024omat24}, we iteratively train NEP89 and build the corresponding training dataset, incorporating structures from many other public datasets \cite{dengchgnet2023, eastman2023spice, zhang2024exploring, wang2024pretrained, unke2024biomolecular, song2024general, Ibragimova2025unifying, zhai2023short}.
Moreover, we add the D3 dispersion correction \cite{Grimme2010consistent, Grimme2011effect} to some datasets such that all of them have proper dispersion interactions. 
Furthermore, the relative energies between different datasets are also optimized during the training process, which is a prerequisite for constructing a unified single-task model.
The NEP89 model achieves competitive accuracy in predicting both static and dynamic properties. 
It also enables large-scale atomistic simulations of both inorganic and organic systems out of the box, while supporting rapid and cost-effective fine-tuning across a broad range of applications.

\section{Results}

\noindent{\textbf{The architecture of the NEP89 model and its finetuning.}}
The NEP89 model is built on the \gls{nep} approach \cite{fan2021neuroevolution, song2024general}. 
We begin by briefly introducing the \gls{nep} architecture and then discuss the extensions that enable the fine-tuning of NEP89.
The \gls{nep} approach is a many-body neural network potential that uses an atom-decomposed total energy and atom-centered descriptors constructed with Chebyshev and Legendre polynomials, along with an advanced training algorithm that incorporates regularization and employs the separable natural evolution strategy \cite{schaul2011high}.
The latter evolves a set of mean and variance values for the trainable parameters that also form the basis for fine-tuning. 
The descriptors are defined within a specified cutoff radius to ensure that the computational cost scales linearly with the number of atoms. 
The input layer neurons of the neural network correspond to the descriptors, while the output layer neuron represents the site energy $U_i$ of the central atom $i$.
Force and virial stress are calculated via analytical derivatives of the energy (see \autoref{snote:nep-formulation} for details).
Atomic species are encoded in the expansion coefficients of a set of radial functions.
For each pair of species, there is an independent set of $N_{\rm ec}$ expansion coefficients optimized during training.
Additionally, each species has its own independent set of $N_{\rm nn}$ trainable weight and bias parameters within the neural network.
Therefore, for a system with $N_{\rm spe}$ species, there are $N_{\rm spe}^2 N_{\rm ec} + N_{\rm spe} N_{\rm nn}$ trainable parameters (refer to \autoref{snote:nep-para} for the hyperparameters used when training NEP89).

Thanks to the combinatorial architecture of \gls{nep}, the trained NEP89 model can be conveniently fine-tuned into smaller models tailored to subsets of species using a new training dataset. 
During the fine-tuning process, it is crucial to maintain the descriptor normalization. 
To this end, for the targeted species, the mean and variance of their trainable parameters, used by the separable natural evolution strategy, are extracted from NEP89 and reused during continued training.
Fine-tuning typically requires only a small fraction of the training steps compared to training from scratch to achieve optimal results.
Additionally, to prevent the model from undesirably losing memory, the variance values of the descriptor parameters can be set to zero, ensuring that only the neural network parameters are updated during fine-tuning.

\vspace{0.5cm}

\noindent{\textbf{Iterative training of NEP89.}}
To train a universal potential model across 89 species for both inorganic and organic materials, the first task is to construct a training set that is diverse, reliable, and consistent. 
So far, no single available dataset meets these three conditions simultaneously.
Fortunately, there are already many individual datasets, and we aim to combine them to form a consistent dataset with enhanced diversity and reliability.
A notable one is the OMat24 dataset \cite{barrosoluque2024omat24} containing over 110 million structures for inorganic bulk materials.
To supplement the OMat24 dataset, we also include structures from other important datasets, including the MPtrj dataset for relaxation trajectories of inorganic bulk materials across 89 elements \cite{dengchgnet2023}, the SPICE dataset for drug-like small molecules, peptides, and solvated amino acids \cite{eastman2023spice}, the ANI-1xnr dataset for reactive organics \cite{zhang2024exploring}, the SSE-ABACUS and SSE-VASP datasets for solid-state electrolytes \cite{wang2024pretrained}, the solvated protein fragments (Protein for short) dataset \cite{unke2024biomolecular}, the UNEP-v1 dataset for 16 metals and their alloys \cite{song2024general}, the CH dataset for general and reactive CH systems \cite{Ibragimova2025unifying}, the water dataset with liquid, gas, and vapor-liquid phases \cite{zhai2023short}, and a reactive CHONPS dataset we created.
More details regarding these datasets are provided in \autoref{stable:dataset}.
The successful training of NEP89 was not accomplished by providing all the aforementioned datasets to the optimizer simultaneously, but rather through an iterative training strategy, as described below.
 
Initially, we note that these datasets are inconsistent in their incorporation of dispersion interactions. 
While some datasets (SPICE, Protein, ANI-1xnr) include the D3 dispersion \cite{Grimme2010consistent, Grimme2011effect}, we added this dispersion interaction to the remaining datasets for uniformity. 
The water dataset is unique, as it was labeled using the MB-pol method \cite{Babin2013JCTC}, delivering accuracy close to CCSD(T). 
Although not all datasets are calculated at the same level of quantum-mechanical theory, the differences in force and stress are minor and can be disregarded in training the NEP89 model.
On the other hand, these datasets have substantially different energy references for a given species, which hinders the development of a unified model. 
To address this challenge, we initially focused exclusively on the energies in the OMAT24 dataset, disregarding reference energies from all other datasets.

Another challenge in training a unified model based on extensive public datasets is managing the vast number of training structures. 
A good training dataset does not necessarily require a large number of structures but rather a diverse and well-balanced selection.
To address this, we first created a randomly sub-sampled dataset from the OMAT24 database to train an initial model. 
Using this model, we then selected a subset of the OMAT24 database with uniformly distributed descriptor values (see \autoref{snote:subsampling} for details) to train an improved model.
We then performed model training and data curation for more than ten iterations. 
During each iteration, the current model predicted unseen structures across all datasets. 
Structures with high force and stress errors were identified, and a subset with evenly distributed descriptor values was sampled and added to the training dataset. 
This refined dataset was then used to train the next model in the sequence.

An essential aspect of dataset refinement is identifying inaccuracies in the reference data. 
Several datasets, such as OMAT24 and MPtrj, contain erroneous entries or unconverged \gls{dft} calculations. 
To mitigate the impact of noisy reference data, prior studies have employed the Huber loss function \cite{chen2022nuniversal, dengchgnet2023, merchant2023scaling, batatia2023foundation, yang2024MatterSim}.
Our approach involves iteratively removing outliers from the training dataset, thereby gradually reducing their influence on the resulting models.
In \sautoref{fig:recalculate-mp}, we show results for 113 typical structures in the MPtrj dataset identified as outliers during our training, where the original \gls{dft} reference values were found to be inaccurate, while predictions by NEP89 closely aligned with our recalculated reference values.


\begin{figure}[ht]
\centering
\includegraphics[width=1.0\columnwidth]{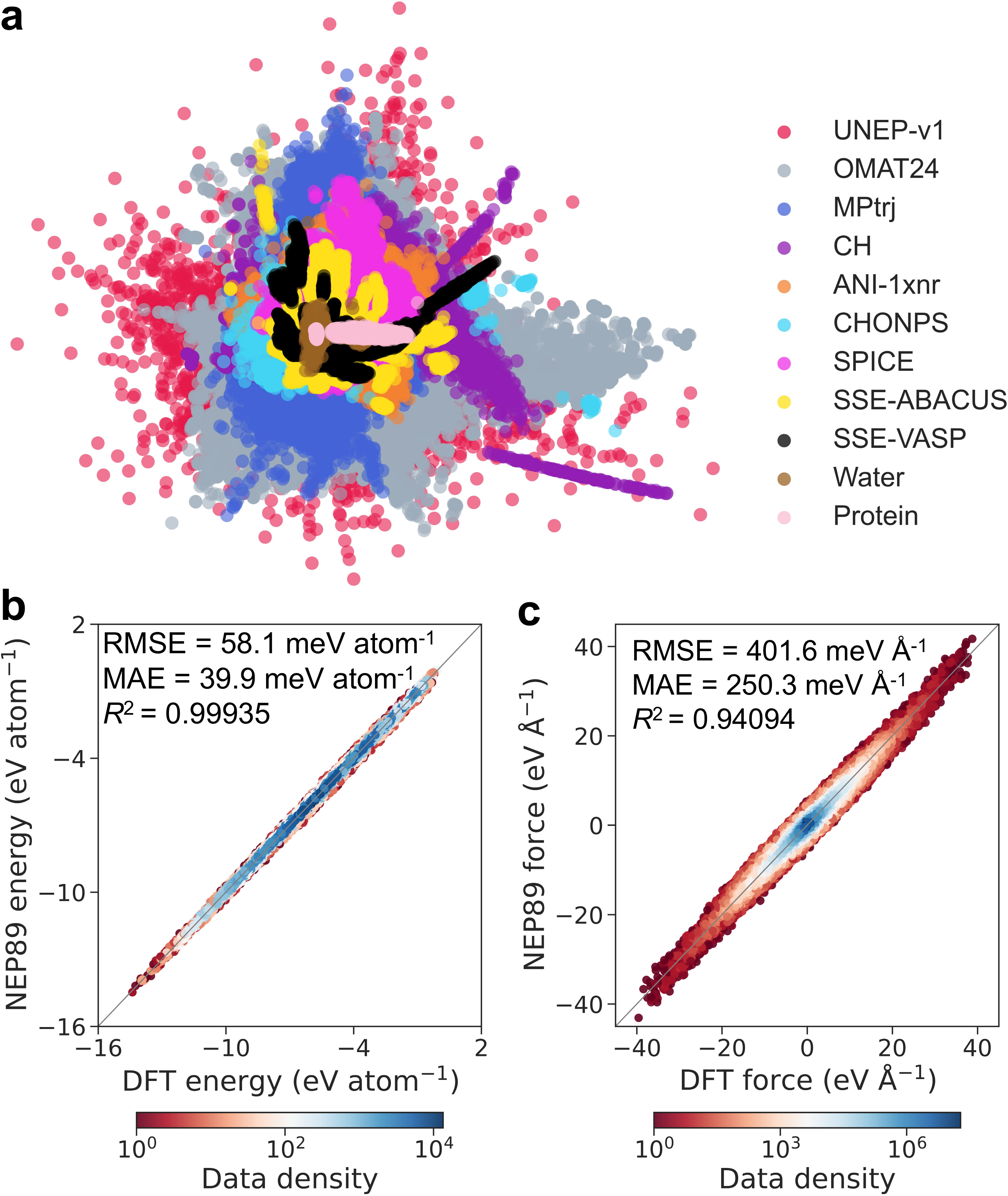}
\caption{
\textbf{Dataset composition for NEP89 and overall training accuracy.} 
\textbf{a}, Distribution of the NEP89 datasets in the reduced descriptor space spanned by the first two principal components.
The individual datasets were subsampled from the UNEP-v1 \cite{song2024general}, OMAT24 \cite{barrosoluque2024omat24}, MPtrj \cite{dengchgnet2023}, CH \cite{Ibragimova2025unifying}, ANI-1xnr \cite{zhang2024exploring}, SPICE \cite{eastman2023spice}, SSE-ABACUS \cite{wang2024pretrained}, SSE-VASP \cite{wang2024pretrained}, water \cite{zhai2023short}, and Protein \cite{unke2024biomolecular} datasets, or specifically prepared in this work (CHONPS).
\textbf{b},\textbf{c}, Parity plots for energy and force comparing NEP89 predictions with reference values (DFT or high-level quantum chemistry).
The color intensity visualizes the distribution and density of the NEP89 datasets.
}
\label{figure:pca_rmse}
\end{figure}

Through this iterative training and refinement process, we developed a diverse and balanced training dataset, which is visualized in the reduced descriptor space spanned by the two principal components in \autoref{figure:pca_rmse}a.
However, the model obtained in this process was exclusively trained on the OMAT24 dataset in terms of energy, introducing a possible bias toward this dataset.
To ensure balanced training across different datasets, we accordingly adjusted the reference energies for the other datasets.
These adjustments minimized differences between the model predictions and the shifted reference energies (see \autoref{snote:shifting} for details).
We excluded the reference energies for the MPtrj dataset from training, since they were incompatible with those in the OMAT24 dataset due to differing treatments of the Hubbard \(U\) correction. 
We then performed the energy adjustment for a few iterations and trained the final model, which we refer to as the NEP89 model.
The parity plot for the energies (\autoref{figure:pca_rmse}b) indicates that all the considered reference energies are indeed quite consistent after our adjustment. 
The parity plot for the forces (\autoref{figure:pca_rmse}c) also indicates a well-trained model for the combined dataset.
(\sautoref{fig:dataset-MAE-RMSE} shows error metrics for energy, force, and stress for the individual datasets.)


\begin{figure}[ht]
\centering
\includegraphics[width=1\columnwidth]{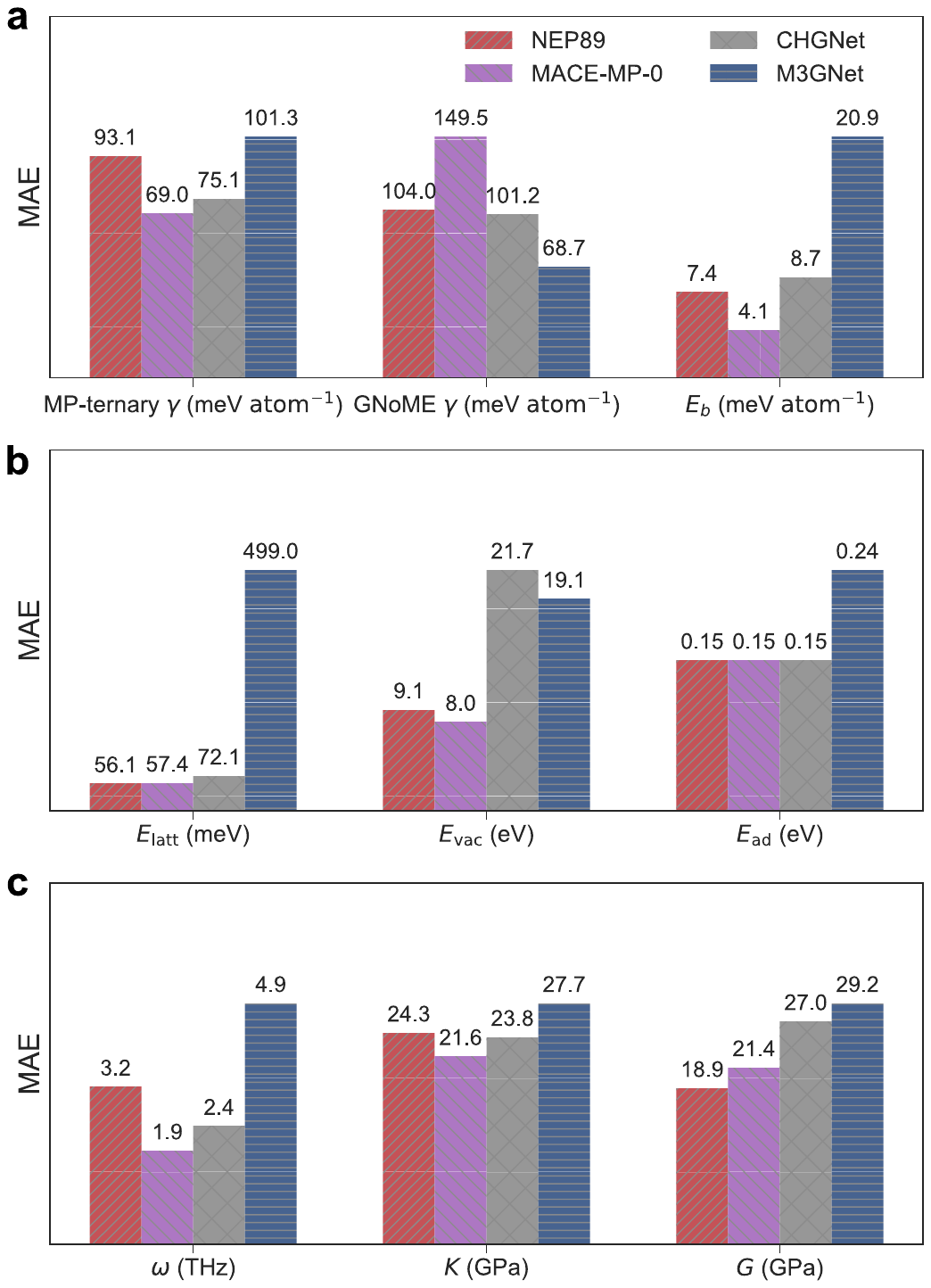}
\caption{\textbf{Various benchmarks on static properties for validating the foundation models.} Mean absolute errors (MAEs) of several foundation models, NEP89 (this work), MACE-MP-0 \cite{batatia2023foundation}, CHGNet \cite{dengchgnet2023}, and M3GNet \cite{chen2022nuniversal}, with respect to DFT reference data for nine evaluated properties. \textbf{a}, Formation energies ($\gamma$) for structures sampled from the Materials Project (MP-ternary) \cite{Jain2013commentary} and the GNoME dataset \cite{merchant2023scaling}, and binding energies ($E_{\rm b}$) of the S66 dimer set \cite{Jan2011S66} with DFT (PBE+D3) references \cite{batatia2023foundation}. \textbf{b}, Lattice energies ($E_{\rm latt}$) for the DMC-ICE13 dataset \cite{Della2022DMC-ICE13} with DFT references \cite{batatia2023foundation}, formation energies ($E_{\rm vac}$) of iron vacancy clusters of different sizes, and adhesion energy ($E_{\rm ad}$) of hydrogen atoms in iron nanopores with DFT references (PBE+D3). \textbf{c}, Highest phonon band frequencies ($\omega$) of 97 materials with DFT (PBE) references from the PhononDB database \cite{Togo2023First-principles}, and bulk ($K$) and shear ($G$) moduli with DFT references from Materials Project \cite{Jain2013commentary} covering more than 11,000 materials.}
\label{figure:benchmark}
\end{figure}

\vspace{0.5cm}

\noindent{\textbf{Benchmarks on static properties.}}
\autoref{figure:benchmark} presents a number of benchmarks on static properties predicted by NEP89 and various foundation models, including MACE-MP-0 (medium version) \cite{batatia2023foundation}, CHGNet \cite{dengchgnet2023} (version 0.3.0), and M3GNet \cite{chen2022nuniversal} (version 2021.2.8-DIRECT-PES); see `Benchmark calculations for static properties' section in Methods, \snautorefRange{snote:bench-formation}{snote:bench-Moduli}\unskip, and \sautorefRange{fig:bench-formation}{fig:bench-moduli}\unskip~for calculation details. 
Overall, NEP89 demonstrates comparable accuracy to other foundation models, based on mean absolute errors across nine evaluated properties relative to \gls{dft} data. 
For the formation energies ($\gamma$) of the ternary structures sampled from the Materials Project (MP-ternary) \cite{Jain2013commentary}, MACE-MP-0 and CHGNet are slightly more accurate, with NEP89 ranking third, despite not being trained on the MPtrj energies.
For the 2-component to 5-component structures from the GNoME dataset \cite{merchant2023scaling}, on which none of the models were trained, MACE-MP-0 shows the worst accuracy, while NEP89 performs very close to CHGNet as the second-best model. 
For the binding energies ($E_{\rm b}$) of the S66 dimer set \cite{Jan2011S66}, NEP89 ranks as the second-best model, even though it does not require a separate D3 correction as for the other models. 
NEP89 demonstrates superior accuracy for lattice energies ($E_{\rm latt}$) on the DMC-ICE13 dataset \cite{Della2022DMC-ICE13, batatia2023foundation}, second-best accuracy for formation energies ($E_{\rm vac}$) of iron vacancy clusters of different sizes, and comparable accuracy for adhesion energy ($E_{\rm ad}$) of hydrogen atoms in iron nanopores (see \autoref{snote:bench-vacancy-energy} for details on our iron vacancy cluster and nanopore datasets).
For the highest phonon band frequencies ($\omega$) for 97 materials from the PhononDB database \cite{Togo2023First-principles}, NEP89 achieves the third-best performance.
For more than \num{11000} materials from Materials Project \cite{Jain2013commentary}, NEP89 demonstrates comparable accuracy in predicting the bulk modulus ($K$) and the best accuracy in predicting the shear modulus ($G$).


\begin{figure*}
\centering
\includegraphics[width=1.8\columnwidth]{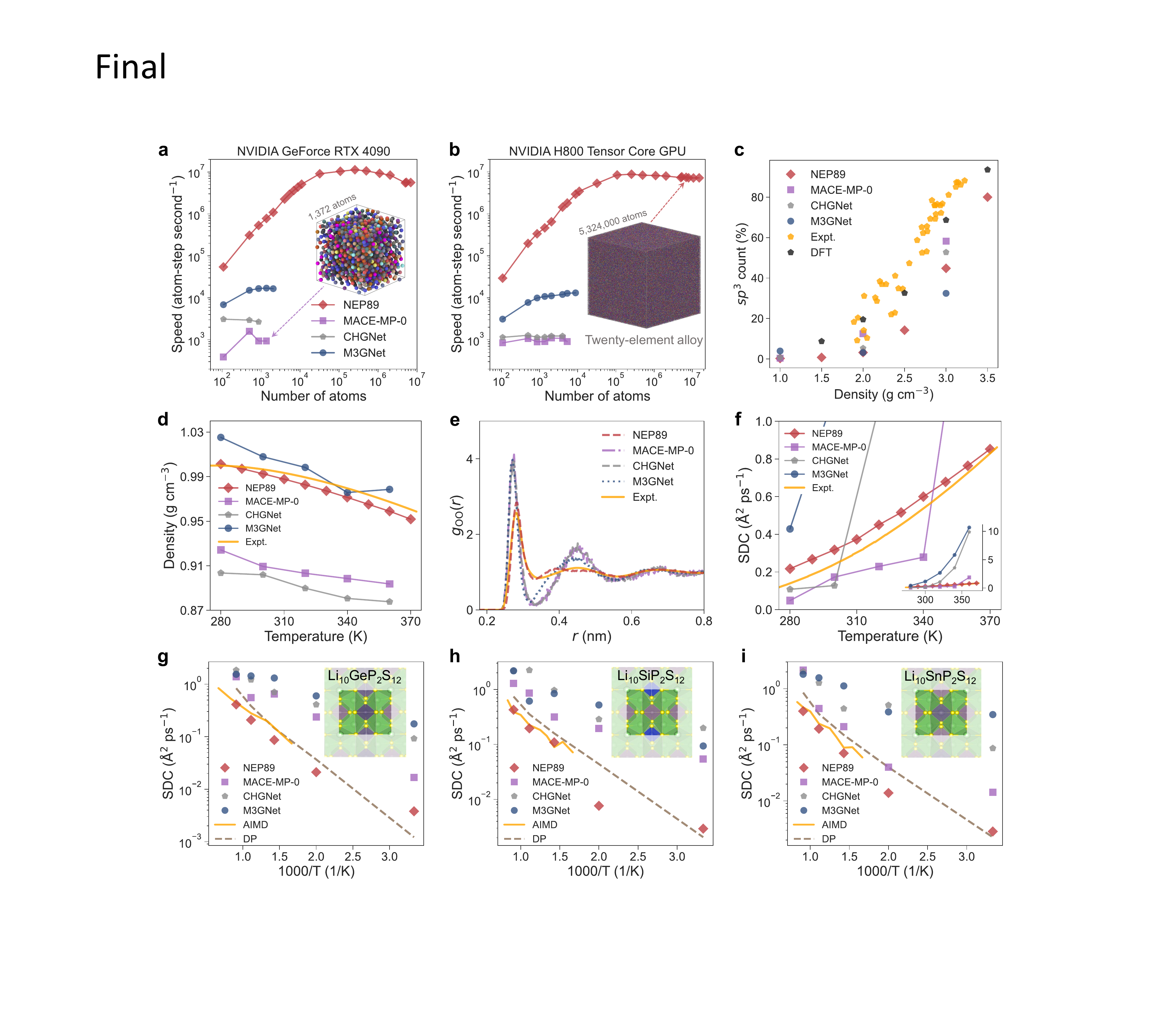}
\caption{
\textbf{Computational performance and benchmarks of dynamical properties}.
\textbf{a},\textbf{b}, Benchmark results for computational efficiency of NEP89 and various foundation models in twenty-element alloys.
Insets show representative atomic snapshots of the alloy model used in the \gls{md} simulations.
NEP89 achieves about a three-order-of-magnitude improvement in computational efficiency compared to the other foundation models.
\textbf{c}, Comparison of experimental and predicted bonding statistics of amorphous carbon. \textbf{d--f}, Experimental and predicted density (\textbf{d}), radial distribution function (\textbf{e}), and self-diffusion coefficient (SDC) (\textbf{f}) of liquid water simulated using NEP89 and other foundation models.
\textbf{g}--\textbf{i}, Experimental and calculated lithium-ion SDC in (\textbf{g}) Li$_{10}$GeP$_{2}$S$_{12}$, (\textbf{h}) Li$_{10}$SiP$_{2}$S$_{12}$, and (\textbf{i}) Li$_{10}$SnP$_{2}$S$_{12}$ solid-state electrolytes using NEP89, three other foundation models, a specialized deep potential (DP) model, and AIMD.
Overall, predictions from NEP89 show good agreement with experimental data or AIMD results.
}
\label{figure:MD-benchmark-performance}
\end{figure*}

\vspace{0.5cm}

\noindent{\textbf{Evaluation of computational performance.}}
Since the primary motivation for developing machine-learned interatomic potentials is to extend the spatiotemporal scales of ab initio calculations, computational efficiency, in terms of both memory usage and computational speed, is a crucial factor. 
As shown in \autoref{figure:MD-benchmark-performance}a--b, NEP89 consistently outperforms other foundation models in terms of speed and memory usage across a wide range of system sizes, using a 20-element high-entropy alloy as a representative example (see `Computational performance evaluation' section in Methods for details). 
On the 24-GB RTX 4090 and 80-GB H800 GPUs considered here, NEP89 can simulate up to \num{8e6} atoms at a speed of \num{5e6} atom-steps per second and \num{1.5e7} atoms at a speed of \num{7e6} atom-steps per second, respectively. 
Overall, NEP89 extends the accessible spatiotemporal scale of atomistic simulations by at least three orders of magnitude compared to other foundation models, while maintaining competitive accuracy for both static properties (as shown above) and dynamic properties (as we will discuss next).
In addition, NEP89 exhibits efficient multi-GPU scalability (\sautoref{fig:Speed_vs_gpu_count}). Benchmarks on water systems (\sautoref{fig:Compare_water_speed}) shows that NEP89 attains simulation speeds approaching those of empirical water potentials, substantially narrowing the performance gap between empirical potentials and foundation models.

\vspace{0.5cm}

\noindent{\textbf{Benchmarks of dynamical properties.}}
To further evaluate the performance of NEP89 in atomistic \gls{md} simulations, we compare dynamical properties of three typical systems with different structural phases, including amorphous carbon, liquid water, and three solid-state electrolytes, obtained using NEP89 and other foundation models (see `MD simulation of amorphous carbon', `MD simulation of water', and `Lithium-ion diffusion in solid-state electrolytes' sections in Methods for computational details).
\autoref{figure:MD-benchmark-performance}c shows the bonding statistics of amorphous carbon, comparing experimental data \cite{Wang2025Density} with predictions from NEP89 and other foundation models. 
NEP89 achieves comparable accuracy to other foundation models and aligns reasonably well with experimental results.
\autoref{figure:MD-benchmark-performance}d--f shows a comparative analysis of the density, radial distribution function, and self-diffusion coefficient of liquid water as obtained from experiments \cite{Xu2025NEPMBPOL}, NEP89, and other foundation models. 
The results show that the NEP89 predictions closely match experimental data and that NEP89 outperforms the other foundation models, particularly in reproducing the density (\autoref{figure:MD-benchmark-performance}d) and the oxygen-oxygen radial distribution function (\autoref{figure:MD-benchmark-performance}e).
As further illustrated in \sautoref{fig:compare-water-density}, NEP89 produces balanced water density predictions that, at certain temperatures, align more closely with experimental data than MB-pol, one of the most accurate models for water \cite{Xu2025NEPMBPOL}. While this behavior is not systematic across all conditions, it suggests the potential role of partial error compensation across heterogeneous data sources enabled by the reference-energy adjustment strategy (see Supplementary Sections \ref{snote:shifting} and \ref{snote:Reference-energy-adjustment}).
In \autoref{figure:MD-benchmark-performance}e, a minor discrepancy near $r = 0.4\,\mathrm{nm}$ is likely attributable to slight underfitting during the global optimization of NEP89; this discrepancy can be further reduced by fine-tuning on a water dataset (see \autoref{stable:dataset}), resulting in predictions that are in good agreement with experiment (see \sautoref{fig:rdf-ooo}). 
Next, \autoref{figure:MD-benchmark-performance}g--i present a comparative analysis of calculated versus experimental lithium-ion self-diffusion coefficients for three thiophosphate solid-state electrolytes: Li$_{10}$GeP$_{2}$S$_{12}$, Li$_{10}$SiP$_{2}$S$_{12}$, and Li$_{10}$SnP$_{2}$S$_{12}$, spanning temperatures from 300 to \qty{1100}{\kelvin}. 
Notably, NEP89 achieves excellent agreement with \gls{aimd} simulations and a specialized force field \cite{wang2024pretrained}, while outperforming other foundation models across the full temperature range. 


\begin{figure*}[!]
\centering
\includegraphics[width=1.75\columnwidth]{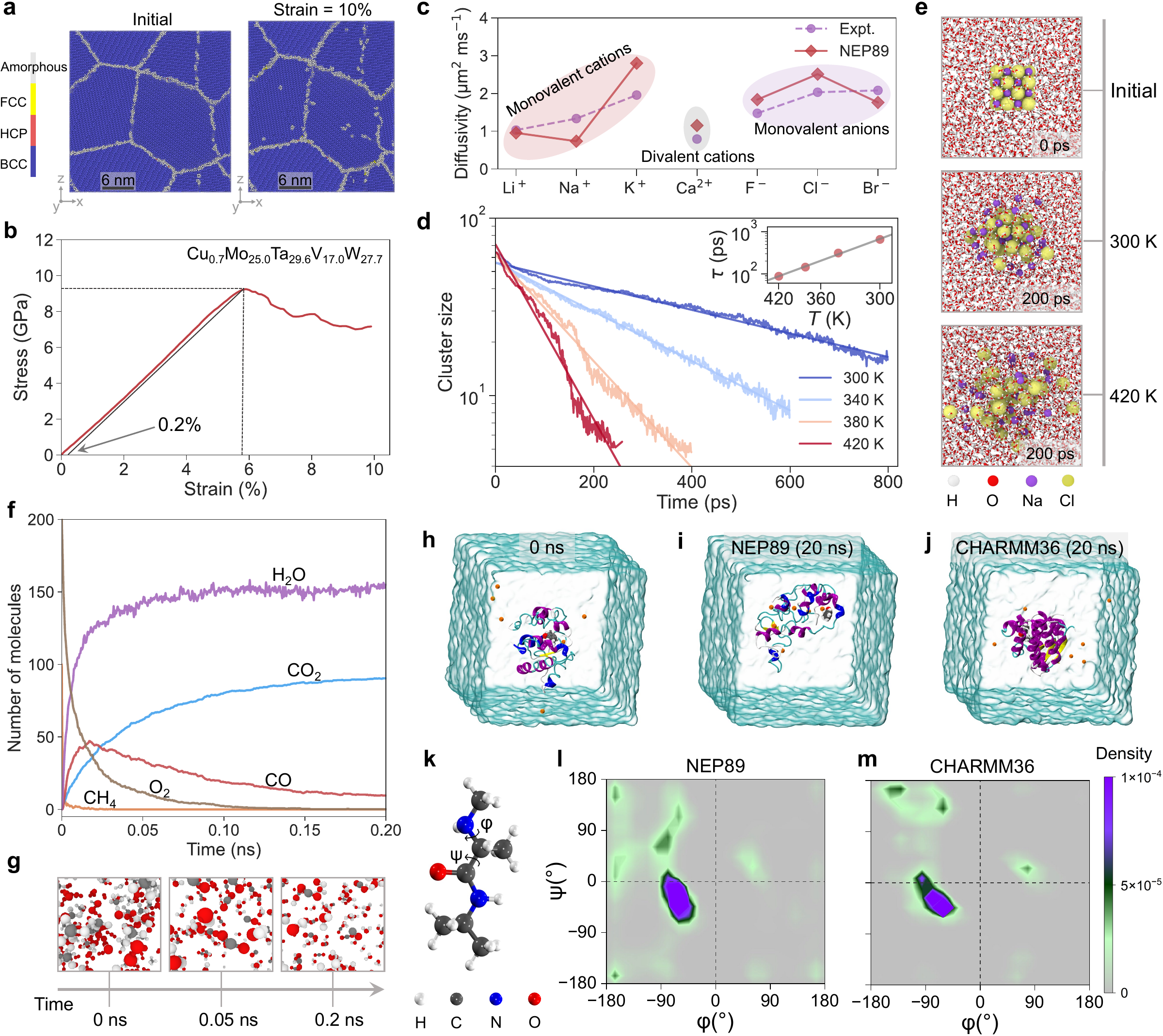}
\caption{
\textbf{NEP89 applications in inorganic and organic systems.}
\textbf{a}, Snapshots of atomic configurations of Cu$_{0.7}$Mo$_{25.0}$Ta$_{29.6}$V$_{17.0}$W$_{27.7}$ alloy at different stages during MD simulations (\num{1221240} atoms) using NEP89.
The left panel shows the initial configuration; the right panel shows the configuration under 10\% strain, colored by common neighbor analysis.
Dark blue, red, and yellow indicate atoms in body-centered cubic, hexagon close-packed, and face-centered cubic environments, respectively; off-white indicates atoms in amorphous environments.
\textbf{b}, Stress-strain curve and yield strength predicted using NEP89.
The dashed line indicates the 0.2\% offset method used to determine the yield strength. 
\textbf{c}, Diffusivity of various ions in water at \qty{300}{\kelvin} predicted by NEP89, compared to experimental values from Ref.~\cite{Lid06Ionic}. 
\textbf{d}, Dissolution of NaCl nanocrystal in water at different temperatures calculated by NEP89. The dissolution rate $\tau$ is extracted from the power law fits (straight lines) at each temperature, and collected in the inset, illustrating that NEP89 captures the expected Arrhenius behavior of the dissolution process.
\textbf{e}, Atomic snapshots of the spatial distribution of NaCl during dissolution in water at different temperatures.
\textbf{f}, Time evolution of O$_2$, CH$_4$, and the major products (CO, CO$_2$, H$_2$O) during a methane combustion simulation at 3000 K.
\textbf{g}, Atomic snapshots of initial reactants, intermediate species, and final products.
\textbf{h}, Initial protein-ligand model; \textbf{i},\textbf{j}, Configurations obtained after \qty{20}{\nano\second} of MD simulations using NEP89 (\textbf{i}) and CHARMM36 (\textbf{j}).
\textbf{k}, Schematic diagram of the backbone rotation angle ($\phi$ and $\psi$) used in the Ramachandran diagram.
\textbf{l},\textbf{m}, Ramachandran diagram showing the distribution of $\phi$ and $\psi$ angles from simulations using NEP89 (\textbf{l}) and CHARMM36 (\textbf{m}).
}
\label{figure:oob}
\end{figure*}

\vspace{0.5cm}

\noindent{\textbf{Out-of-the-box large-scale \gls{md} simulations.}}
After demonstrating the capabilities of NEP89 in dynamical simulations using small systems designed for benchmarking purposes, we next showcase the out-of-the-box applicability of NEP89 for large-scale atomistic \gls{md} simulations of inorganic and organic materials, which are typically impractical for other foundation models.
The examples include a million-atom-scale compression simulation of the complex multicomponent Cu$_{0.7}$Mo$_{25.0}$Ta$_{29.6}$V$_{17.0}$W$_{27.7}$ (CuMoTaVW) alloy, calculations of ion diffusivities in water, a study of rocksalt dissolution kinetics, a simulation of methane combustion, and a dynamical simulation of a protein-ligand system (see Methods Sections `MD simulation of CuMoTaVW alloy'--`MD simulation of protein dynamics' for computational details). 

For the CuMoTaVW alloy, common neighbor analysis (\autoref{figure:oob}a) shows that NEP89 successfully maintains structural stability during isothermal equilibration and compression. 
The polycrystalline body-centered cubic structure remains intact throughout the deformation process, with no observable phase transformation.  
The yield strength and yield strain calculated by NEP89 are \qty{9.2}{\giga\pascal} and \num{5.8}\%, respectively (\autoref{figure:oob}b), which closely reproduce the experimentally measured yield strength ($\sim$\qty{10.0}{\giga\pascal}) and yield strain ($\sim$\num{6}\%) \cite{Alvi2020Synthesis}.
This quantitative agreement illustrates the ability of NEP89 to accurately capture atomic interactions in complex multicomponent alloy systems.

Moreover, the diffusivities of various monovalent and divalent ions in water predicted by the NEP89 model (\autoref{figure:oob}c), and their corresponding activation energies (\sautoref{fig:ion-activation-energy}), are in good agreement with experiments \cite{Lid06Ionic, Lon1954Temperature}.
In combination with computational efficiency, this enables the prediction of, for example, the dissolution kinetics of a small salt crystallite in water (\autoref{figure:oob}d,e, \sautoref{fig:NaCl-dissolution-comparison}).
The expected Arrhenius behavior of the dissolution process is captured out of the box with NEP89 (\autoref{figure:oob}d, inset).

The methane combustion simulation at 3000~K (\autoref{figure:oob}f) shows that NEP89 generates primary reaction products and product distributions that are remarkably similar to those obtained with the ANI-1xnr model \cite{zhang2024exploring}, indicating the capability of NEP89 to capture the underlying reaction physics and mechanisms. 
Atomic snapshots during combustion (\autoref{figure:oob}g) confirm that NEP89 also provides reasonable predictions of the reaction chemistry. 
Furthermore, NEP89 reproduces the near-exponential decay profiles of CH$_4$ and O$_2$ concentrations originally observed in ANI-1xnr simulations \cite{zhang2024exploring}, demonstrating its robust performance in modeling organic reactions. While the overall reaction pathways predicted by NEP89 and ANI-1xnr \cite{zhang2024exploring} are consistent, NEP89 yields faster combustion kinetics. 
At 3000~K, the calculated reaction rate constant $k$  is $2.90 \times 10^{15}\,\mathrm{cm^3\, mol^{-1}\, s^{-1}}$ for NEP89, approximately six times larger than the value obtained with ANI-1xnr ($4.99 \times 10^{14}\,\mathrm{cm^3\, mol^{-1}\, s^{-1}}$). 
As discussed in \autoref{snote:methane_combustion_discussion} and \sautorefRange{fig:Temperature_for_methane_combustion}{fig:Activation_energy_for_methane_combustion}\unskip, this kinetic acceleration is likely associated with a lower C--H bond dissociation energy for NEP89 relative to the ANI-1xnr reference, leading to a reduced effective activation barrier for methane combustion.

For the protein-ligand case, we simulate the dynamics of a protein (T4 lysozyme L99A/M102Q, 3HTB) in complex with a ligand (2-propylphenol, JZ4) using NEP89 and compare the results with those obtained using the specialized protein-optimized CHARMM36 force field. 
As shown in \autoref{figure:oob}h--j, both models achieve the stable binding of the ligand JZ4 to the protein 3HTB in long-term simulations. 
The binding energy calculated by NEP89 is \qty{-1.64}{\electronvolt}, while the reference value from CHARMM36 is \qty{-1.24}{\electronvolt}, suggesting that NEP89 can reasonably describe the protein-ligand interactions. 
In addition, we calculated the Ramachandran diagram of the protein obtained by the two potentials, which can be used to analyze whether the conformation of the protein model conforms to the rules of stereochemistry. 
As shown in the \autoref{figure:oob}l,m, the total allowed areas calculated by NEP89 and CHARMM36 are \num{63.1}\% and \num{90.47}\%, respectively. 
While NEP89 shows a lower proportion of allowed areas compared to the specialized CHARMM36 force field, both Ramachandran diagrams \cite{Ramachandran1963JMB} show that the highest probability of protein configuration is around the (\qty{-60}{\degree},
\qty{-30}{\degree}) area, with a higher proportion of $\alpha$ area (bottom left area in \autoref{figure:oob}l,m) than that of $\beta$ area (top left area in \autoref{figure:oob}l,m). 
These observations indicate that NEP89 captures protein conformational preferences at a qualitative level, but its ability to accurately reproduce the Gibbs free energy landscape of protein-ligand systems remains limited (\sautoref{fig:Protein-ligand-FES}). This limitation likely reflects incomplete coverage of protein-ligand configurations in the training data and/or the absence of an explicit treatment of long-range electrostatic interactions in the current NEP89 framework, highlighting a key direction for future development (see \hyperref[discussion]{Discussion} section).
We also calculated the number of protein-water hydrogen bonds, obtaining \num{220} for NEP89 and \num{375} for CHARMM36, which are of the same order of magnitude. 
Overall, these results demonstrate that NEP89 can qualitatively capture certain aspects of protein-ligand interactions and protein structure, even though it is not specifically optimized for these protein systems. 
Further improvements are expected through fine-tuning, which we will introduce next.


\begin{figure*}
\centering
\includegraphics[width=1.9\columnwidth]{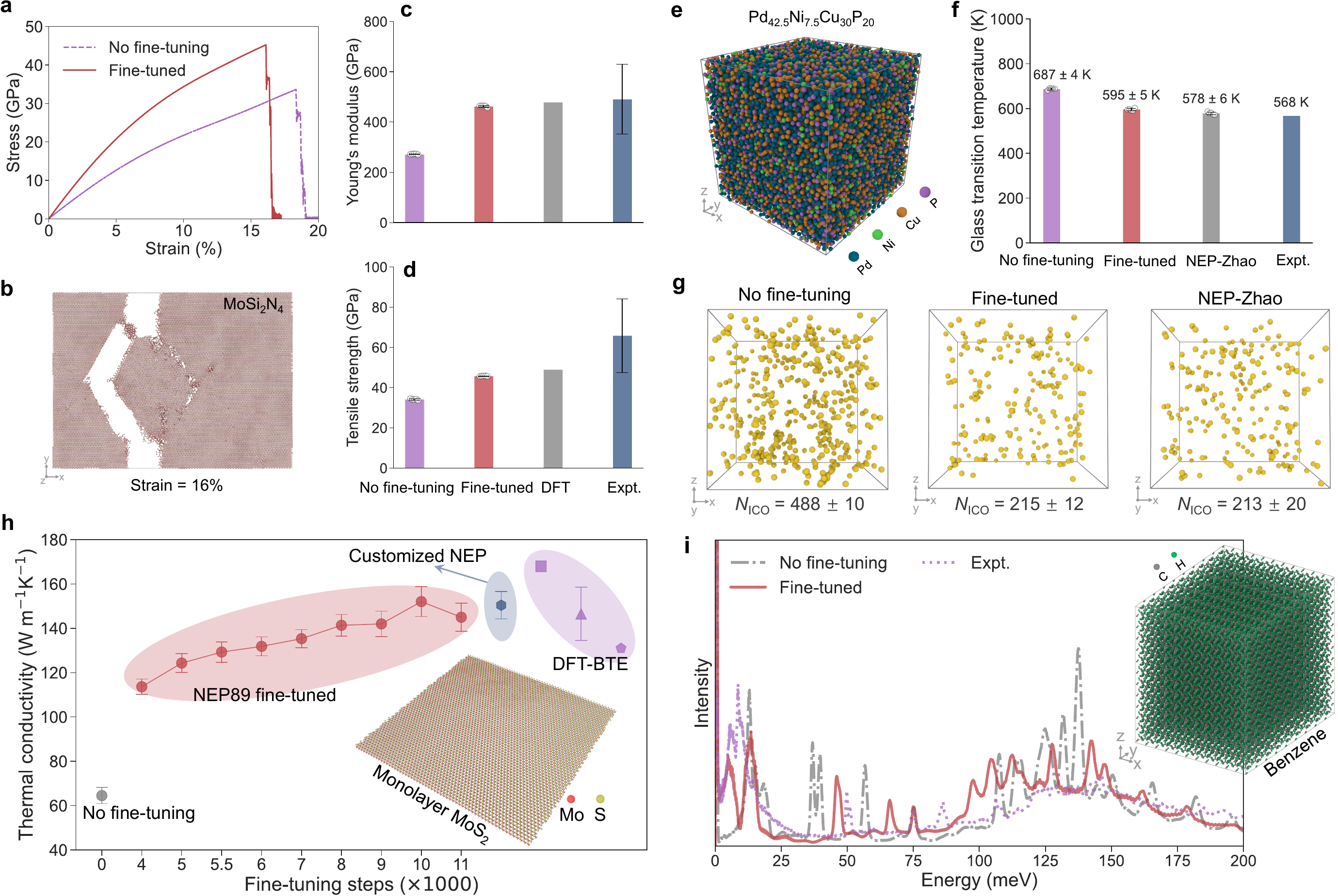}
\caption{
\textbf{Example applications enabled by fine-tuning NEP89.}
\textbf{a--d}, Mechanical properties of monolayer MoSi$_2$N$_4$.
\textbf{a}, Stress-strain curves averaged over five independent \gls{md} simulations at 300 K.
\textbf{b}, Atomic snapshot at 16\% strain after fine-tuning, showing fracture.
\textbf{c},\textbf{d}, Comparison of Young's modulus and tensile strength before and after fine-tuning. Bar heights (here and in panel \textbf{f}) represent the mean of five independent simulations, error bars indicate the standard deviation (s.d.), and individual data points are shown as white circles with black outlines to indicate the data distribution. DFT and experimental data were taken from \cite{hong2020chemical}.
\textbf{e--g}, Thermal and structural properties of Pd$_{42.5}$Cu$_{30}$Ni$_{7.5}$P$_{20}$ metallic glass. 
\textbf{e}, Atomic snapshots of the quenched metallic glass structure generated by a fine-tuned model based on NEP89.
\textbf{f},
Comparison of $T_{\mathrm{g}}$ values before and after fine-tuning, with results obtained using the NEP model by Zhao \textit{et al.} \cite{zhao2023development} and experimental data \cite{Osami2007Thermodynamic} (no uncertainty reported). 
\textbf{g}, Spatial distribution and number ($N_{\mathrm{ICO}}$) of icosahedral structures (colored in gold for central atoms) in a quenched glass predicted by different potentials. Numerical values in the panels denote the mean $\pm$ s.d. from five independent simulations. 
\textbf{h}, Thermal conductivity of monolayer \ce{MoS2}.
Results are compared with predictions from a customized NEP model \cite{jiang2025accurate} and Boltzmann transport equation (BTE) calculations using \gls{dft}-derived force constants \cite{bao2022Bilateral, Gu2016Layer, Cepellotti2015Phonon}. Thermal conductivity values are presented as mean $\pm$ s.e.m. (standard error of the mean) across ten independent simulations. 
\textbf{i}, Simulated inelastic neutron scattering spectra for crystalline benzene at \qty{127}{\kelvin} using NEP89 and a fine-tuned model, compared with experimental data \cite{lindgren2025predicting}.}
\label{figure:finetune}
\end{figure*}

\vspace{0.5cm}

\noindent{\textbf{Fine-tuning applications.}}
For some specific applications, the accuracy of NEP89 may not be sufficient. 
In such cases, one can still fully leverage NEP89 to quickly obtain a reliable model through a simple but effective active learning process. 
We can first use NEP89 to perform atomistic simulations at our target conditions, generating abundant trajectories. 
By sampling structures from the trajectories and performing single-point \gls{dft} calculations, we can create a training dataset to fine-tune NEP89 into a special-purpose model for the material and application in question. 
This process is considerably more efficient than sampling structures using \gls{aimd}. 
In the following, we present four case studies, including the mechanical properties of the 2D material MoSi$_2$N$_4$, thermal and structural properties of Pd$_{42.5}$Cu$_{30}$Ni$_{7.5}$P$_{20}$ metallic glass, thermal conductivity of monolayer \ce{MoS2}, and simulated inelastic neutron scattering spectra for crystalline benzene (see Methods Sections `Fine-tuning for \ce{MoSi2N4} and MD simulations'--`Fine-tuning for crystalline benzene and inelastic neutron scattering calculations' for fine-tuning and simulation details).

MoSi$_2$N$_4$ has been successfully synthesized via chemical vapor deposition in recent years and exhibited promising properties \cite{hong2020chemical}.
Here, we showcase the fine-tuning of NEP89 specifically for MoSi$_2$N$_4$ with only 100 additional configurations. 
The out-of-the-box predictions by NEP89 for the fracture strength and the Young's modulus are \qty{272+-1}{\giga\pascal} and \qty{34.2+-0.4}{\giga\pascal}, respectively.
The fine-tuned model on the other hand yields \qty{463+-4}{\giga\pascal} and \qty{45.7+-0.2}{\giga\pascal}, respectively, in substantially better agreement with the experimental values of \qty{491+-139}{\giga\pascal} and \qty{66+-18}{\giga\pascal}, respectively (\autoref{figure:finetune}a--d).

The \ce{Pd_{42.5}Ni_{7.5}Cu_{30}P_{20}} alloy is among the best-known bulk metallic glasses in terms of glass-forming ability \cite{Shinya2022Relationship}.
Here, we demonstrate that fine-tuning NEP89 with only around 200 additional configurations yields excellent performance.
After fine-tuning, the computed glass transition temperature $T_\mathrm{g}$ matches closely with both the prediction from a specialized \gls{nep} model \cite{zhao2023development} and experimental data \cite{Osami2007Thermodynamic}, as shown in \autoref{figure:finetune}e--f.
Moreover, the icosahedral coordination, which describes the short-range structure, is quantitatively improved after fine-tuning, achieving results comparable to a specialized NEP model (\autoref{figure:finetune}g) \cite{zhao2023development}.

Next, we demonstrate that fine-tuning the NEP89 model with only 104 additional configurations enables accurate prediction of the thermal conductivity of the monolayer \ce{MoS2}. 
\autoref{figure:finetune}h shows the evolution of the predicted thermal conductivity using the homogeneous nonequilibrium \gls{md} method \cite{fan2019Homogeneous} as a function of the number of fine-tuning steps.
As the latter increases, the predicted thermal conductivity gradually rises and eventually converges to the reference values. 
Notably, after \num{10000} fine-tuning steps, the model yields thermal conductivity values that closely match those reported using a customized NEP potential \cite{jiang2025accurate} as well as the Boltzmann transport equation approach using force constants obtained via \gls{dft} calculations \cite{bao2022Bilateral, Gu2016Layer, Cepellotti2015Phonon}.

Finally, we showcase the prediction of scattering experiments using NEP89 and a fine-tuned version trained on \num{119} additional structures obtained via active learning.
While the simulated spectrum from NEP89 is overall in qualitative agreement with experimental data \cite{lindgren2025predicting} (\autoref{figure:finetune}i), the peaks in the spectrum, which correspond to phonon modes, deviate from their experimental counterpart by about \qty{10}{\milli\electronvolt}.
After fine-tuning, the predicted mode frequencies agree even better with the experimental spectrum, barring a general red shift of the broad feature at \qty{100}{\milli\electronvolt}, which can be attributed to the vdW-DF-cx exchange correlation functional used for the \gls{dft} calculations of the structures for fine-tuning.
Moreover, the relative peak intensities are in better agreement with experiments as well.

To assess the robustness of NEP89 under varied fine-tuning tasks, we performed three independent runs with different random seeds across four representative systems (see \sautoref{fig:Fine_tuned_test}). The resulting predictions remain highly consistent, indicating the stability and reliability of the fine-tuning scheme.

\section{Discussion}
\label{discussion}
In summary, we have introduced NEP89, a highly efficient and accurate foundation model for large-scale atomistic simulations of both inorganic and organic materials across 89 elements. 
To the best of our knowledge, no existing machine-learned interatomic potential model offers this combination of elemental and chemical breadth with inference speeds comparable to typical empirical potentials. 
NEP89 represents a key milestone toward the long-standing goal of a unified computational framework that delivers near-ab-initio accuracy and near-empirical-potential computational efficiency across diverse material systems.

To achieve this, we employed an iterative training strategy incorporating descriptor-space subsampling, targeted data curation, and reference energy adjustments. 
The resulting training dataset strikes a balance between diversity and compactness, supporting reliable predictions across a wide range of static and dynamic properties, and remains flexible for future extension.

Comprehensive benchmarks highlight the competitive performance of NEP89 in static properties, including formation energies, binding energies, adhesion energies, lattice energies, vacancy formation energies, phonon frequencies, and elastic moduli. 
NEP89 also demonstrates robust predictions in dynamic properties, such as bonding statistics in amorphous carbon, thermodynamic properties of liquid water, and lithium-ion transport in solid-state electrolytes. 
NEP89 often matches or outperforms representative foundation models, exhibiting good agreement with experimental results and robustness across diverse applications.

Beyond benchmarks on small-scale systems, we demonstrated the out-of-the-box capability of NEP89 in large-scale \gls{md} simulations, enabled by its computational efficiency, which is comparable to that of empirical potentials.
From large-scale simulations of complex alloy systems and solution chemistry to modeling methane combustion and protein-ligand dynamics, NEP89 showcases versatility across a wide range of inorganic and organic systems. 
Notably, NEP89 achieves quantitative agreement with experiments on yield strength and ion diffusivities in water, reproduces the expected Arrhenius behavior in rocksalt dissolution kinetics, and matches specialized model predictions for combustion reaction dynamics. In addition, it qualitatively captures certain structural features and interactions in protein systems.
These results highlight the ability of NEP89 to model complex systems with high fidelity, even without domain-specific optimization.

The fine-tuning capabilities of NEP89 provide a practical pathway for rapidly developing specialized models from limited training data. 
Researchers can efficiently customize NEP89 for specific materials or conditions, as demonstrated in case studies on \ce{MoSi2N4} mechanical properties, thermal and structural properties of \ce{Pd_{42.5}Cu_{30}Ni_{7.5}P_{20}} metallic glass, thermal conductivity of monolayer \ce{MoS2}, and neutron scattering simulations of crystalline benzene. 
Fine-tuning yields substantially improved agreement with results from quantum-mechanical calculations and experiments, demonstrating the utility of NEP89 as both a versatile out-of-the-box and an adaptable foundation model.

While NEP89 offers high computational efficiency and broad applicability, it is important to acknowledge its current limitations and clarify the scope in which its predictions are most reliable.
The current NEP89 foundation model does not explicitly account for long-range interactions, which may affect quantitative accuracy in ionic, aqueous, and large biomolecular systems. 
Although NEP89 qualitatively captures protein conformational preferences, accurately reproducing detailed free energy landscapes for complex biomolecules remains challenging. This limitation likely reflects both the absence of an explicit treatment of long-range electrostatic interactions and incomplete coverage of protein-ligand configurations in the currently available training data. Further efforts to construct more comprehensive training datasets will therefore be important, especially for organic and large-molecule systems.
In addition, the current NEP89 foundation model does not include descriptors associated with variable atomic charges or magnetic degrees of freedom, which may restrict its current applicability to systems with substantial charge transfer, strong magnetism, or coupling to external electric or magnetic fields.

These limitations reflect the scope of the present NEP89 implementation rather than its fundamental constraints. 
Importantly, they identify clear directions for systematic extensions in future work. 
As a concrete example, we have recently developed and released a charge-extended NEP method \cite{fan2026qnep} in the GPUMD package, which introduces environment-dependent atomic charges and explicit long-range electrostatic interactions, demonstrating that such learnable physics can be integrated in a consistent and scalable manner. 
Extensions to include magnetic degrees of freedom or other physics-informed descriptors are natural next steps. 
The architectural extensibility, combined with the near-empirical-potential efficiency of the NEP89 foundation model, provides a solid foundation for the practical development of more specialized and physics-informed models.

While the demonstrated out-of-the-box and fine-tuning applications are not exhaustive, they provide a solid foundation for future explorations in various application domains.
Such efforts will further characterize the capabilities of NEP89 and identify potential areas for improvement. 
We expect that with ongoing extension and refinement, the out-of-the-box performance of NEP89 will continue to improve.
Additionally, we anticipate the development of numerous domain-specific, medium-sized models fine-tuned from NEP89, enabling tailored solutions for a wide range of applications.

\section{Methods}

\noindent{\textbf{Benchmark calculations for static properties.}
Static properties were calculated with the help of the \textsc{gpumd-wizard} \cite{liu2024wizard}, \textsc{ase} \cite{Larsen2017atomic}, \textsc{calorine} \cite{Lindgren2024calorine}, \textsc{phonopy} \cite{Togo2023First-principles}, and \textsc{MatCalc} \cite{matcalc2024} packages.
Detailed computational procedures can be found in \snautorefRange{snote:bench-formation}{snote:bench-Moduli}.
Atomic snapshots were generated using \textsc{ovito} \cite{Stukowski2010Visualization}.\\

\noindent{\textbf{Computational performance evaluation.}}
We performed \gls{md} simulations with \num{1000} steps on a 20-element equiatomic alloy composed of Mn, Cr, Fe, Co, Ni, Cu, Ag, W, Mo, Nb, Al, Cd, Sn, Pb, Bi, Zn, Ge, Si, Sb, and Mg \cite{Cantor2004Microstructural} to evaluate the computational efficiency of different foundation models. 
The initial structure of the alloy is face-centered cubic with randomly distributed atoms and a lattice constant of \qty{3.7}{\angstrom}.
The Nos{\'e}-Hoover chain thermostat \cite{Martyna1992nose} with a coupling time of 0.1 ps was used to control the temperature.
Simulations were conducted on systems ranging from the smallest unit cell to progressively larger supercells until memory limitations were reached.
Benchmarks were performed using GPUMD (version 4.0) \cite{xu2025mega} for NEP89, LAMMPS (version 28 Mar 2023) for MACE-MP-0, and LAMMPS (version 2 Aug 2023) \cite{Thompson2022lammps} for both CHGNet and M3GNet.\\

\noindent{\textbf{MD simulation of amorphous carbon.}}
A system with \num{512} atoms in a diamond structure with a given density undergoes an initial rapid melting process at \qty{9000}{\kelvin} for \qty{25}{\pico\second}, followed by a relaxation at \qty{5000}{\kelvin} for \qty{5}{\pico\second}. 
This is followed by a rapid quenching from \qty{5000}{\kelvin} down to \qty{1000}{\kelvin} in \qty{0.4}{\pico\second}, with further relaxation stages at \qty{1000}{\kelvin} for \qty{5}{\pico\second} and subsequently at \qty{300}{\kelvin} for \qty{5}{\pico\second}. 
The time step for integration is \qty{0.5}{\femto\second}.
The Bussi-Donadio-Parrinello thermostat \cite{Bussi2007canonical} with a coupling time of 0.05 ps was used to control the temperature.\\

\noindent{\textbf{MD simulation of water.}}
A system with \num{699} atoms with an initial density of \qty{1.0}{\gram\per\centi\meter\cubed} was used in the \gls{md} simulations of water.
The system was equilibrated in the isothermal-isobaric ensemble at a given temperature and a pressure of \qty{1}{\bar} for \qty{100}{\pico\second} and then in the canonical ensemble for \qty{50}{\pico\second}, followed by a production run in the microcanonical ensemble for \qty{25}{\pico\second}.
The time step for integration was \qty{0.5}{\femto\second}.
The Bussi-Donadio-Parrinello thermostat \cite{Bussi2007canonical} with a coupling time of 0.1 ps and the stochastic cell rescaling barostat \cite{Bernetti2020pressure} with a coupling time of 1 ps and an elastic modulus of 2 GPa, were used to control the temperature and isotropic pressure, respectively.
For each temperature, five independent runs were performed.\\

\noindent{\textbf{Lithium-ion diffusion in solid-state electrolytes.}} 
\gls{md} simulations were performed on systems with \num{900} atoms for each of the three thiophosphate solid-state electrolytes: \ce{Li_{10}GeP2S_{12}}, \ce{Li_{10}SiP2S_{12}}, and \ce{Li_{10}SnP2S_{12}}.
Each system was equilibrated in the isothermal-isobaric ensemble at a given temperature and a pressure of \qty{1}{\bar} for \qty{50}{\pico\second} and then in the canonical ensemble for \qty{100}{\pico\second}, followed by a production run in the microcanonical ensemble for \qty{200}{\pico\second}.
The time step for integration is \qty{0.5}{\femto\second}.
The Bussi-Donadio-Parrinello thermostat \cite{Bussi2007canonical} with a coupling time of 0.1 ps and the stochastic cell rescaling barostat \cite{Bernetti2020pressure} with a coupling time of 1 ps and an elastic modulus of 100 GPa, were used to control the temperature and diagonal pressure, respectively.
For each material and each temperature, three independent runs were performed.\\

\noindent{\textbf{MD simulation of CuMoTaVW alloy.}}
The initial configuration comprises a body-centered cubic polycrystalline structure with 10 grains (average grain size about \qty{12}{\nano\meter}), forming a \qtyproduct{27x27x27}{\nano\meter} simulation box containing \num{1221240} atoms.
Using a hybrid MC/MD method, we performed 4 million Monte Carlo attempts to obtain a low-energy configuration.
The system was then equilibrated at \qty{300}{\kelvin} for \qty{100}{\pico\second} under the isothermal-isobaric ensemble. Subsequently, uniaxial compression was applied along the $x$-axis at an engineering strain rate of \qty{2e8}{\per\second}, up to a maximum strain of 10\%.
The Bussi-Donadio-Parrinello thermostat \cite{Bussi2007canonical} with a coupling time of 0.1 ps and the stochastic cell rescaling barostat \cite{Bernetti2020pressure} with a coupling time of 1 ps and an elastic modulus of 100 GPa, were used to control the temperature and lateral pressure, respectively.
Yield strength is determined using the commonly adopted 0.2\% offset method, which identifies the point where the stress-strain curve intersects a line parallel to the linear (elastic) region, offset by 0.2\% strain.\\

\noindent{\textbf{Ion diffusivity in water.}}
To compute the diffusivity of \ce{Li+}, \ce{Na+}, \ce{K+}, \ce{Ca^{2+}}, \ce{F-}, \ce{Cl-}, and \ce{Br-}, \num{14} (\num{7}) monovalent (divalent) ions along with the corresponding number of \ce{OH-} or \ce{H+} ions were randomly mixed with \num{14000} \ce{H2O} molecules at an initial density of \qty{1.0}{\gram\per\centi\meter\cubed}.
The systems were equilibrated in the isothermal-isobaric ensemble at a given temperature and a pressure of \qty{0}{\bar} for \qty{200}{\pico\second} using the Bussi-Donadio-Parrinello thermostat \cite{Bussi2007canonical} with a coupling time of \qty{0.02}{\pico\second} to control the temperature and the stochastic cell rescaling barostat \cite{Bernetti2020pressure} with a coupling time of \qty{0.08}{\pico\second} to control the isotropic pressure, followed by a production run in the microcanonical ensemble for \qty{400}{\pico\second} using a time step of \qty{0.2}{\femto\second}.
During the latter run, the mean-square displacement was recorded, from which the diffusivity was calculated.\\

\noindent{\textbf{Dissolution kinetics of NaCl in water.}}
A nanocrystal comprising 32 \ce{NaCl} units was placed in a cell surrounded by \num{14000} \ce{H2O} molecules at a density of \qty{1.0}{\gram\per\centi\meter\cubed}.
The system was then evolved in the canonical ensemble at a given temperature for up to \qty{2}{\nano\second} using a time step of \qty{0.2}{\femto\second}.
The Langevin thermostat \cite{Bussi2007Langevin} with a coupling time of \qty{0.02}{\pico\second} was used to control the temperature.
We confirmed that the dissolution kinetics remain consistent when using a larger time step of \qty{0.5}{\femto\second} (see \sautoref{fig:salt-dissolution-timestep-comparison}).
The size of the largest cluster was tracked over time, where the clusters were defined by connectivity using a maximum Na--Cl bond cutoff of \qty{4}{\angstrom} calculated using \textsc{ovito} \cite{Stukowski2010Visualization}.
\num{10} runs were carried out at each temperature, and the crystal size was averaged over these runs.
The dissolution rate for each temperature was determined by fitting the time dependence of the cluster size to a simple exponential law.\\

\noindent{\textbf{MD simulation of methane combustion.}}
\ce{CH4} and \ce{O2} molecules were mixed into a reactive system containing 100 \ce{CH4} and 200 \ce{O2} molecules, enclosed in a cubic cell of dimensions \qtyproduct{37.5x37.5x37.5}{\angstrom}, corresponding to a density of \qty{0.25}{\gram\per\centi\meter\cubed}. 
Each trajectory began with isothermal equilibration at 300~K; subsequently, the temperature was ramped to trigger combustion.  
The integration was performed with a timestep of \qty{0.1}{\femto\second} to align with the settings used for the ANI-1xnr model \cite{zhang2024exploring}.
As shown in \sautoref{fig:Timestep_for_methane_combustion}, the evolution of combustion products remains consistent when using a larger timestep of \qty{0.5}{\femto\second}. 
The Langevin thermostat \cite{Bussi2007Langevin} with a coupling time of 0.01 ps was used to control the temperature.
The time-dependent combustion products are averaged over five independent MD trajectories.\\

\noindent{\textbf{MD simulation of protein dynamics.}}
Dynamics of a protein (T4 lysozyme L99A/M102Q, 3HTB) in complex with a ligand (2-propylphenol, JZ4) was simulated using both GROMACS \cite{Berendsen1995GROMACS} with the CHARMM36 \cite{charmm36} (for 3HTB) and CGenFF \cite{cgenff} (for JZ4), as well as GPUMD with NEP89. 
The system consists of \num{33917} atoms, and the simulation was performed for \qty{20}{\nano\second} in the isothermal-isobaric ensemble using a timestep of \qty{0.5}{\femto\second}. 
The last \qty{15}{\nano\second} were used for data analysis. 
The Bussi-Donadio-Parrinello thermostat \cite{Bussi2007canonical} with a coupling time of 0.05 ps and the stochastic cell rescaling barostat \cite{Bernetti2020pressure} with a coupling time of 0.5 ps and an elastic modulus of 2.2 GPa, were used to control the temperature and isotropic pressure, respectively.\\

\noindent{\textbf{Fine-tuning for \ce{MoSi2N4} and MD simulations.}}
Using the NEP89 model, we conducted tensile loading \gls{md} simulations of \ce{MoSi2N4} using an orthorhombic cell with 210 atoms at \qty{300}{\kelvin} and evenly sampled 100 structures from the trajectory. 
Then we performed single-point \gls{dft}-D3(BJ) calculations using the \textsc{vasp} package (version 6.5.1) \cite{blochl1994projector, kresse1996efficient} with the projected augmented wave method and the generalized gradient approximation, along with the Perdew-Burke-Ernzerhof (PBE) functional \cite{perdew1996generalized}.
A vacuum layer of \qty{20}{\angstrom} was used to model the two-dimensional systems.
The plane-wave truncation energy was set to \qty{500}{\electronvolt}, and a $\Gamma$-centered $k$-point grid with \numproduct{2x2x1} divisions was used.
Using the new training dataset, we trained 5000 steps to obtain the fine-tuned model for \ce{MoSi2N4}.
Using both the NEP89 and fine-tuned models, we performed tensile loading simulations for the monolayer \ce{MoSi2N4}, using an orthorhombic cell with \num{44800} atoms.
The engineering strain rate in the herringbone direction was 
\qty{8.6e7}{\per\second}.
The thickness of the monolayer was taken as \qty{10.7}{\angstrom} in calculating the volume, following earlier literature \cite{hong2020chemical}. 
The Bussi-Donadio-Parrinello thermostat \cite{Bussi2007canonical} with a coupling time of 0.1 ps and the stochastic cell rescaling barostat \cite{Bernetti2020pressure} with a coupling time of 1 ps and an elastic modulus of 500 GPa, were used to control the temperature and lateral pressure, respectively.
We conducted five independent tensile loading simulations and averaged them to report the stress-strain curves and standard deviations.\\

\noindent{\textbf{Fine-tuning for PdCuNiP metallic glass and MD simulations.}}
To perform fine-tuning for the \ce{Pd_{42.5}Cu_{30}Ni_{7.5}P_{20}} metallic glass, we carried out quenching simulations for three specific compositions: \ce{Pd_{40}Cu_{30}Ni_{10}P_{20}}, \ce{Pd_{40}Ni_{40}P_{20}}, and \ce{Pd_{42.5}Cu_{30}Ni_{7.5}P_{20}} using the NEP89 model.
The simulation protocol involved constructing a 108-atom supercell equilibrated at \qty{1800}{\kelvin} in the isothermal-isobaric ensemble for \qty{1}{\nano\second}, followed by rapid quenching to \qty{300}{\kelvin} at a cooling rate of \qty{5e10}{\kelvin\per\second}.
The structure was then equilibrated at \qty{300}{\kelvin} for \qty{1}{\nano\second}.
From the resulting trajectory, 209 configurations were sampled.
Single-point calculations were performed using \textsc{vasp} package (version 6.5.1) \cite{blochl1994projector, kresse1996efficient} with the PBE functional \cite{perdew1996generalized}, a $\Gamma$-centered $k$-point grid with a spacing of \qty{0.2}{\per\angstrom}, a plane wave energy cutoff of \qty{600}{\electronvolt}, and a threshold for the self-consistency loop of \qty{e-6}{\electronvolt}.
After labeling the 209 structures with \gls{dft} reference data, the NEP89 model was fine-tuned for \num{5000} steps.
For the calculation of the glass transition temperature ($T_g$) and short-range order analysis of \ce{Pd_{42.5}Cu_{30}Ni_{7.5}P_{20}}, a system containing \num{32000} atoms randomly arranged in a cubic box with dimensions of \qtyproduct{7.6x7.6x7.6}{\nano\meter} was equilibrated in the isothermal-isobaric ensemble (zero pressure) at \qty{1800}{\kelvin} for \qty{1}{\nano\second} with a timestep of \qty{1}{\femto\second}, followed by rapid cooling to \qty{300}{\kelvin} at a rate of \qty{5e10}{\kelvin\per\second}, and further equilibrated at \qty{300}{\kelvin} for \qty{1}{\nano\second} (see \sautoref{fig:fine-tuned-Tg} for potential energy curves and $T_g$).
The Bussi-Donadio-Parrinello thermostat \cite{Bussi2007canonical} with a coupling time of 0.1 ps and the stochastic cell rescaling barostat \cite{Bernetti2020pressure} with a coupling time of 1 ps and an elastic modulus of 100 GPa, were used to control the temperature and isotropic pressure, respectively.
The polyhedral template matching method \cite{Larsen2016Robust} with a root-mean-square deviation cutoff of 0.12 was employed to identify local structural motifs such as icosahedral clusters and other short-range order structures. The reported $T_{\rm g}$ and icosahedral cluster counts are averages of five independent simulations, with standard deviations included.\\

\noindent{\textbf{Fine-tuning for \ce{MoS2} and thermal conductivity calculations.}}
Using the NEP89 model, we performed a \qty{6}{\nano\second} isothermal-isobaric simulation of \ce{MoS2} in an orthorhombic box containing 90 atoms at \qty{300}{\kelvin}, and applied the farthest point sampling method to sample 104 configurations from the trajectory.
Single-point \gls{dft}-D3(BJ) calculations were performed with the \textsc{vasp} package (version 6.5.1) \cite{blochl1994projector, kresse1996efficient} with the PBE functional \cite{perdew1996generalized}, a $k$-space with a spacing of \qty{0.15}{\per\angstrom}, a threshold for the self-consistency loop of \qty{e-6}{\electronvolt}, and a plane wave energy cutoff of \qty{520}{\electronvolt}.
The dataset with 104 structures was used to fine-tune the NEP89 model, yielding models for monolayer \ce{MoS2} with different fine-tuned steps.
Using the homogeneous nonequilibrium \gls{md} method \cite{fan2019Homogeneous}, we calculated the thermal conductivity of a \ce{MoS2} model containing \num{11484} atoms. 
The Bussi-Donadio-Parrinello thermostat \cite{Bussi2007canonical} with a coupling time of 0.1 ps and the stochastic cell rescaling barostat \cite{Bernetti2020pressure} with a coupling time of 1 ps and an elastic modulus of 20 GPa, were used to control the temperature and in-plane pressure, respectively.
Ten independent \qty{10}{\nano\second} simulations were conducted, and the standard error was calculated from the results.\\

\noindent{\textbf{Fine-tuning for crystalline benzene and inelastic neutron scattering calculations.}}
Energies and forces for the active-learning training structures were evaluated using the vdW-DF-cx exchange-correlation functional \cite{DioRydSch04}. 
Crystalline benzene was simulated using both the NEP89 and fine-tuned models in a \num{57024}-atom supercell at \qty{127}{\kelvin}, with a timestep of \qty{0.5}{\femto\second}.
First, the system was pre-equilibrated in the canonical ensemble for \SI{5}{\pico\second} using the Berendsen thermostat \cite{BerPosvan84}, with a temperature-coupling time of \SI{0.05}{\pico\second}.
Next, a second pre-equilibration step in the isothermal-isobaric ensemble using the stochastic cell rescaling method \cite{Bernetti2020pressure} for \SI{500}{\pico\second} was performed, with a temperature-coupling time of \SI{0.05}{\pico\second}, a pressure-coupling time of \SI{0.1}{\pico\second}, an external pressure of \SI{0}{\giga\pascal}, and elastic moduli of $C_{xx}=C_{yy}=C_{zz}=$ \SI{40}{\giga\pascal}.
Finally, a path-integral \gls{md} simulation with 32 beads for \SI{50}{\pico\second} performed with the same settings as the isothermal-isobaric equilibration step ensured accurate cell volumes at low temperature \cite{CerParMar10, YinZhoSve25}.
The trajectory was written to file every \SI{50}{\femto\second} during the path-integral \gls{md} simulation, and the last \num{100} frames were subsequently extracted and averaged to yield the average, equilibrated simulation cell that was used for the production simulations.
A short pre-production run of \SI{5}{\pico\second} in the canonical ensemble using the Bussi-Donadio-Parrinello thermostat \cite{Bussi2007canonical} with a coupling time of \SI{0.05}{\pico\second}, followed by another pre-production simulation of \SI{5}{\pico\second} in the microcanonical ensemble was then performed to ensure the dynamics were equilibrated to the averaged simulation cell.
Finally, a \qty{1}{\nano\second} microcanonical production run was performed, with positions saved every \qty{3}{\femto\second} to avoid aliasing artifacts when computing the dynamic structure factor.
The dynamic structure factor $S(\boldsymbol{q}, \omega)$ was computed from the resulting trajectories following Ref.~\cite{lindgren2025predicting}, briefly summarized here.
$S(\boldsymbol{q}, \omega)$ was computed using \textsc{dynasor} \cite{FraSlaErh21, BerFraEri25} for \num{2116} random $\boldsymbol{q}$-points up to $|\boldsymbol{q}| = \qty{14}{\per\angstrom}$.
Results were smoothed with a Gaussian of width \qty{0.01}{\per\angstrom} and averaged over spherical $|\boldsymbol{q}|$ shells to obtain $S(q, \omega)$.
$S(q, \omega)$ was then weighted for neutron probes using species-specific scattering lengths and convolved with the kinematic constraint and resolution function of the TOSCA spectrometer at the ISIS Neutron and Muon Source, UK.
Finally, the weighted $S(q, \omega)$ was integrated to yield the spectrum $S(\omega) = \int S(q, \omega) \mathrm{d}q$.
The TOSCA resolution function, implemented via \textsc{euphonic} \cite{FaiJacVon22} and \textsc{ResINS} \cite{TurJacWil25}, is based on the \textsc{AbINS} module in \textsc{Mantid} \cite{ArnBilBor14}.

\vspace{0.5cm}
\noindent{\textbf{Data availability:}}
The NEP89 model has been made available as part of the GPUMD 5.0 distribution \cite{xu2025mega}.
The associated training data and the source data for Figure 1 are available via Zenodo at \url{https://doi.org/10.5281/zenodo.19440423} (Ref. \cite{liang2026NEP89Dataset}).
Source data for Figures 2--5 are provided with this paper.

\vspace{0.5cm}
\noindent{\textbf{Code availability:}}
The source code for GPUMD (version 5.0) is available at the Zenodo repository \url{https://doi.org/10.5281/zenodo.18977569} \cite{gpumd40} and the GitHub repository \url{https://github.com/brucefan1983/GPUMD/releases/tag/v5.0}. Demonstration examples for both out-of-the-box applications and fine-tuning of the NEP89 model are available via Zenodo at \url{https://doi.org/10.5281/zenodo.19440423} \cite{liang2026NEP89Dataset}. 
A step-by-step fine-tuning tutorial is available at \url{https://github.com/brucefan1983/GPUMD-Tutorials/tree/main/examples/26_fine_tune_NEP89}.

\vspace{0.5cm}
\begin{acknowledgments}
Z.F. was supported by the National Science and Technology Advanced Materials Major Program of China (No. 2024ZD0606900).
T.L., K.X., and J.X. acknowledge support from the National Key R\&D Program of China (No. 2022YFA1203100), RGC Grants (No. 14220022,  JLFS/E-402/24), ITF Grant (No. GHP/134/22SZ), and the CUHK PhD Studentship and Postdoctoral Fellowship.
E.L., E.B., and P.E. gratefully acknowledge funding from the Swedish Research Council (Nos. 2020-04935 and 2021-05072), the Swedish Foundation for Strategic Research via the SwedNESS graduate school (GSn15-0008) as well as computational resources provided by the National Academic Infrastructure for Supercomputing in Sweden at NSC, PDC, and C3SE partially funded by the Swedish Research Council through grant agreement No.~2022-06725, as well as the Berzelius resource provided by the Knut and Alice Wallenberg Foundation at NSC.
T.A-N. and Y.W. have been supported in part by the Academy of Finland through its QTF Center of Excellence program (project No. 312298) and European Union -- NextGenerationEU instrument grant 353298, and grants Nos. 370057 and 373647. 
Computational resources by the CSC IT Centre for Finland and the Aalto Science-IT are also gratefully acknowledged. The authors also acknowledge the support of the Open Source Supercomputing Centre of S-A-I.

\end{acknowledgments}

\vspace{0.5cm}
\noindent{\textbf{Author Contributions:}}
T.L., K.X., S.C., and Z.F. conceived the original ideas and implementation. 
T.L. and K.X. created the visualizations for figures and performed comprehensive benchmarking. 
Z.F. and S.C. led training dataset curation, descriptor-space subsampling, iterative refinement, and model development, with T.L. contributing to parts of these efforts.
Specific application cases were contributed by the following authors: E.L. for crystalline benzene; Z.C. and B.Z. for methane combustion and protein-ligand cases; R.Z. for multicomponent alloys; J.L. for energy benchmarking; E.B. for ion diffusivity and NaCl dissolution; B.T. for MoSi$_2$N$_4$; Y.W. for amorphous carbon; and K.S. for iron vacancy clusters. T.L. also participated in preparing and organizing all application cases.
T.L., K.X., S.C., and Z.F. fully participated in and led the preparation and revision of the manuscript. 
All authors participated in the review and editing process and engaged in extensive discussions. 
Z.F., P.E., T.A-N., and J.X. managed funding acquisition, project administration, resources, and provided supervision throughout the project.

\vspace{0.5cm}
\noindent{\textbf{Competing interests:}}
The authors declare no competing interests.



\bibliography{ref.bib}

\begin{thebibliography}{81}%
\makeatletter
\providecommand \@ifxundefined [1]{%
 \@ifx{#1\undefined}
}%
\providecommand \@ifnum [1]{%
 \ifnum #1\expandafter \@firstoftwo
 \else \expandafter \@secondoftwo
 \fi
}%
\providecommand \@ifx [1]{%
 \ifx #1\expandafter \@firstoftwo
 \else \expandafter \@secondoftwo
 \fi
}%
\providecommand \natexlab [1]{#1}%
\providecommand \enquote  [1]{``#1''}%
\providecommand \bibnamefont  [1]{#1}%
\providecommand \bibfnamefont [1]{#1}%
\providecommand \citenamefont [1]{#1}%
\providecommand \href@noop [0]{\@secondoftwo}%
\providecommand \href [0]{\begingroup \@sanitize@url \@href}%
\providecommand \@href[1]{\@@startlink{#1}\@@href}%
\providecommand \@@href[1]{\endgroup#1\@@endlink}%
\providecommand \@sanitize@url [0]{\catcode `\\12\catcode `\$12\catcode `\&12\catcode `\#12\catcode `\^12\catcode `\_12\catcode `\%12\relax}%
\providecommand \@@startlink[1]{}%
\providecommand \@@endlink[0]{}%
\providecommand \url  [0]{\begingroup\@sanitize@url \@url }%
\providecommand \@url [1]{\endgroup\@href {#1}{\urlprefix }}%
\providecommand \urlprefix  [0]{URL }%
\providecommand \Eprint [0]{\href }%
\providecommand \doibase [0]{https://doi.org/}%
\providecommand \selectlanguage [0]{\@gobble}%
\providecommand \bibinfo  [0]{\@secondoftwo}%
\providecommand \bibfield  [0]{\@secondoftwo}%
\providecommand \translation [1]{[#1]}%
\providecommand \BibitemOpen [0]{}%
\providecommand \bibitemStop [0]{}%
\providecommand \bibitemNoStop [0]{.\EOS\space}%
\providecommand \EOS [0]{\spacefactor3000\relax}%
\providecommand \BibitemShut  [1]{\csname bibitem#1\endcsname}%
\let\auto@bib@innerbib\@empty
\bibitem [{\citenamefont {Behler}\ and\ \citenamefont {Parrinello}(2007)}]{behler2007generalized}%
  \BibitemOpen
  \bibfield  {author} {\bibinfo {author} {\bibfnamefont {J.}~\bibnamefont {Behler}}\ and\ \bibinfo {author} {\bibfnamefont {M.}~\bibnamefont {Parrinello}},\ }\bibfield  {title} {\bibinfo {title} {Generalized neural-network representation of high-dimensional potential-energy surfaces},\ }\href {https://doi.org/10.1103/PhysRevLett.98.146401} {\bibfield  {journal} {\bibinfo  {journal} {Phys. Rev. Lett.}\ }\textbf {\bibinfo {volume} {98}},\ \bibinfo {pages} {146401} (\bibinfo {year} {2007})}\BibitemShut {NoStop}%
\bibitem [{\citenamefont {Bart{\'o}k}\ \emph {et~al.}(2010)\citenamefont {Bart{\'o}k}, \citenamefont {Payne}, \citenamefont {Kondor},\ and\ \citenamefont {Cs{\'a}nyi}}]{bartok2010gaussian}%
  \BibitemOpen
  \bibfield  {author} {\bibinfo {author} {\bibfnamefont {A.~P.}\ \bibnamefont {Bart{\'o}k}}, \bibinfo {author} {\bibfnamefont {M.~C.}\ \bibnamefont {Payne}}, \bibinfo {author} {\bibfnamefont {R.}~\bibnamefont {Kondor}},\ and\ \bibinfo {author} {\bibfnamefont {G.}~\bibnamefont {Cs{\'a}nyi}},\ }\bibfield  {title} {\bibinfo {title} {{Gaussian approximation potentials: The accuracy of quantum mechanics, without the electrons}},\ }\href {https://doi.org/10.1103/PhysRevLett.104.136403} {\bibfield  {journal} {\bibinfo  {journal} {Physical Review Letters}\ }\textbf {\bibinfo {volume} {104}},\ \bibinfo {pages} {136403} (\bibinfo {year} {2010})}\BibitemShut {NoStop}%
\bibitem [{\citenamefont {Behler}(2016)}]{Behler2016perspective}%
  \BibitemOpen
  \bibfield  {author} {\bibinfo {author} {\bibfnamefont {J.}~\bibnamefont {Behler}},\ }\bibfield  {title} {\bibinfo {title} {{Perspective: Machine learning potentials for atomistic simulations}},\ }\href {https://doi.org/10.1063/1.4966192} {\bibfield  {journal} {\bibinfo  {journal} {The Journal of Chemical Physics}\ }\textbf {\bibinfo {volume} {145}},\ \bibinfo {pages} {170901} (\bibinfo {year} {2016})}\BibitemShut {NoStop}%
\bibitem [{\citenamefont {No\'{e}}\ \emph {et~al.}(2020)\citenamefont {No\'{e}}, \citenamefont {Tkatchenko}, \citenamefont {M\"{u}ller},\ and\ \citenamefont {Clementi}}]{Frank2020Machine}%
  \BibitemOpen
  \bibfield  {author} {\bibinfo {author} {\bibfnamefont {F.}~\bibnamefont {No\'{e}}}, \bibinfo {author} {\bibfnamefont {A.}~\bibnamefont {Tkatchenko}}, \bibinfo {author} {\bibfnamefont {K.-R.}\ \bibnamefont {M\"{u}ller}},\ and\ \bibinfo {author} {\bibfnamefont {C.}~\bibnamefont {Clementi}},\ }\bibfield  {title} {\bibinfo {title} {Machine learning for molecular simulation},\ }\href {https://doi.org/10.1146/annurev-physchem-042018-052331} {\bibfield  {journal} {\bibinfo  {journal} {Annual Review of Physical Chemistry}\ }\textbf {\bibinfo {volume} {71}},\ \bibinfo {pages} {361} (\bibinfo {year} {2020})}\BibitemShut {NoStop}%
\bibitem [{\citenamefont {Unke}\ \emph {et~al.}(2021)\citenamefont {Unke}, \citenamefont {Chmiela}, \citenamefont {Sauceda}, \citenamefont {Gastegger}, \citenamefont {Poltavsky}, \citenamefont {Schütt}, \citenamefont {Tkatchenko},\ and\ \citenamefont {Müller}}]{unke2021machine}%
  \BibitemOpen
  \bibfield  {author} {\bibinfo {author} {\bibfnamefont {O.~T.}\ \bibnamefont {Unke}}, \bibinfo {author} {\bibfnamefont {S.}~\bibnamefont {Chmiela}}, \bibinfo {author} {\bibfnamefont {H.~E.}\ \bibnamefont {Sauceda}}, \bibinfo {author} {\bibfnamefont {M.}~\bibnamefont {Gastegger}}, \bibinfo {author} {\bibfnamefont {I.}~\bibnamefont {Poltavsky}}, \bibinfo {author} {\bibfnamefont {K.~T.}\ \bibnamefont {Schütt}}, \bibinfo {author} {\bibfnamefont {A.}~\bibnamefont {Tkatchenko}},\ and\ \bibinfo {author} {\bibfnamefont {K.-R.}\ \bibnamefont {Müller}},\ }\bibfield  {title} {\bibinfo {title} {Machine learning force fields},\ }\href {https://doi.org/10.1021/acs.chemrev.0c01111} {\bibfield  {journal} {\bibinfo  {journal} {Chemical Reviews}\ }\textbf {\bibinfo {volume} {121}},\ \bibinfo {pages} {10142} (\bibinfo {year} {2021})}\BibitemShut {NoStop}%
\bibitem [{\citenamefont {Deringer}\ \emph {et~al.}(2019)\citenamefont {Deringer}, \citenamefont {Caro},\ and\ \citenamefont {Cs\'anyi}}]{Deringer2021Machine}%
  \BibitemOpen
  \bibfield  {author} {\bibinfo {author} {\bibfnamefont {V.~L.}\ \bibnamefont {Deringer}}, \bibinfo {author} {\bibfnamefont {M.~A.}\ \bibnamefont {Caro}},\ and\ \bibinfo {author} {\bibfnamefont {G.}~\bibnamefont {Cs\'anyi}},\ }\bibfield  {title} {\bibinfo {title} {Machine learning interatomic potentials as emerging tools for materials science},\ }\href {https://doi.org/10.1002/adma.201902765} {\bibfield  {journal} {\bibinfo  {journal} {Advanced Materials}\ }\textbf {\bibinfo {volume} {31}},\ \bibinfo {pages} {1902765} (\bibinfo {year} {2019})}\BibitemShut {NoStop}%
\bibitem [{\citenamefont {Mishin}(2021)}]{Mishin2021Machine}%
  \BibitemOpen
  \bibfield  {author} {\bibinfo {author} {\bibfnamefont {Y.}~\bibnamefont {Mishin}},\ }\bibfield  {title} {\bibinfo {title} {Machine-learning interatomic potentials for materials science},\ }\href {https://doi.org/10.1016/j.actamat.2021.116980} {\bibfield  {journal} {\bibinfo  {journal} {Acta Materialia}\ }\textbf {\bibinfo {volume} {214}},\ \bibinfo {pages} {116980} (\bibinfo {year} {2021})}\BibitemShut {NoStop}%
\bibitem [{\citenamefont {Bart\'ok}\ \emph {et~al.}(2018)\citenamefont {Bart\'ok}, \citenamefont {Kermode}, \citenamefont {Bernstein},\ and\ \citenamefont {Cs\'anyi}}]{bartok2018machine}%
  \BibitemOpen
  \bibfield  {author} {\bibinfo {author} {\bibfnamefont {A.~P.}\ \bibnamefont {Bart\'ok}}, \bibinfo {author} {\bibfnamefont {J.}~\bibnamefont {Kermode}}, \bibinfo {author} {\bibfnamefont {N.}~\bibnamefont {Bernstein}},\ and\ \bibinfo {author} {\bibfnamefont {G.}~\bibnamefont {Cs\'anyi}},\ }\bibfield  {title} {\bibinfo {title} {{Machine Learning a General-Purpose Interatomic Potential for Silicon}},\ }\href {https://doi.org/10.1103/PhysRevX.8.041048} {\bibfield  {journal} {\bibinfo  {journal} {Phys. Rev. X}\ }\textbf {\bibinfo {volume} {8}},\ \bibinfo {pages} {041048} (\bibinfo {year} {2018})}\BibitemShut {NoStop}%
\bibitem [{\citenamefont {Song}\ \emph {et~al.}(2024)\citenamefont {Song}, \citenamefont {Zhao}, \citenamefont {Liu}, \citenamefont {Wang}, \citenamefont {Lindgren}, \citenamefont {Wang}, \citenamefont {Chen}, \citenamefont {Xu}, \citenamefont {Liang}, \citenamefont {Ying}, \citenamefont {Xu}, \citenamefont {Zhao}, \citenamefont {Shi}, \citenamefont {Wang}, \citenamefont {Lyu}, \citenamefont {Zeng}, \citenamefont {Liang}, \citenamefont {Dong}, \citenamefont {Sun}, \citenamefont {Chen}, \citenamefont {Zhang}, \citenamefont {Guo}, \citenamefont {Qian}, \citenamefont {Sun}, \citenamefont {Erhart}, \citenamefont {Ala-Nissila}, \citenamefont {Su},\ and\ \citenamefont {Fan}}]{song2024general}%
  \BibitemOpen
  \bibfield  {author} {\bibinfo {author} {\bibfnamefont {K.}~\bibnamefont {Song}}, \bibinfo {author} {\bibfnamefont {R.}~\bibnamefont {Zhao}}, \bibinfo {author} {\bibfnamefont {J.}~\bibnamefont {Liu}}, \bibinfo {author} {\bibfnamefont {Y.}~\bibnamefont {Wang}}, \bibinfo {author} {\bibfnamefont {E.}~\bibnamefont {Lindgren}}, \bibinfo {author} {\bibfnamefont {Y.}~\bibnamefont {Wang}}, \bibinfo {author} {\bibfnamefont {S.}~\bibnamefont {Chen}}, \bibinfo {author} {\bibfnamefont {K.}~\bibnamefont {Xu}}, \bibinfo {author} {\bibfnamefont {T.}~\bibnamefont {Liang}}, \bibinfo {author} {\bibfnamefont {P.}~\bibnamefont {Ying}}, \bibinfo {author} {\bibfnamefont {N.}~\bibnamefont {Xu}}, \bibinfo {author} {\bibfnamefont {Z.}~\bibnamefont {Zhao}}, \bibinfo {author} {\bibfnamefont {J.}~\bibnamefont {Shi}}, \bibinfo {author} {\bibfnamefont {J.}~\bibnamefont {Wang}}, \bibinfo {author} {\bibfnamefont {S.}~\bibnamefont {Lyu}}, \bibinfo {author} {\bibfnamefont {Z.}~\bibnamefont {Zeng}}, \bibinfo {author} {\bibfnamefont
  {S.}~\bibnamefont {Liang}}, \bibinfo {author} {\bibfnamefont {H.}~\bibnamefont {Dong}}, \bibinfo {author} {\bibfnamefont {L.}~\bibnamefont {Sun}}, \bibinfo {author} {\bibfnamefont {Y.}~\bibnamefont {Chen}}, \bibinfo {author} {\bibfnamefont {Z.}~\bibnamefont {Zhang}}, \bibinfo {author} {\bibfnamefont {W.}~\bibnamefont {Guo}}, \bibinfo {author} {\bibfnamefont {P.}~\bibnamefont {Qian}}, \bibinfo {author} {\bibfnamefont {J.}~\bibnamefont {Sun}}, \bibinfo {author} {\bibfnamefont {P.}~\bibnamefont {Erhart}}, \bibinfo {author} {\bibfnamefont {T.}~\bibnamefont {Ala-Nissila}}, \bibinfo {author} {\bibfnamefont {Y.}~\bibnamefont {Su}},\ and\ \bibinfo {author} {\bibfnamefont {Z.}~\bibnamefont {Fan}},\ }\bibfield  {title} {\bibinfo {title} {General-purpose machine-learned potential for 16 elemental metals and their alloys},\ }\href {https://doi.org/https://doi.org/10.1038/s41467-024-54554-x} {\bibfield  {journal} {\bibinfo  {journal} {Nature Communications}\ }\textbf {\bibinfo {volume} {15}},\ \bibinfo {pages} {10208}
  (\bibinfo {year} {2024})}\BibitemShut {NoStop}%
\bibitem [{\citenamefont {Takamoto}\ \emph {et~al.}(2022)\citenamefont {Takamoto}, \citenamefont {Shinagawa}, \citenamefont {Motoki}, \citenamefont {Nakago}, \citenamefont {Li}, \citenamefont {Kurata}, \citenamefont {Watanabe}, \citenamefont {Yayama}, \citenamefont {Iriguchi}, \citenamefont {Asano} \emph {et~al.}}]{takamoto2022towards}%
  \BibitemOpen
  \bibfield  {author} {\bibinfo {author} {\bibfnamefont {S.}~\bibnamefont {Takamoto}}, \bibinfo {author} {\bibfnamefont {C.}~\bibnamefont {Shinagawa}}, \bibinfo {author} {\bibfnamefont {D.}~\bibnamefont {Motoki}}, \bibinfo {author} {\bibfnamefont {K.}~\bibnamefont {Nakago}}, \bibinfo {author} {\bibfnamefont {W.}~\bibnamefont {Li}}, \bibinfo {author} {\bibfnamefont {I.}~\bibnamefont {Kurata}}, \bibinfo {author} {\bibfnamefont {T.}~\bibnamefont {Watanabe}}, \bibinfo {author} {\bibfnamefont {Y.}~\bibnamefont {Yayama}}, \bibinfo {author} {\bibfnamefont {H.}~\bibnamefont {Iriguchi}}, \bibinfo {author} {\bibfnamefont {Y.}~\bibnamefont {Asano}}, \emph {et~al.},\ }\bibfield  {title} {\bibinfo {title} {Towards universal neural network potential for material discovery applicable to arbitrary combination of 45 elements},\ }\href {https://doi.org/10.1038/s41467-022-30687-9} {\bibfield  {journal} {\bibinfo  {journal} {Nature Communications}\ }\textbf {\bibinfo {volume} {13}},\ \bibinfo {pages} {1} (\bibinfo {year}
  {2022})}\BibitemShut {NoStop}%
\bibitem [{\citenamefont {Zhang}\ \emph {et~al.}(2024{\natexlab{a}})\citenamefont {Zhang}, \citenamefont {Liu}, \citenamefont {Zhang}, \citenamefont {Zhang}, \citenamefont {Cai}, \citenamefont {Bi}, \citenamefont {Du}, \citenamefont {Qin}, \citenamefont {Huang}, \citenamefont {Li}, \citenamefont {Shan}, \citenamefont {Zeng}, \citenamefont {Zhang}, \citenamefont {Liu}, \citenamefont {Li}, \citenamefont {Chang}, \citenamefont {Wang}, \citenamefont {Zhou}, \citenamefont {Liu}, \citenamefont {Luo}, \citenamefont {Wang}, \citenamefont {Jiang}, \citenamefont {Wu}, \citenamefont {Yang}, \citenamefont {Yang}, \citenamefont {Yang}, \citenamefont {Gong}, \citenamefont {Zhang}, \citenamefont {Shi}, \citenamefont {Dai}, \citenamefont {York}, \citenamefont {Liu}, \citenamefont {Zhu}, \citenamefont {Zhong}, \citenamefont {Lv}, \citenamefont {Cheng}, \citenamefont {Jia}, \citenamefont {Chen}, \citenamefont {Ke}, \citenamefont {E}, \citenamefont {Zhang},\ and\ \citenamefont {Wang}}]{zhang2024dpa2}%
  \BibitemOpen
  \bibfield  {author} {\bibinfo {author} {\bibfnamefont {D.}~\bibnamefont {Zhang}}, \bibinfo {author} {\bibfnamefont {X.}~\bibnamefont {Liu}}, \bibinfo {author} {\bibfnamefont {X.}~\bibnamefont {Zhang}}, \bibinfo {author} {\bibfnamefont {C.}~\bibnamefont {Zhang}}, \bibinfo {author} {\bibfnamefont {C.}~\bibnamefont {Cai}}, \bibinfo {author} {\bibfnamefont {H.}~\bibnamefont {Bi}}, \bibinfo {author} {\bibfnamefont {Y.}~\bibnamefont {Du}}, \bibinfo {author} {\bibfnamefont {X.}~\bibnamefont {Qin}}, \bibinfo {author} {\bibfnamefont {J.}~\bibnamefont {Huang}}, \bibinfo {author} {\bibfnamefont {B.}~\bibnamefont {Li}}, \bibinfo {author} {\bibfnamefont {Y.}~\bibnamefont {Shan}}, \bibinfo {author} {\bibfnamefont {J.}~\bibnamefont {Zeng}}, \bibinfo {author} {\bibfnamefont {Y.}~\bibnamefont {Zhang}}, \bibinfo {author} {\bibfnamefont {S.}~\bibnamefont {Liu}}, \bibinfo {author} {\bibfnamefont {Y.}~\bibnamefont {Li}}, \bibinfo {author} {\bibfnamefont {J.}~\bibnamefont {Chang}}, \bibinfo {author} {\bibfnamefont
  {X.}~\bibnamefont {Wang}}, \bibinfo {author} {\bibfnamefont {S.}~\bibnamefont {Zhou}}, \bibinfo {author} {\bibfnamefont {J.}~\bibnamefont {Liu}}, \bibinfo {author} {\bibfnamefont {X.}~\bibnamefont {Luo}}, \bibinfo {author} {\bibfnamefont {Z.}~\bibnamefont {Wang}}, \bibinfo {author} {\bibfnamefont {W.}~\bibnamefont {Jiang}}, \bibinfo {author} {\bibfnamefont {J.}~\bibnamefont {Wu}}, \bibinfo {author} {\bibfnamefont {Y.}~\bibnamefont {Yang}}, \bibinfo {author} {\bibfnamefont {J.}~\bibnamefont {Yang}}, \bibinfo {author} {\bibfnamefont {M.}~\bibnamefont {Yang}}, \bibinfo {author} {\bibfnamefont {F.-Q.}\ \bibnamefont {Gong}}, \bibinfo {author} {\bibfnamefont {L.}~\bibnamefont {Zhang}}, \bibinfo {author} {\bibfnamefont {M.}~\bibnamefont {Shi}}, \bibinfo {author} {\bibfnamefont {F.-Z.}\ \bibnamefont {Dai}}, \bibinfo {author} {\bibfnamefont {D.~M.}\ \bibnamefont {York}}, \bibinfo {author} {\bibfnamefont {S.}~\bibnamefont {Liu}}, \bibinfo {author} {\bibfnamefont {T.}~\bibnamefont {Zhu}}, \bibinfo {author}
  {\bibfnamefont {Z.}~\bibnamefont {Zhong}}, \bibinfo {author} {\bibfnamefont {J.}~\bibnamefont {Lv}}, \bibinfo {author} {\bibfnamefont {J.}~\bibnamefont {Cheng}}, \bibinfo {author} {\bibfnamefont {W.}~\bibnamefont {Jia}}, \bibinfo {author} {\bibfnamefont {M.}~\bibnamefont {Chen}}, \bibinfo {author} {\bibfnamefont {G.}~\bibnamefont {Ke}}, \bibinfo {author} {\bibfnamefont {W.}~\bibnamefont {E}}, \bibinfo {author} {\bibfnamefont {L.}~\bibnamefont {Zhang}},\ and\ \bibinfo {author} {\bibfnamefont {H.}~\bibnamefont {Wang}},\ }\bibfield  {title} {\bibinfo {title} {{DPA-2: a large atomic model as a multi-task learner}},\ }\href {https://doi.org/10.1038/s41524-024-01493-2} {\bibfield  {journal} {\bibinfo  {journal} {npj Computational Materials}\ }\textbf {\bibinfo {volume} {10}},\ \bibinfo {pages} {293} (\bibinfo {year} {2024}{\natexlab{a}})}\BibitemShut {NoStop}%
\bibitem [{\citenamefont {Chen}\ and\ \citenamefont {Ong}(2022)}]{chen2022nuniversal}%
  \BibitemOpen
  \bibfield  {author} {\bibinfo {author} {\bibfnamefont {C.}~\bibnamefont {Chen}}\ and\ \bibinfo {author} {\bibfnamefont {S.~P.}\ \bibnamefont {Ong}},\ }\bibfield  {title} {\bibinfo {title} {A universal graph deep learning interatomic potential for the periodic table},\ }\href {https://doi.org/10.1038/s43588-022-00349-3} {\bibfield  {journal} {\bibinfo  {journal} {Nature Computational Science}\ }\textbf {\bibinfo {volume} {2}},\ \bibinfo {pages} {718} (\bibinfo {year} {2022})}\BibitemShut {NoStop}%
\bibitem [{\citenamefont {Deng}\ \emph {et~al.}(2023)\citenamefont {Deng}, \citenamefont {Zhong}, \citenamefont {Jun}, \citenamefont {Riebesell}, \citenamefont {Han}, \citenamefont {Bartel},\ and\ \citenamefont {Ceder}}]{dengchgnet2023}%
  \BibitemOpen
  \bibfield  {author} {\bibinfo {author} {\bibfnamefont {B.}~\bibnamefont {Deng}}, \bibinfo {author} {\bibfnamefont {P.}~\bibnamefont {Zhong}}, \bibinfo {author} {\bibfnamefont {K.}~\bibnamefont {Jun}}, \bibinfo {author} {\bibfnamefont {J.}~\bibnamefont {Riebesell}}, \bibinfo {author} {\bibfnamefont {K.}~\bibnamefont {Han}}, \bibinfo {author} {\bibfnamefont {C.~J.}\ \bibnamefont {Bartel}},\ and\ \bibinfo {author} {\bibfnamefont {G.}~\bibnamefont {Ceder}},\ }\bibfield  {title} {\bibinfo {title} {{{CHGNet}} as a pretrained universal neural network potential for charge-informed atomistic modeling},\ }\href {https://doi.org/10.1038/s42256-023-00716-3} {\bibfield  {journal} {\bibinfo  {journal} {Nature Machine Intelligence}\ }\textbf {\bibinfo {volume} {5}},\ \bibinfo {pages} {1031} (\bibinfo {year} {2023})}\BibitemShut {NoStop}%
\bibitem [{\citenamefont {Merchant}\ \emph {et~al.}(2023)\citenamefont {Merchant}, \citenamefont {Batzner}, \citenamefont {Schoenholz}, \citenamefont {Aykol}, \citenamefont {Cheon},\ and\ \citenamefont {Cubuk}}]{merchant2023scaling}%
  \BibitemOpen
  \bibfield  {author} {\bibinfo {author} {\bibfnamefont {A.}~\bibnamefont {Merchant}}, \bibinfo {author} {\bibfnamefont {S.}~\bibnamefont {Batzner}}, \bibinfo {author} {\bibfnamefont {S.~S.}\ \bibnamefont {Schoenholz}}, \bibinfo {author} {\bibfnamefont {M.}~\bibnamefont {Aykol}}, \bibinfo {author} {\bibfnamefont {G.}~\bibnamefont {Cheon}},\ and\ \bibinfo {author} {\bibfnamefont {E.~D.}\ \bibnamefont {Cubuk}},\ }\bibfield  {title} {\bibinfo {title} {Scaling deep learning for materials discovery},\ }\href {https://doi.org/10.1038/s41586-023-06735-9} {\bibfield  {journal} {\bibinfo  {journal} {Nature}\ }\textbf {\bibinfo {volume} {624}},\ \bibinfo {pages} {80} (\bibinfo {year} {2023})}\BibitemShut {NoStop}%
\bibitem [{\citenamefont {Xie}\ \emph {et~al.}(2024)\citenamefont {Xie}, \citenamefont {Lu}, \citenamefont {Meng},\ and\ \citenamefont {Liu}}]{Xie2024GPTFF}%
  \BibitemOpen
  \bibfield  {author} {\bibinfo {author} {\bibfnamefont {F.}~\bibnamefont {Xie}}, \bibinfo {author} {\bibfnamefont {T.}~\bibnamefont {Lu}}, \bibinfo {author} {\bibfnamefont {S.}~\bibnamefont {Meng}},\ and\ \bibinfo {author} {\bibfnamefont {M.}~\bibnamefont {Liu}},\ }\bibfield  {title} {\bibinfo {title} {{GPTFF: A high-accuracy out-of-the-box universal AI force field for arbitrary inorganic materials}},\ }\href {https://doi.org/10.1016/j.scib.2024.08.039} {\bibfield  {journal} {\bibinfo  {journal} {Science Bulletin}\ }\textbf {\bibinfo {volume} {69}},\ \bibinfo {pages} {3525} (\bibinfo {year} {2024})}\BibitemShut {NoStop}%
\bibitem [{\citenamefont {Batatia}\ \emph {et~al.}(2025)\citenamefont {Batatia}, \citenamefont {Benner}, \citenamefont {Chiang}, \citenamefont {Elena}, \citenamefont {Kovács}, \citenamefont {Riebesell}, \citenamefont {Advincula}, \citenamefont {Asta}, \citenamefont {Avaylon}, \citenamefont {Baldwin}, \citenamefont {Berger}, \citenamefont {Bernstein}, \citenamefont {Bhowmik}, \citenamefont {Bigi}, \citenamefont {Blau}, \citenamefont {Cărare}, \citenamefont {Ceriotti}, \citenamefont {Chong}, \citenamefont {Darby}, \citenamefont {De}, \citenamefont {Della~Pia}, \citenamefont {Deringer}, \citenamefont {Elijošius}, \citenamefont {El-Machachi}, \citenamefont {Fako}, \citenamefont {Falcioni}, \citenamefont {Ferrari}, \citenamefont {Gardner}, \citenamefont {Gawkowski}, \citenamefont {Genreith-Schriever}, \citenamefont {George}, \citenamefont {Goodall}, \citenamefont {Grandel}, \citenamefont {Grey}, \citenamefont {Grigorev}, \citenamefont {Han}, \citenamefont {Handley}, \citenamefont {Heenen}, \citenamefont
  {Hermansson}, \citenamefont {Ho}, \citenamefont {Hofmann}, \citenamefont {Holm}, \citenamefont {Jaafar}, \citenamefont {Jakob}, \citenamefont {Jung}, \citenamefont {Kapil}, \citenamefont {Kaplan}, \citenamefont {Karimitari}, \citenamefont {Kermode}, \citenamefont {Kourtis}, \citenamefont {Kroupa}, \citenamefont {Kullgren}, \citenamefont {Kuner}, \citenamefont {Kuryla}, \citenamefont {Liepuoniute}, \citenamefont {Lin}, \citenamefont {Margraf}, \citenamefont {Magdău}, \citenamefont {Michaelides}, \citenamefont {Moore}, \citenamefont {Naik}, \citenamefont {Niblett}, \citenamefont {Norwood}, \citenamefont {O’Neill}, \citenamefont {Ortner}, \citenamefont {Persson}, \citenamefont {Reuter}, \citenamefont {Rosen}, \citenamefont {Rosset}, \citenamefont {Schaaf}, \citenamefont {Schran}, \citenamefont {Shi}, \citenamefont {Sivonxay}, \citenamefont {Stenczel}, \citenamefont {Sutton}, \citenamefont {Svahn}, \citenamefont {Swinburne}, \citenamefont {Tilly}, \citenamefont {van~der Oord}, \citenamefont {Vargas},
  \citenamefont {Varga-Umbrich}, \citenamefont {Vegge}, \citenamefont {Vondrák}, \citenamefont {Wang}, \citenamefont {Witt}, \citenamefont {Wolf}, \citenamefont {Zills},\ and\ \citenamefont {Csányi}}]{batatia2023foundation}%
  \BibitemOpen
  \bibfield  {author} {\bibinfo {author} {\bibfnamefont {I.}~\bibnamefont {Batatia}}, \bibinfo {author} {\bibfnamefont {P.}~\bibnamefont {Benner}}, \bibinfo {author} {\bibfnamefont {Y.}~\bibnamefont {Chiang}}, \bibinfo {author} {\bibfnamefont {A.~M.}\ \bibnamefont {Elena}}, \bibinfo {author} {\bibfnamefont {D.~P.}\ \bibnamefont {Kovács}}, \bibinfo {author} {\bibfnamefont {J.}~\bibnamefont {Riebesell}}, \bibinfo {author} {\bibfnamefont {X.~R.}\ \bibnamefont {Advincula}}, \bibinfo {author} {\bibfnamefont {M.}~\bibnamefont {Asta}}, \bibinfo {author} {\bibfnamefont {M.}~\bibnamefont {Avaylon}}, \bibinfo {author} {\bibfnamefont {W.~J.}\ \bibnamefont {Baldwin}}, \bibinfo {author} {\bibfnamefont {F.}~\bibnamefont {Berger}}, \bibinfo {author} {\bibfnamefont {N.}~\bibnamefont {Bernstein}}, \bibinfo {author} {\bibfnamefont {A.}~\bibnamefont {Bhowmik}}, \bibinfo {author} {\bibfnamefont {F.}~\bibnamefont {Bigi}}, \bibinfo {author} {\bibfnamefont {S.~M.}\ \bibnamefont {Blau}}, \bibinfo {author} {\bibfnamefont
  {V.}~\bibnamefont {Cărare}}, \bibinfo {author} {\bibfnamefont {M.}~\bibnamefont {Ceriotti}}, \bibinfo {author} {\bibfnamefont {S.}~\bibnamefont {Chong}}, \bibinfo {author} {\bibfnamefont {J.~P.}\ \bibnamefont {Darby}}, \bibinfo {author} {\bibfnamefont {S.}~\bibnamefont {De}}, \bibinfo {author} {\bibfnamefont {F.}~\bibnamefont {Della~Pia}}, \bibinfo {author} {\bibfnamefont {V.~L.}\ \bibnamefont {Deringer}}, \bibinfo {author} {\bibfnamefont {R.}~\bibnamefont {Elijošius}}, \bibinfo {author} {\bibfnamefont {Z.}~\bibnamefont {El-Machachi}}, \bibinfo {author} {\bibfnamefont {E.}~\bibnamefont {Fako}}, \bibinfo {author} {\bibfnamefont {F.}~\bibnamefont {Falcioni}}, \bibinfo {author} {\bibfnamefont {A.~C.}\ \bibnamefont {Ferrari}}, \bibinfo {author} {\bibfnamefont {J.~L.~A.}\ \bibnamefont {Gardner}}, \bibinfo {author} {\bibfnamefont {M.~J.}\ \bibnamefont {Gawkowski}}, \bibinfo {author} {\bibfnamefont {A.}~\bibnamefont {Genreith-Schriever}}, \bibinfo {author} {\bibfnamefont {J.}~\bibnamefont {George}}, \bibinfo
  {author} {\bibfnamefont {R.~E.~A.}\ \bibnamefont {Goodall}}, \bibinfo {author} {\bibfnamefont {J.}~\bibnamefont {Grandel}}, \bibinfo {author} {\bibfnamefont {C.~P.}\ \bibnamefont {Grey}}, \bibinfo {author} {\bibfnamefont {P.}~\bibnamefont {Grigorev}}, \bibinfo {author} {\bibfnamefont {S.}~\bibnamefont {Han}}, \bibinfo {author} {\bibfnamefont {W.}~\bibnamefont {Handley}}, \bibinfo {author} {\bibfnamefont {H.~H.}\ \bibnamefont {Heenen}}, \bibinfo {author} {\bibfnamefont {K.}~\bibnamefont {Hermansson}}, \bibinfo {author} {\bibfnamefont {C.~H.}\ \bibnamefont {Ho}}, \bibinfo {author} {\bibfnamefont {S.}~\bibnamefont {Hofmann}}, \bibinfo {author} {\bibfnamefont {C.}~\bibnamefont {Holm}}, \bibinfo {author} {\bibfnamefont {J.}~\bibnamefont {Jaafar}}, \bibinfo {author} {\bibfnamefont {K.~S.}\ \bibnamefont {Jakob}}, \bibinfo {author} {\bibfnamefont {H.}~\bibnamefont {Jung}}, \bibinfo {author} {\bibfnamefont {V.}~\bibnamefont {Kapil}}, \bibinfo {author} {\bibfnamefont {A.~D.}\ \bibnamefont {Kaplan}}, \bibinfo {author}
  {\bibfnamefont {N.}~\bibnamefont {Karimitari}}, \bibinfo {author} {\bibfnamefont {J.~R.}\ \bibnamefont {Kermode}}, \bibinfo {author} {\bibfnamefont {P.}~\bibnamefont {Kourtis}}, \bibinfo {author} {\bibfnamefont {N.}~\bibnamefont {Kroupa}}, \bibinfo {author} {\bibfnamefont {J.}~\bibnamefont {Kullgren}}, \bibinfo {author} {\bibfnamefont {M.~C.}\ \bibnamefont {Kuner}}, \bibinfo {author} {\bibfnamefont {D.}~\bibnamefont {Kuryla}}, \bibinfo {author} {\bibfnamefont {G.}~\bibnamefont {Liepuoniute}}, \bibinfo {author} {\bibfnamefont {C.}~\bibnamefont {Lin}}, \bibinfo {author} {\bibfnamefont {J.~T.}\ \bibnamefont {Margraf}}, \bibinfo {author} {\bibfnamefont {I.-B.}\ \bibnamefont {Magdău}}, \bibinfo {author} {\bibfnamefont {A.}~\bibnamefont {Michaelides}}, \bibinfo {author} {\bibfnamefont {J.~H.}\ \bibnamefont {Moore}}, \bibinfo {author} {\bibfnamefont {A.~A.}\ \bibnamefont {Naik}}, \bibinfo {author} {\bibfnamefont {S.~P.}\ \bibnamefont {Niblett}}, \bibinfo {author} {\bibfnamefont {S.~W.}\ \bibnamefont {Norwood}},
  \bibinfo {author} {\bibfnamefont {N.}~\bibnamefont {O’Neill}}, \bibinfo {author} {\bibfnamefont {C.}~\bibnamefont {Ortner}}, \bibinfo {author} {\bibfnamefont {K.~A.}\ \bibnamefont {Persson}}, \bibinfo {author} {\bibfnamefont {K.}~\bibnamefont {Reuter}}, \bibinfo {author} {\bibfnamefont {A.~S.}\ \bibnamefont {Rosen}}, \bibinfo {author} {\bibfnamefont {L.~A.~M.}\ \bibnamefont {Rosset}}, \bibinfo {author} {\bibfnamefont {L.~L.}\ \bibnamefont {Schaaf}}, \bibinfo {author} {\bibfnamefont {C.}~\bibnamefont {Schran}}, \bibinfo {author} {\bibfnamefont {B.~X.}\ \bibnamefont {Shi}}, \bibinfo {author} {\bibfnamefont {E.}~\bibnamefont {Sivonxay}}, \bibinfo {author} {\bibfnamefont {T.~K.}\ \bibnamefont {Stenczel}}, \bibinfo {author} {\bibfnamefont {C.}~\bibnamefont {Sutton}}, \bibinfo {author} {\bibfnamefont {V.}~\bibnamefont {Svahn}}, \bibinfo {author} {\bibfnamefont {T.~D.}\ \bibnamefont {Swinburne}}, \bibinfo {author} {\bibfnamefont {J.}~\bibnamefont {Tilly}}, \bibinfo {author} {\bibfnamefont {C.}~\bibnamefont
  {van~der Oord}}, \bibinfo {author} {\bibfnamefont {S.}~\bibnamefont {Vargas}}, \bibinfo {author} {\bibfnamefont {E.}~\bibnamefont {Varga-Umbrich}}, \bibinfo {author} {\bibfnamefont {T.}~\bibnamefont {Vegge}}, \bibinfo {author} {\bibfnamefont {M.}~\bibnamefont {Vondrák}}, \bibinfo {author} {\bibfnamefont {Y.}~\bibnamefont {Wang}}, \bibinfo {author} {\bibfnamefont {W.~C.}\ \bibnamefont {Witt}}, \bibinfo {author} {\bibfnamefont {T.}~\bibnamefont {Wolf}}, \bibinfo {author} {\bibfnamefont {F.}~\bibnamefont {Zills}},\ and\ \bibinfo {author} {\bibfnamefont {G.}~\bibnamefont {Csányi}},\ }\bibfield  {title} {\bibinfo {title} {A foundation model for atomistic materials chemistry},\ }\href {https://doi.org/10.1063/5.0297006} {\bibfield  {journal} {\bibinfo  {journal} {The Journal of Chemical Physics}\ }\textbf {\bibinfo {volume} {163}},\ \bibinfo {pages} {184110} (\bibinfo {year} {2025})}\BibitemShut {NoStop}%
\bibitem [{\citenamefont {Yang}\ \emph {et~al.}(2024)\citenamefont {Yang}, \citenamefont {Hu}, \citenamefont {Zhou}, \citenamefont {Liu}, \citenamefont {Shi}, \citenamefont {Li}, \citenamefont {Li}, \citenamefont {Chen}, \citenamefont {Chen}, \citenamefont {Zeni}, \citenamefont {Horton}, \citenamefont {Pinsler}, \citenamefont {Fowler}, \citenamefont {Zügner}, \citenamefont {Xie}, \citenamefont {Smith}, \citenamefont {Sun}, \citenamefont {Wang}, \citenamefont {Kong}, \citenamefont {Liu}, \citenamefont {Hao},\ and\ \citenamefont {Lu}}]{yang2024MatterSim}%
  \BibitemOpen
  \bibfield  {author} {\bibinfo {author} {\bibfnamefont {H.}~\bibnamefont {Yang}}, \bibinfo {author} {\bibfnamefont {C.}~\bibnamefont {Hu}}, \bibinfo {author} {\bibfnamefont {Y.}~\bibnamefont {Zhou}}, \bibinfo {author} {\bibfnamefont {X.}~\bibnamefont {Liu}}, \bibinfo {author} {\bibfnamefont {Y.}~\bibnamefont {Shi}}, \bibinfo {author} {\bibfnamefont {J.}~\bibnamefont {Li}}, \bibinfo {author} {\bibfnamefont {G.}~\bibnamefont {Li}}, \bibinfo {author} {\bibfnamefont {Z.}~\bibnamefont {Chen}}, \bibinfo {author} {\bibfnamefont {S.}~\bibnamefont {Chen}}, \bibinfo {author} {\bibfnamefont {C.}~\bibnamefont {Zeni}}, \bibinfo {author} {\bibfnamefont {M.}~\bibnamefont {Horton}}, \bibinfo {author} {\bibfnamefont {R.}~\bibnamefont {Pinsler}}, \bibinfo {author} {\bibfnamefont {A.}~\bibnamefont {Fowler}}, \bibinfo {author} {\bibfnamefont {D.}~\bibnamefont {Zügner}}, \bibinfo {author} {\bibfnamefont {T.}~\bibnamefont {Xie}}, \bibinfo {author} {\bibfnamefont {J.}~\bibnamefont {Smith}}, \bibinfo {author} {\bibfnamefont
  {L.}~\bibnamefont {Sun}}, \bibinfo {author} {\bibfnamefont {Q.}~\bibnamefont {Wang}}, \bibinfo {author} {\bibfnamefont {L.}~\bibnamefont {Kong}}, \bibinfo {author} {\bibfnamefont {C.}~\bibnamefont {Liu}}, \bibinfo {author} {\bibfnamefont {H.}~\bibnamefont {Hao}},\ and\ \bibinfo {author} {\bibfnamefont {Z.}~\bibnamefont {Lu}},\ }\href {https://arxiv.org/abs/2405.04967} {\bibinfo {title} {Mattersim: A deep learning atomistic model across elements, temperatures and pressures}} (\bibinfo {year} {2024}),\ \Eprint {https://arxiv.org/abs/2405.04967} {arXiv:2405.04967 [cond-mat.mtrl-sci]} \BibitemShut {NoStop}%
\bibitem [{\citenamefont {Fan}\ \emph {et~al.}(2021)\citenamefont {Fan}, \citenamefont {Zeng}, \citenamefont {Zhang}, \citenamefont {Wang}, \citenamefont {Song}, \citenamefont {Dong}, \citenamefont {Chen},\ and\ \citenamefont {Ala-Nissila}}]{fan2021neuroevolution}%
  \BibitemOpen
  \bibfield  {author} {\bibinfo {author} {\bibfnamefont {Z.}~\bibnamefont {Fan}}, \bibinfo {author} {\bibfnamefont {Z.}~\bibnamefont {Zeng}}, \bibinfo {author} {\bibfnamefont {C.}~\bibnamefont {Zhang}}, \bibinfo {author} {\bibfnamefont {Y.}~\bibnamefont {Wang}}, \bibinfo {author} {\bibfnamefont {K.}~\bibnamefont {Song}}, \bibinfo {author} {\bibfnamefont {H.}~\bibnamefont {Dong}}, \bibinfo {author} {\bibfnamefont {Y.}~\bibnamefont {Chen}},\ and\ \bibinfo {author} {\bibfnamefont {T.}~\bibnamefont {Ala-Nissila}},\ }\bibfield  {title} {\bibinfo {title} {Neuroevolution machine learning potentials: Combining high accuracy and low cost in atomistic simulations and application to heat transport},\ }\href {https://doi.org/10.1103/PhysRevB.104.104309} {\bibfield  {journal} {\bibinfo  {journal} {Phys. Rev. B}\ }\textbf {\bibinfo {volume} {104}},\ \bibinfo {pages} {104309} (\bibinfo {year} {2021})}\BibitemShut {NoStop}%
\bibitem [{\citenamefont {Unke}\ \emph {et~al.}(2024)\citenamefont {Unke}, \citenamefont {Stöhr}, \citenamefont {Ganscha}, \citenamefont {Unterthiner}, \citenamefont {Maennel}, \citenamefont {Kashubin}, \citenamefont {Ahlin}, \citenamefont {Gastegger}, \citenamefont {Sandonas}, \citenamefont {Berryman}, \citenamefont {Tkatchenko},\ and\ \citenamefont {Müller}}]{unke2024biomolecular}%
  \BibitemOpen
  \bibfield  {author} {\bibinfo {author} {\bibfnamefont {O.~T.}\ \bibnamefont {Unke}}, \bibinfo {author} {\bibfnamefont {M.}~\bibnamefont {Stöhr}}, \bibinfo {author} {\bibfnamefont {S.}~\bibnamefont {Ganscha}}, \bibinfo {author} {\bibfnamefont {T.}~\bibnamefont {Unterthiner}}, \bibinfo {author} {\bibfnamefont {H.}~\bibnamefont {Maennel}}, \bibinfo {author} {\bibfnamefont {S.}~\bibnamefont {Kashubin}}, \bibinfo {author} {\bibfnamefont {D.}~\bibnamefont {Ahlin}}, \bibinfo {author} {\bibfnamefont {M.}~\bibnamefont {Gastegger}}, \bibinfo {author} {\bibfnamefont {L.~M.}\ \bibnamefont {Sandonas}}, \bibinfo {author} {\bibfnamefont {J.~T.}\ \bibnamefont {Berryman}}, \bibinfo {author} {\bibfnamefont {A.}~\bibnamefont {Tkatchenko}},\ and\ \bibinfo {author} {\bibfnamefont {K.-R.}\ \bibnamefont {Müller}},\ }\bibfield  {title} {\bibinfo {title} {Biomolecular dynamics with machine-learned quantum-mechanical force fields trained on diverse chemical fragments},\ }\href {https://doi.org/10.1126/sciadv.adn4397} {\bibfield
  {journal} {\bibinfo  {journal} {Science Advances}\ }\textbf {\bibinfo {volume} {10}},\ \bibinfo {pages} {eadn4397} (\bibinfo {year} {2024})}\BibitemShut {NoStop}%
\bibitem [{\citenamefont {Eastman}\ \emph {et~al.}(2023)\citenamefont {Eastman}, \citenamefont {Behara}, \citenamefont {Dotson}, \citenamefont {Galvelis}, \citenamefont {Herr}, \citenamefont {Horton}, \citenamefont {Mao}, \citenamefont {Chodera}, \citenamefont {Pritchard}, \citenamefont {Wang}, \citenamefont {De~Fabritiis},\ and\ \citenamefont {Markland}}]{eastman2023spice}%
  \BibitemOpen
  \bibfield  {author} {\bibinfo {author} {\bibfnamefont {P.}~\bibnamefont {Eastman}}, \bibinfo {author} {\bibfnamefont {P.~K.}\ \bibnamefont {Behara}}, \bibinfo {author} {\bibfnamefont {D.~L.}\ \bibnamefont {Dotson}}, \bibinfo {author} {\bibfnamefont {R.}~\bibnamefont {Galvelis}}, \bibinfo {author} {\bibfnamefont {J.~E.}\ \bibnamefont {Herr}}, \bibinfo {author} {\bibfnamefont {J.~T.}\ \bibnamefont {Horton}}, \bibinfo {author} {\bibfnamefont {Y.}~\bibnamefont {Mao}}, \bibinfo {author} {\bibfnamefont {J.~D.}\ \bibnamefont {Chodera}}, \bibinfo {author} {\bibfnamefont {B.~P.}\ \bibnamefont {Pritchard}}, \bibinfo {author} {\bibfnamefont {Y.}~\bibnamefont {Wang}}, \bibinfo {author} {\bibfnamefont {G.}~\bibnamefont {De~Fabritiis}},\ and\ \bibinfo {author} {\bibfnamefont {T.~E.}\ \bibnamefont {Markland}},\ }\bibfield  {title} {\bibinfo {title} {{SPICE}, a dataset of drug-like molecules and peptides for training machine learning potentials},\ }\href {https://doi.org/10.1038/s41597-022-01882-6} {\bibfield  {journal}
  {\bibinfo  {journal} {Scientific Data}\ }\textbf {\bibinfo {volume} {10}},\ \bibinfo {pages} {11} (\bibinfo {year} {2023})}\BibitemShut {NoStop}%
\bibitem [{\citenamefont {Zhang}\ \emph {et~al.}(2024{\natexlab{b}})\citenamefont {Zhang}, \citenamefont {Mako{\'s}}, \citenamefont {Jadrich}, \citenamefont {Kraka}, \citenamefont {Barros}, \citenamefont {Nebgen}, \citenamefont {Tretiak}, \citenamefont {Isayev}, \citenamefont {Lubbers}, \citenamefont {Messerly},\ and\ \citenamefont {Smith}}]{zhang2024exploring}%
  \BibitemOpen
  \bibfield  {author} {\bibinfo {author} {\bibfnamefont {S.}~\bibnamefont {Zhang}}, \bibinfo {author} {\bibfnamefont {M.~Z.}\ \bibnamefont {Mako{\'s}}}, \bibinfo {author} {\bibfnamefont {R.~B.}\ \bibnamefont {Jadrich}}, \bibinfo {author} {\bibfnamefont {E.}~\bibnamefont {Kraka}}, \bibinfo {author} {\bibfnamefont {K.}~\bibnamefont {Barros}}, \bibinfo {author} {\bibfnamefont {B.~T.}\ \bibnamefont {Nebgen}}, \bibinfo {author} {\bibfnamefont {S.}~\bibnamefont {Tretiak}}, \bibinfo {author} {\bibfnamefont {O.}~\bibnamefont {Isayev}}, \bibinfo {author} {\bibfnamefont {N.}~\bibnamefont {Lubbers}}, \bibinfo {author} {\bibfnamefont {R.~A.}\ \bibnamefont {Messerly}},\ and\ \bibinfo {author} {\bibfnamefont {J.~S.}\ \bibnamefont {Smith}},\ }\bibfield  {title} {\bibinfo {title} {Exploring the frontiers of condensed-phase chemistry with a general reactive machine learning potential},\ }\href {https://doi.org/10.1038/s41557-023-01427-3} {\bibfield  {journal} {\bibinfo  {journal} {Nature Chemistry}\ }\textbf {\bibinfo
  {volume} {16}},\ \bibinfo {pages} {727} (\bibinfo {year} {2024}{\natexlab{b}})}\BibitemShut {NoStop}%
\bibitem [{\citenamefont {Kov{\'a}cs}\ \emph {et~al.}(2025)\citenamefont {Kov{\'a}cs}, \citenamefont {Moore}, \citenamefont {Browning}, \citenamefont {Batatia}, \citenamefont {Horton}, \citenamefont {Pu}, \citenamefont {Kapil}, \citenamefont {Witt}, \citenamefont {Magdău}, \citenamefont {Cole},\ and\ \citenamefont {Cs{\'a}nyi}}]{kovacs2025maceoff}%
  \BibitemOpen
  \bibfield  {author} {\bibinfo {author} {\bibfnamefont {D.~P.}\ \bibnamefont {Kov{\'a}cs}}, \bibinfo {author} {\bibfnamefont {J.~H.}\ \bibnamefont {Moore}}, \bibinfo {author} {\bibfnamefont {N.~J.}\ \bibnamefont {Browning}}, \bibinfo {author} {\bibfnamefont {I.}~\bibnamefont {Batatia}}, \bibinfo {author} {\bibfnamefont {J.~T.}\ \bibnamefont {Horton}}, \bibinfo {author} {\bibfnamefont {Y.}~\bibnamefont {Pu}}, \bibinfo {author} {\bibfnamefont {V.}~\bibnamefont {Kapil}}, \bibinfo {author} {\bibfnamefont {W.~C.}\ \bibnamefont {Witt}}, \bibinfo {author} {\bibfnamefont {I.-B.}\ \bibnamefont {Magdău}}, \bibinfo {author} {\bibfnamefont {D.~J.}\ \bibnamefont {Cole}},\ and\ \bibinfo {author} {\bibfnamefont {G.}~\bibnamefont {Cs{\'a}nyi}},\ }\bibfield  {title} {\bibinfo {title} {{MACE-OFF: Short-Range Transferable Machine Learning Force Fields for Organic Molecules}},\ }\href {https://doi.org/10.1021/jacs.4c07099} {\bibfield  {journal} {\bibinfo  {journal} {Journal of the American Chemical Society}\ }\textbf {\bibinfo
  {volume} {147}},\ \bibinfo {pages} {17598} (\bibinfo {year} {2025})}\BibitemShut {NoStop}%
\bibitem [{\citenamefont {Barroso-Luque}\ \emph {et~al.}(2024)\citenamefont {Barroso-Luque}, \citenamefont {Shuaibi}, \citenamefont {Fu}, \citenamefont {Wood}, \citenamefont {Dzamba}, \citenamefont {Gao}, \citenamefont {Rizvi}, \citenamefont {Zitnick},\ and\ \citenamefont {Ulissi}}]{barrosoluque2024omat24}%
  \BibitemOpen
  \bibfield  {author} {\bibinfo {author} {\bibfnamefont {L.}~\bibnamefont {Barroso-Luque}}, \bibinfo {author} {\bibfnamefont {M.}~\bibnamefont {Shuaibi}}, \bibinfo {author} {\bibfnamefont {X.}~\bibnamefont {Fu}}, \bibinfo {author} {\bibfnamefont {B.~M.}\ \bibnamefont {Wood}}, \bibinfo {author} {\bibfnamefont {M.}~\bibnamefont {Dzamba}}, \bibinfo {author} {\bibfnamefont {M.}~\bibnamefont {Gao}}, \bibinfo {author} {\bibfnamefont {A.}~\bibnamefont {Rizvi}}, \bibinfo {author} {\bibfnamefont {C.~L.}\ \bibnamefont {Zitnick}},\ and\ \bibinfo {author} {\bibfnamefont {Z.~W.}\ \bibnamefont {Ulissi}},\ }\href {https://arxiv.org/abs/2410.12771} {\bibinfo {title} {Open materials 2024 {(OMat24)} inorganic materials dataset and models}} (\bibinfo {year} {2024}),\ \Eprint {https://arxiv.org/abs/2410.12771} {arXiv:2410.12771 [cond-mat.mtrl-sci]} \BibitemShut {NoStop}%
\bibitem [{\citenamefont {Wang}\ \emph {et~al.}(2025{\natexlab{a}})\citenamefont {Wang}, \citenamefont {Guo}, \citenamefont {Gao}, \citenamefont {Wang}, \citenamefont {Zhang}, \citenamefont {Deng}, \citenamefont {Shi}, \citenamefont {Zhang},\ and\ \citenamefont {Zhong}}]{wang2024pretrained}%
  \BibitemOpen
  \bibfield  {author} {\bibinfo {author} {\bibfnamefont {R.}~\bibnamefont {Wang}}, \bibinfo {author} {\bibfnamefont {M.}~\bibnamefont {Guo}}, \bibinfo {author} {\bibfnamefont {Y.}~\bibnamefont {Gao}}, \bibinfo {author} {\bibfnamefont {X.}~\bibnamefont {Wang}}, \bibinfo {author} {\bibfnamefont {Y.}~\bibnamefont {Zhang}}, \bibinfo {author} {\bibfnamefont {B.}~\bibnamefont {Deng}}, \bibinfo {author} {\bibfnamefont {M.}~\bibnamefont {Shi}}, \bibinfo {author} {\bibfnamefont {L.}~\bibnamefont {Zhang}},\ and\ \bibinfo {author} {\bibfnamefont {Z.}~\bibnamefont {Zhong}},\ }\bibfield  {title} {\bibinfo {title} {A pre-trained deep potential model for sulfide solid electrolytes with broad coverage and high accuracy},\ }\href {https://doi.org/10.1038/s41524-025-01764-6} {\bibfield  {journal} {\bibinfo  {journal} {npj Computational Materials}\ }\textbf {\bibinfo {volume} {11}},\ \bibinfo {pages} {266} (\bibinfo {year} {2025}{\natexlab{a}})}\BibitemShut {NoStop}%
\bibitem [{\citenamefont {Ibragimova}\ \emph {et~al.}(2025)\citenamefont {Ibragimova}, \citenamefont {Kuklin}, \citenamefont {Zarrouk},\ and\ \citenamefont {Caro}}]{Ibragimova2025unifying}%
  \BibitemOpen
  \bibfield  {author} {\bibinfo {author} {\bibfnamefont {R.}~\bibnamefont {Ibragimova}}, \bibinfo {author} {\bibfnamefont {M.~S.}\ \bibnamefont {Kuklin}}, \bibinfo {author} {\bibfnamefont {T.}~\bibnamefont {Zarrouk}},\ and\ \bibinfo {author} {\bibfnamefont {M.~A.}\ \bibnamefont {Caro}},\ }\bibfield  {title} {\bibinfo {title} {Unifying the description of hydrocarbons and hydrogenated carbon materials with a chemically reactive machine learning interatomic potential},\ }\href {https://doi.org/10.1021/acs.chemmater.4c02905} {\bibfield  {journal} {\bibinfo  {journal} {Chemistry of Materials}\ }\textbf {\bibinfo {volume} {37}},\ \bibinfo {pages} {1094} (\bibinfo {year} {2025})}\BibitemShut {NoStop}%
\bibitem [{\citenamefont {Zhai}\ \emph {et~al.}(2023)\citenamefont {Zhai}, \citenamefont {Caruso}, \citenamefont {Bore}, \citenamefont {Luo},\ and\ \citenamefont {Paesani}}]{zhai2023short}%
  \BibitemOpen
  \bibfield  {author} {\bibinfo {author} {\bibfnamefont {Y.}~\bibnamefont {Zhai}}, \bibinfo {author} {\bibfnamefont {A.}~\bibnamefont {Caruso}}, \bibinfo {author} {\bibfnamefont {S.~L.}\ \bibnamefont {Bore}}, \bibinfo {author} {\bibfnamefont {Z.}~\bibnamefont {Luo}},\ and\ \bibinfo {author} {\bibfnamefont {F.}~\bibnamefont {Paesani}},\ }\bibfield  {title} {\bibinfo {title} {A “short blanket” dilemma for a state-of-the-art neural network potential for water: Reproducing experimental properties or the physics of the underlying many-body interactions?},\ }\href {https://doi.org/10.1063/5.0142843} {\bibfield  {journal} {\bibinfo  {journal} {The Journal of Chemical Physics}\ }\textbf {\bibinfo {volume} {158}},\ \bibinfo {pages} {084111} (\bibinfo {year} {2023})}\BibitemShut {NoStop}%
\bibitem [{\citenamefont {Grimme}\ \emph {et~al.}(2010)\citenamefont {Grimme}, \citenamefont {Antony}, \citenamefont {Ehrlich},\ and\ \citenamefont {Krieg}}]{Grimme2010consistent}%
  \BibitemOpen
  \bibfield  {author} {\bibinfo {author} {\bibfnamefont {S.}~\bibnamefont {Grimme}}, \bibinfo {author} {\bibfnamefont {J.}~\bibnamefont {Antony}}, \bibinfo {author} {\bibfnamefont {S.}~\bibnamefont {Ehrlich}},\ and\ \bibinfo {author} {\bibfnamefont {H.}~\bibnamefont {Krieg}},\ }\bibfield  {title} {\bibinfo {title} {{A consistent and accurate ab initio parametrization of density functional dispersion correction (DFT-D) for the 94 elements H-Pu}},\ }\href {https://doi.org/10.1063/1.3382344} {\bibfield  {journal} {\bibinfo  {journal} {The Journal of Chemical Physics}\ }\textbf {\bibinfo {volume} {132}},\ \bibinfo {pages} {154104} (\bibinfo {year} {2010})}\BibitemShut {NoStop}%
\bibitem [{\citenamefont {Grimme}\ \emph {et~al.}(2011)\citenamefont {Grimme}, \citenamefont {Ehrlich},\ and\ \citenamefont {Goerigk}}]{Grimme2011effect}%
  \BibitemOpen
  \bibfield  {author} {\bibinfo {author} {\bibfnamefont {S.}~\bibnamefont {Grimme}}, \bibinfo {author} {\bibfnamefont {S.}~\bibnamefont {Ehrlich}},\ and\ \bibinfo {author} {\bibfnamefont {L.}~\bibnamefont {Goerigk}},\ }\bibfield  {title} {\bibinfo {title} {Effect of the damping function in dispersion corrected density functional theory},\ }\href {https://doi.org/10.1002/jcc.21759} {\bibfield  {journal} {\bibinfo  {journal} {Journal of Computational Chemistry}\ }\textbf {\bibinfo {volume} {32}},\ \bibinfo {pages} {1456} (\bibinfo {year} {2011})}\BibitemShut {NoStop}%
\bibitem [{\citenamefont {Schaul}\ \emph {et~al.}(2011)\citenamefont {Schaul}, \citenamefont {Glasmachers},\ and\ \citenamefont {Schmidhuber}}]{schaul2011high}%
  \BibitemOpen
  \bibfield  {author} {\bibinfo {author} {\bibfnamefont {T.}~\bibnamefont {Schaul}}, \bibinfo {author} {\bibfnamefont {T.}~\bibnamefont {Glasmachers}},\ and\ \bibinfo {author} {\bibfnamefont {J.}~\bibnamefont {Schmidhuber}},\ }\bibfield  {title} {\bibinfo {title} {{High Dimensions and Heavy Tails for Natural Evolution Strategies}},\ }in\ \href {https://doi.org/10.1145/2001576.2001692} {\emph {\bibinfo {booktitle} {Proceedings of the 13th Annual Conference on Genetic and Evolutionary Computation}}},\ \bibinfo {series and number} {GECCO '11}\ (\bibinfo  {publisher} {Association for Computing Machinery},\ \bibinfo {address} {New York, NY, USA},\ \bibinfo {year} {2011})\ pp.\ \bibinfo {pages} {845--852}\BibitemShut {NoStop}%
\bibitem [{\citenamefont {Babin}\ \emph {et~al.}(2013)\citenamefont {Babin}, \citenamefont {Leforestier},\ and\ \citenamefont {Paesani}}]{Babin2013JCTC}%
  \BibitemOpen
  \bibfield  {author} {\bibinfo {author} {\bibfnamefont {V.}~\bibnamefont {Babin}}, \bibinfo {author} {\bibfnamefont {C.}~\bibnamefont {Leforestier}},\ and\ \bibinfo {author} {\bibfnamefont {F.}~\bibnamefont {Paesani}},\ }\bibfield  {title} {\bibinfo {title} {Development of a “first principles” water potential with flexible monomers: {Dimer} potential energy surface, {VRT} spectrum, and second virial coefficient},\ }\href {https://doi.org/10.1021/ct400863t} {\bibfield  {journal} {\bibinfo  {journal} {Journal of Chemical Theory and Computation}\ }\textbf {\bibinfo {volume} {9}},\ \bibinfo {pages} {5395} (\bibinfo {year} {2013})}\BibitemShut {NoStop}%
\bibitem [{\citenamefont {Jain}\ \emph {et~al.}(2013)\citenamefont {Jain}, \citenamefont {Ong}, \citenamefont {Hautier}, \citenamefont {Chen}, \citenamefont {Richards}, \citenamefont {Dacek}, \citenamefont {Cholia}, \citenamefont {Gunter}, \citenamefont {Skinner}, \citenamefont {Ceder},\ and\ \citenamefont {Persson}}]{Jain2013commentary}%
  \BibitemOpen
  \bibfield  {author} {\bibinfo {author} {\bibfnamefont {A.}~\bibnamefont {Jain}}, \bibinfo {author} {\bibfnamefont {S.~P.}\ \bibnamefont {Ong}}, \bibinfo {author} {\bibfnamefont {G.}~\bibnamefont {Hautier}}, \bibinfo {author} {\bibfnamefont {W.}~\bibnamefont {Chen}}, \bibinfo {author} {\bibfnamefont {W.~D.}\ \bibnamefont {Richards}}, \bibinfo {author} {\bibfnamefont {S.}~\bibnamefont {Dacek}}, \bibinfo {author} {\bibfnamefont {S.}~\bibnamefont {Cholia}}, \bibinfo {author} {\bibfnamefont {D.}~\bibnamefont {Gunter}}, \bibinfo {author} {\bibfnamefont {D.}~\bibnamefont {Skinner}}, \bibinfo {author} {\bibfnamefont {G.}~\bibnamefont {Ceder}},\ and\ \bibinfo {author} {\bibfnamefont {K.~A.}\ \bibnamefont {Persson}},\ }\bibfield  {title} {\bibinfo {title} {{Commentary: The Materials Project: A materials genome approach to accelerating materials innovation}},\ }\href {https://doi.org/10.1063/1.4812323} {\bibfield  {journal} {\bibinfo  {journal} {APL Materials}\ }\textbf {\bibinfo {volume} {1}},\ \bibinfo {pages}
  {011002} (\bibinfo {year} {2013})}\BibitemShut {NoStop}%
\bibitem [{\citenamefont {Rezac}\ \emph {et~al.}(2011)\citenamefont {Rezac}, \citenamefont {Riley},\ and\ \citenamefont {Hobza}}]{Jan2011S66}%
  \BibitemOpen
  \bibfield  {author} {\bibinfo {author} {\bibfnamefont {J.}~\bibnamefont {Rezac}}, \bibinfo {author} {\bibfnamefont {K.~E.}\ \bibnamefont {Riley}},\ and\ \bibinfo {author} {\bibfnamefont {P.}~\bibnamefont {Hobza}},\ }\bibfield  {title} {\bibinfo {title} {S66: A well-balanced database of benchmark interaction energies relevant to biomolecular structures},\ }\href {https://doi.org/10.1021/ct2002946} {\bibfield  {journal} {\bibinfo  {journal} {Journal of Chemical Theory and Computation}\ }\textbf {\bibinfo {volume} {7}},\ \bibinfo {pages} {2427} (\bibinfo {year} {2011})}\BibitemShut {NoStop}%
\bibitem [{\citenamefont {Della~Pia}\ \emph {et~al.}(2022)\citenamefont {Della~Pia}, \citenamefont {Zen}, \citenamefont {Alfè},\ and\ \citenamefont {Michaelides}}]{Della2022DMC-ICE13}%
  \BibitemOpen
  \bibfield  {author} {\bibinfo {author} {\bibfnamefont {F.}~\bibnamefont {Della~Pia}}, \bibinfo {author} {\bibfnamefont {A.}~\bibnamefont {Zen}}, \bibinfo {author} {\bibfnamefont {D.}~\bibnamefont {Alfè}},\ and\ \bibinfo {author} {\bibfnamefont {A.}~\bibnamefont {Michaelides}},\ }\bibfield  {title} {\bibinfo {title} {{DMC-ICE13: Ambient and high pressure polymorphs of ice from diffusion Monte Carlo and density functional theory}},\ }\href {https://doi.org/10.1063/5.0102645} {\bibfield  {journal} {\bibinfo  {journal} {The Journal of Chemical Physics}\ }\textbf {\bibinfo {volume} {157}},\ \bibinfo {pages} {134701} (\bibinfo {year} {2022})}\BibitemShut {NoStop}%
\bibitem [{\citenamefont {Togo}(2023)}]{Togo2023First-principles}%
  \BibitemOpen
  \bibfield  {author} {\bibinfo {author} {\bibfnamefont {A.}~\bibnamefont {Togo}},\ }\bibfield  {title} {\bibinfo {title} {First-principles phonon calculations with {Phonopy} and {Phono3py}},\ }\href {https://doi.org/10.7566/JPSJ.92.012001} {\bibfield  {journal} {\bibinfo  {journal} {Journal of the Physical Society of Japan}\ }\textbf {\bibinfo {volume} {92}},\ \bibinfo {pages} {012001} (\bibinfo {year} {2023})}\BibitemShut {NoStop}%
\bibitem [{\citenamefont {Wang}\ \emph {et~al.}(2025{\natexlab{b}})\citenamefont {Wang}, \citenamefont {Fan}, \citenamefont {Qian}, \citenamefont {Caro},\ and\ \citenamefont {Ala-Nissila}}]{Wang2025Density}%
  \BibitemOpen
  \bibfield  {author} {\bibinfo {author} {\bibfnamefont {Y.}~\bibnamefont {Wang}}, \bibinfo {author} {\bibfnamefont {Z.}~\bibnamefont {Fan}}, \bibinfo {author} {\bibfnamefont {P.}~\bibnamefont {Qian}}, \bibinfo {author} {\bibfnamefont {M.~A.}\ \bibnamefont {Caro}},\ and\ \bibinfo {author} {\bibfnamefont {T.}~\bibnamefont {Ala-Nissila}},\ }\bibfield  {title} {\bibinfo {title} {Density dependence of thermal conductivity in nanoporous and amorphous carbon with machine-learned molecular dynamics},\ }\href {https://doi.org/10.1103/PhysRevB.111.094205} {\bibfield  {journal} {\bibinfo  {journal} {Phys. Rev. B}\ }\textbf {\bibinfo {volume} {111}},\ \bibinfo {pages} {094205} (\bibinfo {year} {2025}{\natexlab{b}})}\BibitemShut {NoStop}%
\bibitem [{\citenamefont {Xu}\ \emph {et~al.}(2025{\natexlab{a}})\citenamefont {Xu}, \citenamefont {Liang}, \citenamefont {Xu}, \citenamefont {Ying}, \citenamefont {Chen}, \citenamefont {Wei}, \citenamefont {Xu},\ and\ \citenamefont {Fan}}]{Xu2025NEPMBPOL}%
  \BibitemOpen
  \bibfield  {author} {\bibinfo {author} {\bibfnamefont {K.}~\bibnamefont {Xu}}, \bibinfo {author} {\bibfnamefont {T.}~\bibnamefont {Liang}}, \bibinfo {author} {\bibfnamefont {N.}~\bibnamefont {Xu}}, \bibinfo {author} {\bibfnamefont {P.}~\bibnamefont {Ying}}, \bibinfo {author} {\bibfnamefont {S.}~\bibnamefont {Chen}}, \bibinfo {author} {\bibfnamefont {N.}~\bibnamefont {Wei}}, \bibinfo {author} {\bibfnamefont {J.}~\bibnamefont {Xu}},\ and\ \bibinfo {author} {\bibfnamefont {Z.}~\bibnamefont {Fan}},\ }\bibfield  {title} {\bibinfo {title} {{NEP-MB-pol}: a unified machine-learned framework for fast and accurate prediction of water\'s thermodynamic and transport properties},\ }\href {https://doi.org/10.1038/s41524-025-01777-1} {\bibfield  {journal} {\bibinfo  {journal} {npj Computational Materials}\ }\textbf {\bibinfo {volume} {11}},\ \bibinfo {pages} {279} (\bibinfo {year} {2025}{\natexlab{a}})}\BibitemShut {NoStop}%
\bibitem [{\citenamefont {Vanýsek}(2006)}]{Lid06Ionic}%
  \BibitemOpen
  \bibfield  {author} {\bibinfo {author} {\bibfnamefont {P.}~\bibnamefont {Vanýsek}},\ }\bibinfo {title} {Ionic {{Conductivity}} and {{Diffusion}} at {{Infinite Dilution}}},\ in\ \href@noop {} {\emph {\bibinfo {booktitle} {{{CRC Handbook}} of {{Chemistry}} and {{Physics}}}}},\ \bibinfo {editor} {edited by\ \bibinfo {editor} {\bibfnamefont {D.~R.}\ \bibnamefont {Lide}}}\ (\bibinfo  {publisher} {{Taylor and Francis}},\ \bibinfo {address} {Boca Raton, FL},\ \bibinfo {year} {2006})\ pp.\ \bibinfo {pages} {5--76 to 5--78}\BibitemShut {NoStop}%
\bibitem [{\citenamefont {Alvi}\ \emph {et~al.}(2020)\citenamefont {Alvi}, \citenamefont {Jarzabek}, \citenamefont {Kohan}, \citenamefont {Hedman}, \citenamefont {Jenczyk}, \citenamefont {Natile}, \citenamefont {Vomiero},\ and\ \citenamefont {Akhtar}}]{Alvi2020Synthesis}%
  \BibitemOpen
  \bibfield  {author} {\bibinfo {author} {\bibfnamefont {S.}~\bibnamefont {Alvi}}, \bibinfo {author} {\bibfnamefont {D.~M.}\ \bibnamefont {Jarzabek}}, \bibinfo {author} {\bibfnamefont {M.~G.}\ \bibnamefont {Kohan}}, \bibinfo {author} {\bibfnamefont {D.}~\bibnamefont {Hedman}}, \bibinfo {author} {\bibfnamefont {P.}~\bibnamefont {Jenczyk}}, \bibinfo {author} {\bibfnamefont {M.~M.}\ \bibnamefont {Natile}}, \bibinfo {author} {\bibfnamefont {A.}~\bibnamefont {Vomiero}},\ and\ \bibinfo {author} {\bibfnamefont {F.}~\bibnamefont {Akhtar}},\ }\bibfield  {title} {\bibinfo {title} {Synthesis and mechanical characterization of a cumotawv high-entropy film by magnetron sputtering},\ }\href {https://doi.org/10.1021/acsami.0c02156} {\bibfield  {journal} {\bibinfo  {journal} {ACS Applied Materials \& Interfaces}\ }\textbf {\bibinfo {volume} {12}},\ \bibinfo {pages} {21070} (\bibinfo {year} {2020})}\BibitemShut {NoStop}%
\bibitem [{\citenamefont {Longsworth}(1954)}]{Lon1954Temperature}%
  \BibitemOpen
  \bibfield  {author} {\bibinfo {author} {\bibfnamefont {L.~G.}\ \bibnamefont {Longsworth}},\ }\bibfield  {title} {\bibinfo {title} {Temperature {{Dependence}} of {{Diffusion}} in {{Aqueous Solutions}}},\ }\href {https://doi.org/10.1021/j150519a017} {\bibfield  {journal} {\bibinfo  {journal} {The Journal of Physical Chemistry}\ }\textbf {\bibinfo {volume} {58}},\ \bibinfo {pages} {770} (\bibinfo {year} {1954})}\BibitemShut {NoStop}%
\bibitem [{\citenamefont {Ramachandran}\ \emph {et~al.}(1963)\citenamefont {Ramachandran}, \citenamefont {Ramakrishnan},\ and\ \citenamefont {Sasisekharan}}]{Ramachandran1963JMB}%
  \BibitemOpen
  \bibfield  {author} {\bibinfo {author} {\bibfnamefont {G.}~\bibnamefont {Ramachandran}}, \bibinfo {author} {\bibfnamefont {C.}~\bibnamefont {Ramakrishnan}},\ and\ \bibinfo {author} {\bibfnamefont {V.}~\bibnamefont {Sasisekharan}},\ }\bibfield  {title} {\bibinfo {title} {Stereochemistry of polypeptide chain configurations},\ }\href {https://doi.org/10.1016/S0022-2836(63)80023-6} {\bibfield  {journal} {\bibinfo  {journal} {Journal of Molecular Biology}\ }\textbf {\bibinfo {volume} {7}},\ \bibinfo {pages} {95} (\bibinfo {year} {1963})}\BibitemShut {NoStop}%
\bibitem [{\citenamefont {Hong}\ \emph {et~al.}(2020)\citenamefont {Hong}, \citenamefont {Liu}, \citenamefont {Wang}, \citenamefont {Zhou}, \citenamefont {Ma}, \citenamefont {Xu}, \citenamefont {Feng}, \citenamefont {Chen}, \citenamefont {Chen}, \citenamefont {Sun}, \citenamefont {Chen}, \citenamefont {Cheng},\ and\ \citenamefont {Ren}}]{hong2020chemical}%
  \BibitemOpen
  \bibfield  {author} {\bibinfo {author} {\bibfnamefont {Y.-L.}\ \bibnamefont {Hong}}, \bibinfo {author} {\bibfnamefont {Z.}~\bibnamefont {Liu}}, \bibinfo {author} {\bibfnamefont {L.}~\bibnamefont {Wang}}, \bibinfo {author} {\bibfnamefont {T.}~\bibnamefont {Zhou}}, \bibinfo {author} {\bibfnamefont {W.}~\bibnamefont {Ma}}, \bibinfo {author} {\bibfnamefont {C.}~\bibnamefont {Xu}}, \bibinfo {author} {\bibfnamefont {S.}~\bibnamefont {Feng}}, \bibinfo {author} {\bibfnamefont {L.}~\bibnamefont {Chen}}, \bibinfo {author} {\bibfnamefont {M.-L.}\ \bibnamefont {Chen}}, \bibinfo {author} {\bibfnamefont {D.-M.}\ \bibnamefont {Sun}}, \bibinfo {author} {\bibfnamefont {X.-Q.}\ \bibnamefont {Chen}}, \bibinfo {author} {\bibfnamefont {H.-M.}\ \bibnamefont {Cheng}},\ and\ \bibinfo {author} {\bibfnamefont {W.}~\bibnamefont {Ren}},\ }\bibfield  {title} {\bibinfo {title} {Chemical vapor deposition of layered two-dimensional {MoSi}$_2${N}$_4$ materials},\ }\href {https://doi.org/10.1126/science.abb7023} {\bibfield  {journal}
  {\bibinfo  {journal} {Science}\ }\textbf {\bibinfo {volume} {369}},\ \bibinfo {pages} {670} (\bibinfo {year} {2020})}\BibitemShut {NoStop}%
\bibitem [{\citenamefont {Zhao}\ \emph {et~al.}(2023)\citenamefont {Zhao}, \citenamefont {Wang}, \citenamefont {Kong}, \citenamefont {Xu}, \citenamefont {Fu}, \citenamefont {Peng},\ and\ \citenamefont {Wu}}]{zhao2023development}%
  \BibitemOpen
  \bibfield  {author} {\bibinfo {author} {\bibfnamefont {R.}~\bibnamefont {Zhao}}, \bibinfo {author} {\bibfnamefont {S.}~\bibnamefont {Wang}}, \bibinfo {author} {\bibfnamefont {Z.}~\bibnamefont {Kong}}, \bibinfo {author} {\bibfnamefont {Y.}~\bibnamefont {Xu}}, \bibinfo {author} {\bibfnamefont {K.}~\bibnamefont {Fu}}, \bibinfo {author} {\bibfnamefont {P.}~\bibnamefont {Peng}},\ and\ \bibinfo {author} {\bibfnamefont {C.}~\bibnamefont {Wu}},\ }\bibfield  {title} {\bibinfo {title} {{Development of a neuroevolution machine learning potential of Pd-Cu-Ni-P alloys}},\ }\href {https://doi.org/10.1016/j.matdes.2023.112012} {\bibfield  {journal} {\bibinfo  {journal} {Materials \& Design}\ }\textbf {\bibinfo {volume} {231}},\ \bibinfo {pages} {112012} (\bibinfo {year} {2023})}\BibitemShut {NoStop}%
\bibitem [{\citenamefont {Haruyama}(2007)}]{Osami2007Thermodynamic}%
  \BibitemOpen
  \bibfield  {author} {\bibinfo {author} {\bibfnamefont {O.}~\bibnamefont {Haruyama}},\ }\bibfield  {title} {\bibinfo {title} {{Thermodynamic approach to free volume kinetics during isothermal relaxation in bulk Pd-Cu-Ni-P$_{20}$ glasses}},\ }\href {https://doi.org/10.1016/j.intermet.2006.10.040} {\bibfield  {journal} {\bibinfo  {journal} {Intermetallics}\ }\textbf {\bibinfo {volume} {15}},\ \bibinfo {pages} {659} (\bibinfo {year} {2007})}\BibitemShut {NoStop}%
\bibitem [{\citenamefont {Jiang}\ \emph {et~al.}(2026)\citenamefont {Jiang}, \citenamefont {Bu}, \citenamefont {Liang}, \citenamefont {Ying}, \citenamefont {Fan}, \citenamefont {Xu},\ and\ \citenamefont {Ouyang}}]{jiang2025accurate}%
  \BibitemOpen
  \bibfield  {author} {\bibinfo {author} {\bibfnamefont {W.}~\bibnamefont {Jiang}}, \bibinfo {author} {\bibfnamefont {H.}~\bibnamefont {Bu}}, \bibinfo {author} {\bibfnamefont {T.}~\bibnamefont {Liang}}, \bibinfo {author} {\bibfnamefont {P.}~\bibnamefont {Ying}}, \bibinfo {author} {\bibfnamefont {Z.}~\bibnamefont {Fan}}, \bibinfo {author} {\bibfnamefont {J.}~\bibnamefont {Xu}},\ and\ \bibinfo {author} {\bibfnamefont {W.}~\bibnamefont {Ouyang}},\ }\bibfield  {title} {\bibinfo {title} {Accurate modeling of interfacial thermal transport in van der waals heterostructures via hybrid machine learning and registry-dependent potentials},\ }\href {https://doi.org/10.1021/acs.jctc.5c01950} {\bibfield  {journal} {\bibinfo  {journal} {Journal of Chemical Theory and Computation}\ }\textbf {\bibinfo {volume} {22}},\ \bibinfo {pages} {4699} (\bibinfo {year} {2026})}\BibitemShut {NoStop}%
\bibitem [{\citenamefont {Bao}\ \emph {et~al.}(2022)\citenamefont {Bao}, \citenamefont {Chen}, \citenamefont {Wang},\ and\ \citenamefont {Tang}}]{bao2022Bilateral}%
  \BibitemOpen
  \bibfield  {author} {\bibinfo {author} {\bibfnamefont {W.}~\bibnamefont {Bao}}, \bibinfo {author} {\bibfnamefont {G.}~\bibnamefont {Chen}}, \bibinfo {author} {\bibfnamefont {Z.}~\bibnamefont {Wang}},\ and\ \bibinfo {author} {\bibfnamefont {D.}~\bibnamefont {Tang}},\ }\bibfield  {title} {\bibinfo {title} {{Bilateral phonon transport modulation of Bi-layer TMDCs (MX$_2$, M=Mo, W; X=S)}},\ }\href {https://doi.org/10.1016/j.ijthermalsci.2022.107669} {\bibfield  {journal} {\bibinfo  {journal} {International Journal of Thermal Sciences}\ }\textbf {\bibinfo {volume} {179}},\ \bibinfo {pages} {107669} (\bibinfo {year} {2022})}\BibitemShut {NoStop}%
\bibitem [{\citenamefont {Gu}\ \emph {et~al.}(2016)\citenamefont {Gu}, \citenamefont {Li},\ and\ \citenamefont {Yang}}]{Gu2016Layer}%
  \BibitemOpen
  \bibfield  {author} {\bibinfo {author} {\bibfnamefont {X.}~\bibnamefont {Gu}}, \bibinfo {author} {\bibfnamefont {B.}~\bibnamefont {Li}},\ and\ \bibinfo {author} {\bibfnamefont {R.}~\bibnamefont {Yang}},\ }\bibfield  {title} {\bibinfo {title} {{Layer thickness-dependent phonon properties and thermal conductivity of MoS$_2$}},\ }\href {https://doi.org/10.1063/1.4942827} {\bibfield  {journal} {\bibinfo  {journal} {Journal of Applied Physics}\ }\textbf {\bibinfo {volume} {119}},\ \bibinfo {pages} {085106} (\bibinfo {year} {2016})}\BibitemShut {NoStop}%
\bibitem [{\citenamefont {Cepellotti}\ \emph {et~al.}(2015)\citenamefont {Cepellotti}, \citenamefont {Fugallo}, \citenamefont {Paulatto}, \citenamefont {Lazzeri}, \citenamefont {Mauri},\ and\ \citenamefont {Marzari}}]{Cepellotti2015Phonon}%
  \BibitemOpen
  \bibfield  {author} {\bibinfo {author} {\bibfnamefont {A.}~\bibnamefont {Cepellotti}}, \bibinfo {author} {\bibfnamefont {G.}~\bibnamefont {Fugallo}}, \bibinfo {author} {\bibfnamefont {L.}~\bibnamefont {Paulatto}}, \bibinfo {author} {\bibfnamefont {M.}~\bibnamefont {Lazzeri}}, \bibinfo {author} {\bibfnamefont {F.}~\bibnamefont {Mauri}},\ and\ \bibinfo {author} {\bibfnamefont {N.}~\bibnamefont {Marzari}},\ }\bibfield  {title} {\bibinfo {title} {{Phonon hydrodynamics in two-dimensional materials}},\ }\href {https://doi.org/10.1038/ncomms7400} {\bibfield  {journal} {\bibinfo  {journal} {Nature Communications}\ }\textbf {\bibinfo {volume} {6}},\ \bibinfo {pages} {1} (\bibinfo {year} {2015})}\BibitemShut {NoStop}%
\bibitem [{\citenamefont {Lindgren}\ \emph {et~al.}(2025)\citenamefont {Lindgren}, \citenamefont {Jackson}, \citenamefont {Fransson}, \citenamefont {Berger}, \citenamefont {{\v S}koro}, \citenamefont {Rudi{\'c}}, \citenamefont {Turanyi}, \citenamefont {Mukhopadhyay},\ and\ \citenamefont {Erhart}}]{lindgren2025predicting}%
  \BibitemOpen
  \bibfield  {author} {\bibinfo {author} {\bibfnamefont {E.}~\bibnamefont {Lindgren}}, \bibinfo {author} {\bibfnamefont {A.~J.}\ \bibnamefont {Jackson}}, \bibinfo {author} {\bibfnamefont {E.}~\bibnamefont {Fransson}}, \bibinfo {author} {\bibfnamefont {E.}~\bibnamefont {Berger}}, \bibinfo {author} {\bibfnamefont {G.}~\bibnamefont {{\v S}koro}}, \bibinfo {author} {\bibfnamefont {S.}~\bibnamefont {Rudi{\'c}}}, \bibinfo {author} {\bibfnamefont {R.}~\bibnamefont {Turanyi}}, \bibinfo {author} {\bibfnamefont {S.}~\bibnamefont {Mukhopadhyay}},\ and\ \bibinfo {author} {\bibfnamefont {P.}~\bibnamefont {Erhart}},\ }\bibfield  {title} {\bibinfo {title} {Predicting neutron experiments from first principles: A workflow powered by machine learning},\ }\href {https://doi.org/10.1039/D5TA03325J} {\bibfield  {journal} {\bibinfo  {journal} {Journal of Materials Chemistry A}\ }\textbf {\bibinfo {volume} {13}},\ \bibinfo {pages} {25509} (\bibinfo {year} {2025})}\BibitemShut {NoStop}%
\bibitem [{\citenamefont {Hosokawa}\ \emph {et~al.}(2022)\citenamefont {Hosokawa}, \citenamefont {Bérar}, \citenamefont {Boudet}, \citenamefont {Pilgrim}, \citenamefont {Pusztai}, \citenamefont {Hiroi}, \citenamefont {Kohara}, \citenamefont {Kato}, \citenamefont {Fischer},\ and\ \citenamefont {Zeidler}}]{Shinya2022Relationship}%
  \BibitemOpen
  \bibfield  {author} {\bibinfo {author} {\bibfnamefont {S.}~\bibnamefont {Hosokawa}}, \bibinfo {author} {\bibfnamefont {J.-F.}\ \bibnamefont {Bérar}}, \bibinfo {author} {\bibfnamefont {N.}~\bibnamefont {Boudet}}, \bibinfo {author} {\bibfnamefont {W.-C.}\ \bibnamefont {Pilgrim}}, \bibinfo {author} {\bibfnamefont {L.}~\bibnamefont {Pusztai}}, \bibinfo {author} {\bibfnamefont {S.}~\bibnamefont {Hiroi}}, \bibinfo {author} {\bibfnamefont {S.}~\bibnamefont {Kohara}}, \bibinfo {author} {\bibfnamefont {H.}~\bibnamefont {Kato}}, \bibinfo {author} {\bibfnamefont {H.~E.}\ \bibnamefont {Fischer}},\ and\ \bibinfo {author} {\bibfnamefont {A.}~\bibnamefont {Zeidler}},\ }\bibfield  {title} {\bibinfo {title} {{Relationship between atomic structure and excellent glass forming ability in Pd$_{42.5}$Ni$_{7.5}$Cu$_{30}$P$_{20}$ metallic glass}},\ }\href {https://doi.org/10.1016/j.jnoncrysol.2022.121868} {\bibfield  {journal} {\bibinfo  {journal} {Journal of Non-Crystalline Solids}\ }\textbf {\bibinfo {volume} {596}},\ \bibinfo
  {pages} {121868} (\bibinfo {year} {2022})}\BibitemShut {NoStop}%
\bibitem [{\citenamefont {Fan}\ \emph {et~al.}(2019)\citenamefont {Fan}, \citenamefont {Dong}, \citenamefont {Harju},\ and\ \citenamefont {Ala-Nissila}}]{fan2019Homogeneous}%
  \BibitemOpen
  \bibfield  {author} {\bibinfo {author} {\bibfnamefont {Z.}~\bibnamefont {Fan}}, \bibinfo {author} {\bibfnamefont {H.}~\bibnamefont {Dong}}, \bibinfo {author} {\bibfnamefont {A.}~\bibnamefont {Harju}},\ and\ \bibinfo {author} {\bibfnamefont {T.}~\bibnamefont {Ala-Nissila}},\ }\bibfield  {title} {\bibinfo {title} {{Homogeneous nonequilibrium molecular dynamics method for heat transport and spectral decomposition with many-body potentials}},\ }\href {https://doi.org/10.1103/PhysRevB.99.064308} {\bibfield  {journal} {\bibinfo  {journal} {Phys. Rev. B}\ }\textbf {\bibinfo {volume} {99}},\ \bibinfo {pages} {064308} (\bibinfo {year} {2019})}\BibitemShut {NoStop}%
\bibitem [{\citenamefont {Fan}\ \emph {et~al.}(2026)\citenamefont {Fan}, \citenamefont {Tang}, \citenamefont {Berger}, \citenamefont {Berger}, \citenamefont {Fransson}, \citenamefont {Xu}, \citenamefont {Yan}, \citenamefont {Liu}, \citenamefont {Song}, \citenamefont {Dong}, \citenamefont {Chen}, \citenamefont {Li}, \citenamefont {Wang}, \citenamefont {Zhu}, \citenamefont {Wiktor},\ and\ \citenamefont {Erhart}}]{fan2026qnep}%
  \BibitemOpen
  \bibfield  {author} {\bibinfo {author} {\bibfnamefont {Z.}~\bibnamefont {Fan}}, \bibinfo {author} {\bibfnamefont {B.}~\bibnamefont {Tang}}, \bibinfo {author} {\bibfnamefont {E.}~\bibnamefont {Berger}}, \bibinfo {author} {\bibfnamefont {E.}~\bibnamefont {Berger}}, \bibinfo {author} {\bibfnamefont {E.}~\bibnamefont {Fransson}}, \bibinfo {author} {\bibfnamefont {K.}~\bibnamefont {Xu}}, \bibinfo {author} {\bibfnamefont {Z.}~\bibnamefont {Yan}}, \bibinfo {author} {\bibfnamefont {Z.}~\bibnamefont {Liu}}, \bibinfo {author} {\bibfnamefont {Z.}~\bibnamefont {Song}}, \bibinfo {author} {\bibfnamefont {H.}~\bibnamefont {Dong}}, \bibinfo {author} {\bibfnamefont {S.}~\bibnamefont {Chen}}, \bibinfo {author} {\bibfnamefont {L.}~\bibnamefont {Li}}, \bibinfo {author} {\bibfnamefont {Z.}~\bibnamefont {Wang}}, \bibinfo {author} {\bibfnamefont {Y.}~\bibnamefont {Zhu}}, \bibinfo {author} {\bibfnamefont {J.}~\bibnamefont {Wiktor}},\ and\ \bibinfo {author} {\bibfnamefont {P.}~\bibnamefont {Erhart}},\ }\bibfield  {title} {\bibinfo
  {title} {{qNEP}: A highly efficient neuroevolution potential with dynamic charges for large-scale atomistic simulations},\ }\href {https://doi.org/10.1021/acs.jctc.6c00146} {\bibfield  {journal} {\bibinfo  {journal} {Journal of Chemical Theory and Computation}\ }\textbf {\bibinfo {volume} {22}},\ \bibinfo {pages} {4787} (\bibinfo {year} {2026})}\BibitemShut {NoStop}%
\bibitem [{\citenamefont {Liu}(2024)}]{liu2024wizard}%
  \BibitemOpen
  \bibfield  {author} {\bibinfo {author} {\bibfnamefont {J.}~\bibnamefont {Liu}},\ }\href {https://doi.org/10.5281/zenodo.13948627} {\bibinfo {title} {{GPUMD-Wizard: A python package for generating and evaluating machine learning potentials}}} (\bibinfo {year} {2024})\BibitemShut {NoStop}%
\bibitem [{\citenamefont {Hjorth~Larsen}\ \emph {et~al.}(2017)\citenamefont {Hjorth~Larsen}, \citenamefont {Jørgen~Mortensen}, \citenamefont {Blomqvist}, \citenamefont {Castelli}, \citenamefont {Christensen}, \citenamefont {Dułak}, \citenamefont {Friis}, \citenamefont {Groves}, \citenamefont {Hammer}, \citenamefont {Hargus}, \citenamefont {Hermes}, \citenamefont {Jennings}, \citenamefont {Bjerre~Jensen}, \citenamefont {Kermode}, \citenamefont {Kitchin}, \citenamefont {Leonhard~Kolsbjerg}, \citenamefont {Kubal}, \citenamefont {Kaasbjerg}, \citenamefont {Lysgaard}, \citenamefont {Bergmann~Maronsson}, \citenamefont {Maxson}, \citenamefont {Olsen}, \citenamefont {Pastewka}, \citenamefont {Peterson}, \citenamefont {Rostgaard}, \citenamefont {Schiøtz}, \citenamefont {Schütt}, \citenamefont {Strange}, \citenamefont {Thygesen}, \citenamefont {Vegge}, \citenamefont {Vilhelmsen}, \citenamefont {Walter}, \citenamefont {Zeng},\ and\ \citenamefont {Jacobsen}}]{Larsen2017atomic}%
  \BibitemOpen
  \bibfield  {author} {\bibinfo {author} {\bibfnamefont {A.}~\bibnamefont {Hjorth~Larsen}}, \bibinfo {author} {\bibfnamefont {J.}~\bibnamefont {Jørgen~Mortensen}}, \bibinfo {author} {\bibfnamefont {J.}~\bibnamefont {Blomqvist}}, \bibinfo {author} {\bibfnamefont {I.~E.}\ \bibnamefont {Castelli}}, \bibinfo {author} {\bibfnamefont {R.}~\bibnamefont {Christensen}}, \bibinfo {author} {\bibfnamefont {M.}~\bibnamefont {Dułak}}, \bibinfo {author} {\bibfnamefont {J.}~\bibnamefont {Friis}}, \bibinfo {author} {\bibfnamefont {M.~N.}\ \bibnamefont {Groves}}, \bibinfo {author} {\bibfnamefont {B.}~\bibnamefont {Hammer}}, \bibinfo {author} {\bibfnamefont {C.}~\bibnamefont {Hargus}}, \bibinfo {author} {\bibfnamefont {E.~D.}\ \bibnamefont {Hermes}}, \bibinfo {author} {\bibfnamefont {P.~C.}\ \bibnamefont {Jennings}}, \bibinfo {author} {\bibfnamefont {P.}~\bibnamefont {Bjerre~Jensen}}, \bibinfo {author} {\bibfnamefont {J.}~\bibnamefont {Kermode}}, \bibinfo {author} {\bibfnamefont {J.~R.}\ \bibnamefont {Kitchin}}, \bibinfo
  {author} {\bibfnamefont {E.}~\bibnamefont {Leonhard~Kolsbjerg}}, \bibinfo {author} {\bibfnamefont {J.}~\bibnamefont {Kubal}}, \bibinfo {author} {\bibfnamefont {K.}~\bibnamefont {Kaasbjerg}}, \bibinfo {author} {\bibfnamefont {S.}~\bibnamefont {Lysgaard}}, \bibinfo {author} {\bibfnamefont {J.}~\bibnamefont {Bergmann~Maronsson}}, \bibinfo {author} {\bibfnamefont {T.}~\bibnamefont {Maxson}}, \bibinfo {author} {\bibfnamefont {T.}~\bibnamefont {Olsen}}, \bibinfo {author} {\bibfnamefont {L.}~\bibnamefont {Pastewka}}, \bibinfo {author} {\bibfnamefont {A.}~\bibnamefont {Peterson}}, \bibinfo {author} {\bibfnamefont {C.}~\bibnamefont {Rostgaard}}, \bibinfo {author} {\bibfnamefont {J.}~\bibnamefont {Schiøtz}}, \bibinfo {author} {\bibfnamefont {O.}~\bibnamefont {Schütt}}, \bibinfo {author} {\bibfnamefont {M.}~\bibnamefont {Strange}}, \bibinfo {author} {\bibfnamefont {K.~S.}\ \bibnamefont {Thygesen}}, \bibinfo {author} {\bibfnamefont {T.}~\bibnamefont {Vegge}}, \bibinfo {author} {\bibfnamefont {L.}~\bibnamefont
  {Vilhelmsen}}, \bibinfo {author} {\bibfnamefont {M.}~\bibnamefont {Walter}}, \bibinfo {author} {\bibfnamefont {Z.}~\bibnamefont {Zeng}},\ and\ \bibinfo {author} {\bibfnamefont {K.~W.}\ \bibnamefont {Jacobsen}},\ }\bibfield  {title} {\bibinfo {title} {{The atomic simulation environment--a Python library for working with atoms}},\ }\href {https://doi.org/10.1088/1361-648X/aa680e} {\bibfield  {journal} {\bibinfo  {journal} {Journal of Physics: Condensed Matter}\ }\textbf {\bibinfo {volume} {29}},\ \bibinfo {pages} {273002} (\bibinfo {year} {2017})}\BibitemShut {NoStop}%
\bibitem [{\citenamefont {Lindgren}\ \emph {et~al.}(2024)\citenamefont {Lindgren}, \citenamefont {Rahm}, \citenamefont {Fransson}, \citenamefont {Eriksson}, \citenamefont {Österbacka}, \citenamefont {Fan},\ and\ \citenamefont {Erhart}}]{Lindgren2024calorine}%
  \BibitemOpen
  \bibfield  {author} {\bibinfo {author} {\bibfnamefont {E.}~\bibnamefont {Lindgren}}, \bibinfo {author} {\bibfnamefont {M.}~\bibnamefont {Rahm}}, \bibinfo {author} {\bibfnamefont {E.}~\bibnamefont {Fransson}}, \bibinfo {author} {\bibfnamefont {F.}~\bibnamefont {Eriksson}}, \bibinfo {author} {\bibfnamefont {N.}~\bibnamefont {Österbacka}}, \bibinfo {author} {\bibfnamefont {Z.}~\bibnamefont {Fan}},\ and\ \bibinfo {author} {\bibfnamefont {P.}~\bibnamefont {Erhart}},\ }\bibfield  {title} {\bibinfo {title} {calorine: A python package for constructing and sampling neuroevolution potential models},\ }\href {https://doi.org/10.21105/joss.06264} {\bibfield  {journal} {\bibinfo  {journal} {Journal of Open Source Software}\ }\textbf {\bibinfo {volume} {9}},\ \bibinfo {pages} {6264} (\bibinfo {year} {2024})}\BibitemShut {NoStop}%
\bibitem [{\citenamefont {Liu}\ \emph {et~al.}(2024)\citenamefont {Liu}, \citenamefont {Liu}, \citenamefont {Riebesell}, \citenamefont {Qi}, \citenamefont {Ong},\ and\ \citenamefont {Ko}}]{matcalc2024}%
  \BibitemOpen
  \bibfield  {author} {\bibinfo {author} {\bibfnamefont {R.}~\bibnamefont {Liu}}, \bibinfo {author} {\bibfnamefont {E.}~\bibnamefont {Liu}}, \bibinfo {author} {\bibfnamefont {J.}~\bibnamefont {Riebesell}}, \bibinfo {author} {\bibfnamefont {J.}~\bibnamefont {Qi}}, \bibinfo {author} {\bibfnamefont {S.~P.}\ \bibnamefont {Ong}},\ and\ \bibinfo {author} {\bibfnamefont {T.~W.}\ \bibnamefont {Ko}},\ }\href@noop {} {\bibinfo {title} {{MatCalc: A Python library for calculating materials properties from the potential energy surface (PES)}}},\ \bibinfo {howpublished} {\url{https://github.com/materialsvirtuallab/matcalc}} (\bibinfo {year} {2024})\BibitemShut {NoStop}%
\bibitem [{\citenamefont {Stukowski}(2009)}]{Stukowski2010Visualization}%
  \BibitemOpen
  \bibfield  {author} {\bibinfo {author} {\bibfnamefont {A.}~\bibnamefont {Stukowski}},\ }\bibfield  {title} {\bibinfo {title} {{Visualization and analysis of atomistic simulation data with OVITO--the Open Visualization Tool}},\ }\href {https://doi.org/10.1088/0965-0393/18/1/015012} {\bibfield  {journal} {\bibinfo  {journal} {Modelling and Simulation in Materials Science and Engineering}\ }\textbf {\bibinfo {volume} {18}},\ \bibinfo {pages} {015012} (\bibinfo {year} {2009})}\BibitemShut {NoStop}%
\bibitem [{\citenamefont {Cantor}\ \emph {et~al.}(2004)\citenamefont {Cantor}, \citenamefont {Chang}, \citenamefont {Knight},\ and\ \citenamefont {Vincent}}]{Cantor2004Microstructural}%
  \BibitemOpen
  \bibfield  {author} {\bibinfo {author} {\bibfnamefont {B.}~\bibnamefont {Cantor}}, \bibinfo {author} {\bibfnamefont {I.}~\bibnamefont {Chang}}, \bibinfo {author} {\bibfnamefont {P.}~\bibnamefont {Knight}},\ and\ \bibinfo {author} {\bibfnamefont {A.}~\bibnamefont {Vincent}},\ }\bibfield  {title} {\bibinfo {title} {Microstructural development in equiatomic multicomponent alloys},\ }\href {https://doi.org/10.1016/j.msea.2003.10.257} {\bibfield  {journal} {\bibinfo  {journal} {Materials Science and Engineering: A}\ }\textbf {\bibinfo {volume} {375-377}},\ \bibinfo {pages} {213} (\bibinfo {year} {2004})}\BibitemShut {NoStop}%
\bibitem [{\citenamefont {Martyna}\ \emph {et~al.}(1992)\citenamefont {Martyna}, \citenamefont {Klein},\ and\ \citenamefont {Tuckerman}}]{Martyna1992nose}%
  \BibitemOpen
  \bibfield  {author} {\bibinfo {author} {\bibfnamefont {G.~J.}\ \bibnamefont {Martyna}}, \bibinfo {author} {\bibfnamefont {M.~L.}\ \bibnamefont {Klein}},\ and\ \bibinfo {author} {\bibfnamefont {M.}~\bibnamefont {Tuckerman}},\ }\bibfield  {title} {\bibinfo {title} {{Nosé–Hoover chains: The canonical ensemble via continuous dynamics}},\ }\href {https://doi.org/10.1063/1.463940} {\bibfield  {journal} {\bibinfo  {journal} {The Journal of Chemical Physics}\ }\textbf {\bibinfo {volume} {97}},\ \bibinfo {pages} {2635} (\bibinfo {year} {1992})}\BibitemShut {NoStop}%
\bibitem [{\citenamefont {Xu}\ \emph {et~al.}(2025{\natexlab{b}})\citenamefont {Xu}, \citenamefont {Bu}, \citenamefont {Pan}, \citenamefont {Lindgren}, \citenamefont {Wu}, \citenamefont {Wang}, \citenamefont {Liu}, \citenamefont {Song}, \citenamefont {Xu}, \citenamefont {Li}, \citenamefont {Hainer}, \citenamefont {Svensson}, \citenamefont {Wiktor}, \citenamefont {Zhao}, \citenamefont {Huang}, \citenamefont {Qian}, \citenamefont {Zhang}, \citenamefont {Zeng}, \citenamefont {Zhang}, \citenamefont {Tang}, \citenamefont {Xiao}, \citenamefont {Yan}, \citenamefont {Shi}, \citenamefont {Liang}, \citenamefont {Wang}, \citenamefont {Liang}, \citenamefont {Cao}, \citenamefont {Wang}, \citenamefont {Ying}, \citenamefont {Xu}, \citenamefont {Chen}, \citenamefont {Zhang}, \citenamefont {Chen}, \citenamefont {Wu}, \citenamefont {Jiang}, \citenamefont {Berger}, \citenamefont {Li}, \citenamefont {Chen}, \citenamefont {Gabourie}, \citenamefont {Dong}, \citenamefont {Xiong}, \citenamefont {Wei}, \citenamefont {Chen},
  \citenamefont {Xu}, \citenamefont {Ding}, \citenamefont {Sun}, \citenamefont {Ala-Nissila}, \citenamefont {Harju}, \citenamefont {Zheng}, \citenamefont {Guan}, \citenamefont {Erhart}, \citenamefont {Sun}, \citenamefont {Ouyang}, \citenamefont {Su},\ and\ \citenamefont {Fan}}]{xu2025mega}%
  \BibitemOpen
  \bibfield  {author} {\bibinfo {author} {\bibfnamefont {K.}~\bibnamefont {Xu}}, \bibinfo {author} {\bibfnamefont {H.}~\bibnamefont {Bu}}, \bibinfo {author} {\bibfnamefont {S.}~\bibnamefont {Pan}}, \bibinfo {author} {\bibfnamefont {E.}~\bibnamefont {Lindgren}}, \bibinfo {author} {\bibfnamefont {Y.}~\bibnamefont {Wu}}, \bibinfo {author} {\bibfnamefont {Y.}~\bibnamefont {Wang}}, \bibinfo {author} {\bibfnamefont {J.}~\bibnamefont {Liu}}, \bibinfo {author} {\bibfnamefont {K.}~\bibnamefont {Song}}, \bibinfo {author} {\bibfnamefont {B.}~\bibnamefont {Xu}}, \bibinfo {author} {\bibfnamefont {Y.}~\bibnamefont {Li}}, \bibinfo {author} {\bibfnamefont {T.}~\bibnamefont {Hainer}}, \bibinfo {author} {\bibfnamefont {L.}~\bibnamefont {Svensson}}, \bibinfo {author} {\bibfnamefont {J.}~\bibnamefont {Wiktor}}, \bibinfo {author} {\bibfnamefont {R.}~\bibnamefont {Zhao}}, \bibinfo {author} {\bibfnamefont {H.}~\bibnamefont {Huang}}, \bibinfo {author} {\bibfnamefont {C.}~\bibnamefont {Qian}}, \bibinfo {author} {\bibfnamefont
  {S.}~\bibnamefont {Zhang}}, \bibinfo {author} {\bibfnamefont {Z.}~\bibnamefont {Zeng}}, \bibinfo {author} {\bibfnamefont {B.}~\bibnamefont {Zhang}}, \bibinfo {author} {\bibfnamefont {B.}~\bibnamefont {Tang}}, \bibinfo {author} {\bibfnamefont {Y.}~\bibnamefont {Xiao}}, \bibinfo {author} {\bibfnamefont {Z.}~\bibnamefont {Yan}}, \bibinfo {author} {\bibfnamefont {J.}~\bibnamefont {Shi}}, \bibinfo {author} {\bibfnamefont {Z.}~\bibnamefont {Liang}}, \bibinfo {author} {\bibfnamefont {J.}~\bibnamefont {Wang}}, \bibinfo {author} {\bibfnamefont {T.}~\bibnamefont {Liang}}, \bibinfo {author} {\bibfnamefont {S.}~\bibnamefont {Cao}}, \bibinfo {author} {\bibfnamefont {Y.}~\bibnamefont {Wang}}, \bibinfo {author} {\bibfnamefont {P.}~\bibnamefont {Ying}}, \bibinfo {author} {\bibfnamefont {N.}~\bibnamefont {Xu}}, \bibinfo {author} {\bibfnamefont {C.}~\bibnamefont {Chen}}, \bibinfo {author} {\bibfnamefont {Y.}~\bibnamefont {Zhang}}, \bibinfo {author} {\bibfnamefont {Z.}~\bibnamefont {Chen}}, \bibinfo {author} {\bibfnamefont
  {X.}~\bibnamefont {Wu}}, \bibinfo {author} {\bibfnamefont {W.}~\bibnamefont {Jiang}}, \bibinfo {author} {\bibfnamefont {E.}~\bibnamefont {Berger}}, \bibinfo {author} {\bibfnamefont {Y.}~\bibnamefont {Li}}, \bibinfo {author} {\bibfnamefont {S.}~\bibnamefont {Chen}}, \bibinfo {author} {\bibfnamefont {A.~J.}\ \bibnamefont {Gabourie}}, \bibinfo {author} {\bibfnamefont {H.}~\bibnamefont {Dong}}, \bibinfo {author} {\bibfnamefont {S.}~\bibnamefont {Xiong}}, \bibinfo {author} {\bibfnamefont {N.}~\bibnamefont {Wei}}, \bibinfo {author} {\bibfnamefont {Y.}~\bibnamefont {Chen}}, \bibinfo {author} {\bibfnamefont {J.}~\bibnamefont {Xu}}, \bibinfo {author} {\bibfnamefont {F.}~\bibnamefont {Ding}}, \bibinfo {author} {\bibfnamefont {Z.}~\bibnamefont {Sun}}, \bibinfo {author} {\bibfnamefont {T.}~\bibnamefont {Ala-Nissila}}, \bibinfo {author} {\bibfnamefont {A.}~\bibnamefont {Harju}}, \bibinfo {author} {\bibfnamefont {J.}~\bibnamefont {Zheng}}, \bibinfo {author} {\bibfnamefont {P.}~\bibnamefont {Guan}}, \bibinfo {author}
  {\bibfnamefont {P.}~\bibnamefont {Erhart}}, \bibinfo {author} {\bibfnamefont {J.}~\bibnamefont {Sun}}, \bibinfo {author} {\bibfnamefont {W.}~\bibnamefont {Ouyang}}, \bibinfo {author} {\bibfnamefont {Y.}~\bibnamefont {Su}},\ and\ \bibinfo {author} {\bibfnamefont {Z.}~\bibnamefont {Fan}},\ }\bibfield  {title} {\bibinfo {title} {{GPUMD 4.0: A high-performance molecular dynamics package for versatile materials simulations with machine-learned potentials}},\ }\href {https://doi.org/doi: 10.1002/mgea.70028} {\bibfield  {journal} {\bibinfo  {journal} {Materials Genome Engineering Advances}\ }\textbf {\bibinfo {volume} {3}},\ \bibinfo {pages} {e70028} (\bibinfo {year} {2025}{\natexlab{b}})}\BibitemShut {NoStop}%
\bibitem [{\citenamefont {Thompson}\ \emph {et~al.}(2022)\citenamefont {Thompson}, \citenamefont {Aktulga}, \citenamefont {Berger}, \citenamefont {Bolintineanu}, \citenamefont {Brown}, \citenamefont {Crozier}, \citenamefont {{in 't Veld}}, \citenamefont {Kohlmeyer}, \citenamefont {Moore}, \citenamefont {Nguyen}, \citenamefont {Shan}, \citenamefont {Stevens}, \citenamefont {Tranchida}, \citenamefont {Trott},\ and\ \citenamefont {Plimpton}}]{Thompson2022lammps}%
  \BibitemOpen
  \bibfield  {author} {\bibinfo {author} {\bibfnamefont {A.~P.}\ \bibnamefont {Thompson}}, \bibinfo {author} {\bibfnamefont {H.~M.}\ \bibnamefont {Aktulga}}, \bibinfo {author} {\bibfnamefont {R.}~\bibnamefont {Berger}}, \bibinfo {author} {\bibfnamefont {D.~S.}\ \bibnamefont {Bolintineanu}}, \bibinfo {author} {\bibfnamefont {W.~M.}\ \bibnamefont {Brown}}, \bibinfo {author} {\bibfnamefont {P.~S.}\ \bibnamefont {Crozier}}, \bibinfo {author} {\bibfnamefont {P.~J.}\ \bibnamefont {{in 't Veld}}}, \bibinfo {author} {\bibfnamefont {A.}~\bibnamefont {Kohlmeyer}}, \bibinfo {author} {\bibfnamefont {S.~G.}\ \bibnamefont {Moore}}, \bibinfo {author} {\bibfnamefont {T.~D.}\ \bibnamefont {Nguyen}}, \bibinfo {author} {\bibfnamefont {R.}~\bibnamefont {Shan}}, \bibinfo {author} {\bibfnamefont {M.~J.}\ \bibnamefont {Stevens}}, \bibinfo {author} {\bibfnamefont {J.}~\bibnamefont {Tranchida}}, \bibinfo {author} {\bibfnamefont {C.}~\bibnamefont {Trott}},\ and\ \bibinfo {author} {\bibfnamefont {S.~J.}\ \bibnamefont {Plimpton}},\
  }\bibfield  {title} {\bibinfo {title} {{LAMMPS - a flexible simulation tool for particle-based materials modeling at the atomic, meso, and continuum scales}},\ }\href {https://doi.org/https://doi.org/10.1016/j.cpc.2021.108171} {\bibfield  {journal} {\bibinfo  {journal} {Computer Physics Communications}\ }\textbf {\bibinfo {volume} {271}},\ \bibinfo {pages} {108171} (\bibinfo {year} {2022})}\BibitemShut {NoStop}%
\bibitem [{\citenamefont {Bussi}\ \emph {et~al.}(2007)\citenamefont {Bussi}, \citenamefont {Donadio},\ and\ \citenamefont {Parrinello}}]{Bussi2007canonical}%
  \BibitemOpen
  \bibfield  {author} {\bibinfo {author} {\bibfnamefont {G.}~\bibnamefont {Bussi}}, \bibinfo {author} {\bibfnamefont {D.}~\bibnamefont {Donadio}},\ and\ \bibinfo {author} {\bibfnamefont {M.}~\bibnamefont {Parrinello}},\ }\bibfield  {title} {\bibinfo {title} {Canonical sampling through velocity rescaling},\ }\href {https://doi.org/10.1063/1.2408420} {\bibfield  {journal} {\bibinfo  {journal} {The Journal of Chemical Physics}\ }\textbf {\bibinfo {volume} {126}},\ \bibinfo {pages} {014101} (\bibinfo {year} {2007})}\BibitemShut {NoStop}%
\bibitem [{\citenamefont {Bernetti}\ and\ \citenamefont {Bussi}(2020)}]{Bernetti2020pressure}%
  \BibitemOpen
  \bibfield  {author} {\bibinfo {author} {\bibfnamefont {M.}~\bibnamefont {Bernetti}}\ and\ \bibinfo {author} {\bibfnamefont {G.}~\bibnamefont {Bussi}},\ }\bibfield  {title} {\bibinfo {title} {Pressure control using stochastic cell rescaling},\ }\href {https://doi.org/10.1063/5.0020514} {\bibfield  {journal} {\bibinfo  {journal} {The Journal of Chemical Physics}\ }\textbf {\bibinfo {volume} {153}},\ \bibinfo {pages} {114107} (\bibinfo {year} {2020})}\BibitemShut {NoStop}%
\bibitem [{\citenamefont {Bussi}\ and\ \citenamefont {Parrinello}(2007)}]{Bussi2007Langevin}%
  \BibitemOpen
  \bibfield  {author} {\bibinfo {author} {\bibfnamefont {G.}~\bibnamefont {Bussi}}\ and\ \bibinfo {author} {\bibfnamefont {M.}~\bibnamefont {Parrinello}},\ }\bibfield  {title} {\bibinfo {title} {Accurate sampling using langevin dynamics},\ }\href {https://doi.org/10.1103/PhysRevE.75.056707} {\bibfield  {journal} {\bibinfo  {journal} {Phys. Rev. E}\ }\textbf {\bibinfo {volume} {75}},\ \bibinfo {pages} {056707} (\bibinfo {year} {2007})}\BibitemShut {NoStop}%
\bibitem [{\citenamefont {Berendsen}\ \emph {et~al.}(1995)\citenamefont {Berendsen}, \citenamefont {{van der Spoel}},\ and\ \citenamefont {{van Drunen}}}]{Berendsen1995GROMACS}%
  \BibitemOpen
  \bibfield  {author} {\bibinfo {author} {\bibfnamefont {H.}~\bibnamefont {Berendsen}}, \bibinfo {author} {\bibfnamefont {D.}~\bibnamefont {{van der Spoel}}},\ and\ \bibinfo {author} {\bibfnamefont {R.}~\bibnamefont {{van Drunen}}},\ }\bibfield  {title} {\bibinfo {title} {Gromacs: A message-passing parallel molecular dynamics implementation},\ }\href {https://doi.org/10.1016/0010-4655(95)00042-E} {\bibfield  {journal} {\bibinfo  {journal} {Computer Physics Communications}\ }\textbf {\bibinfo {volume} {91}},\ \bibinfo {pages} {43} (\bibinfo {year} {1995})}\BibitemShut {NoStop}%
\bibitem [{\citenamefont {Huang}\ \emph {et~al.}(2017)\citenamefont {Huang}, \citenamefont {Rauscher}, \citenamefont {Nawrocki}, \citenamefont {Ran}, \citenamefont {Feig}, \citenamefont {de~Groot}, \citenamefont {Grubm{\"u}ller},\ and\ \citenamefont {MacKerell}}]{charmm36}%
  \BibitemOpen
  \bibfield  {author} {\bibinfo {author} {\bibfnamefont {J.}~\bibnamefont {Huang}}, \bibinfo {author} {\bibfnamefont {S.}~\bibnamefont {Rauscher}}, \bibinfo {author} {\bibfnamefont {G.}~\bibnamefont {Nawrocki}}, \bibinfo {author} {\bibfnamefont {T.}~\bibnamefont {Ran}}, \bibinfo {author} {\bibfnamefont {M.}~\bibnamefont {Feig}}, \bibinfo {author} {\bibfnamefont {B.~L.}\ \bibnamefont {de~Groot}}, \bibinfo {author} {\bibfnamefont {H.}~\bibnamefont {Grubm{\"u}ller}},\ and\ \bibinfo {author} {\bibfnamefont {A.~D.}\ \bibnamefont {MacKerell}},\ }\bibfield  {title} {\bibinfo {title} {{CHARMM36m: an improved force field for folded and intrinsically disordered proteins}},\ }\href {https://doi.org/10.1038/nmeth.4067} {\bibfield  {journal} {\bibinfo  {journal} {Nature Methods}\ }\textbf {\bibinfo {volume} {14}},\ \bibinfo {pages} {71} (\bibinfo {year} {2017})}\BibitemShut {NoStop}%
\bibitem [{\citenamefont {{Soteras Gutiérrez}}\ \emph {et~al.}(2016)\citenamefont {{Soteras Gutiérrez}}, \citenamefont {Lin}, \citenamefont {Vanommeslaeghe}, \citenamefont {Lemkul}, \citenamefont {Armacost}, \citenamefont {Brooks},\ and\ \citenamefont {MacKerell}}]{cgenff}%
  \BibitemOpen
  \bibfield  {author} {\bibinfo {author} {\bibfnamefont {I.}~\bibnamefont {{Soteras Gutiérrez}}}, \bibinfo {author} {\bibfnamefont {F.-Y.}\ \bibnamefont {Lin}}, \bibinfo {author} {\bibfnamefont {K.}~\bibnamefont {Vanommeslaeghe}}, \bibinfo {author} {\bibfnamefont {J.~A.}\ \bibnamefont {Lemkul}}, \bibinfo {author} {\bibfnamefont {K.~A.}\ \bibnamefont {Armacost}}, \bibinfo {author} {\bibfnamefont {C.~L.}\ \bibnamefont {Brooks}},\ and\ \bibinfo {author} {\bibfnamefont {A.~D.}\ \bibnamefont {MacKerell}},\ }\bibfield  {title} {\bibinfo {title} {Parametrization of halogen bonds in the charmm general force field: Improved treatment of ligand–protein interactions},\ }\href {https://doi.org/10.1016/j.bmc.2016.06.034} {\bibfield  {journal} {\bibinfo  {journal} {Bioorganic \& Medicinal Chemistry}\ }\textbf {\bibinfo {volume} {24}},\ \bibinfo {pages} {4812} (\bibinfo {year} {2016})}\BibitemShut {NoStop}%
\bibitem [{\citenamefont {Bl\"ochl}(1994)}]{blochl1994projector}%
  \BibitemOpen
  \bibfield  {author} {\bibinfo {author} {\bibfnamefont {P.~E.}\ \bibnamefont {Bl\"ochl}},\ }\bibfield  {title} {\bibinfo {title} {Projector augmented-wave method},\ }\href {https://doi.org/10.1103/PhysRevB.50.17953} {\bibfield  {journal} {\bibinfo  {journal} {Phys. Rev. B}\ }\textbf {\bibinfo {volume} {50}},\ \bibinfo {pages} {17953} (\bibinfo {year} {1994})}\BibitemShut {NoStop}%
\bibitem [{\citenamefont {Kresse}\ and\ \citenamefont {Furthm\"uller}(1996)}]{kresse1996efficient}%
  \BibitemOpen
  \bibfield  {author} {\bibinfo {author} {\bibfnamefont {G.}~\bibnamefont {Kresse}}\ and\ \bibinfo {author} {\bibfnamefont {J.}~\bibnamefont {Furthm\"uller}},\ }\bibfield  {title} {\bibinfo {title} {Efficient iterative schemes for ab initio total-energy calculations using a plane-wave basis set},\ }\href {https://doi.org/10.1103/PhysRevB.54.11169} {\bibfield  {journal} {\bibinfo  {journal} {Phys. Rev. B}\ }\textbf {\bibinfo {volume} {54}},\ \bibinfo {pages} {11169} (\bibinfo {year} {1996})}\BibitemShut {NoStop}%
\bibitem [{\citenamefont {Perdew}\ \emph {et~al.}(1996)\citenamefont {Perdew}, \citenamefont {Burke},\ and\ \citenamefont {Ernzerhof}}]{perdew1996generalized}%
  \BibitemOpen
  \bibfield  {author} {\bibinfo {author} {\bibfnamefont {J.~P.}\ \bibnamefont {Perdew}}, \bibinfo {author} {\bibfnamefont {K.}~\bibnamefont {Burke}},\ and\ \bibinfo {author} {\bibfnamefont {M.}~\bibnamefont {Ernzerhof}},\ }\bibfield  {title} {\bibinfo {title} {Generalized gradient approximation made simple},\ }\href {https://doi.org/10.1103/PhysRevLett.77.3865} {\bibfield  {journal} {\bibinfo  {journal} {Phys. Rev. Lett.}\ }\textbf {\bibinfo {volume} {77}},\ \bibinfo {pages} {3865} (\bibinfo {year} {1996})}\BibitemShut {NoStop}%
\bibitem [{\citenamefont {Larsen}\ \emph {et~al.}(2016)\citenamefont {Larsen}, \citenamefont {Schmidt},\ and\ \citenamefont {Schiøtz}}]{Larsen2016Robust}%
  \BibitemOpen
  \bibfield  {author} {\bibinfo {author} {\bibfnamefont {P.~M.}\ \bibnamefont {Larsen}}, \bibinfo {author} {\bibfnamefont {S.}~\bibnamefont {Schmidt}},\ and\ \bibinfo {author} {\bibfnamefont {J.}~\bibnamefont {Schiøtz}},\ }\bibfield  {title} {\bibinfo {title} {{Robust structural identification via polyhedral template matching}},\ }\href {https://doi.org/10.1088/0965-0393/24/5/055007} {\bibfield  {journal} {\bibinfo  {journal} {Modelling and Simulation in Materials Science and Engineering}\ }\textbf {\bibinfo {volume} {24}},\ \bibinfo {pages} {055007} (\bibinfo {year} {2016})}\BibitemShut {NoStop}%
\bibitem [{\citenamefont {Dion}\ \emph {et~al.}(2004)\citenamefont {Dion}, \citenamefont {Rydberg}, \citenamefont {Schr\"oder}, \citenamefont {Langreth},\ and\ \citenamefont {Lundqvist}}]{DioRydSch04}%
  \BibitemOpen
  \bibfield  {author} {\bibinfo {author} {\bibfnamefont {M.}~\bibnamefont {Dion}}, \bibinfo {author} {\bibfnamefont {H.}~\bibnamefont {Rydberg}}, \bibinfo {author} {\bibfnamefont {E.}~\bibnamefont {Schr\"oder}}, \bibinfo {author} {\bibfnamefont {D.~C.}\ \bibnamefont {Langreth}},\ and\ \bibinfo {author} {\bibfnamefont {B.~I.}\ \bibnamefont {Lundqvist}},\ }\bibfield  {title} {\bibinfo {title} {Van der waals density functional for general geometries},\ }\href {https://doi.org/10.1103/PhysRevLett.92.246401} {\bibfield  {journal} {\bibinfo  {journal} {Physical Review Letters}\ }\textbf {\bibinfo {volume} {92}},\ \bibinfo {pages} {246401} (\bibinfo {year} {2004})}\BibitemShut {NoStop}%
\bibitem [{\citenamefont {Berendsen}\ \emph {et~al.}(1984)\citenamefont {Berendsen}, \citenamefont {Postma}, \citenamefont {{van Gunsteren}}, \citenamefont {DiNola},\ and\ \citenamefont {Haak}}]{BerPosvan84}%
  \BibitemOpen
  \bibfield  {author} {\bibinfo {author} {\bibfnamefont {H.~J.~C.}\ \bibnamefont {Berendsen}}, \bibinfo {author} {\bibfnamefont {J.~P.~M.}\ \bibnamefont {Postma}}, \bibinfo {author} {\bibfnamefont {W.~F.}\ \bibnamefont {{van Gunsteren}}}, \bibinfo {author} {\bibfnamefont {A.}~\bibnamefont {DiNola}},\ and\ \bibinfo {author} {\bibfnamefont {J.~R.}\ \bibnamefont {Haak}},\ }\bibfield  {title} {\bibinfo {title} {Molecular dynamics with coupling to an external bath},\ }\href {https://doi.org/10.1063/1.448118} {\bibfield  {journal} {\bibinfo  {journal} {The Journal of Chemical Physics}\ }\textbf {\bibinfo {volume} {81}},\ \bibinfo {pages} {3684} (\bibinfo {year} {1984})}\BibitemShut {NoStop}%
\bibitem [{\citenamefont {Ceriotti}\ \emph {et~al.}(2010)\citenamefont {Ceriotti}, \citenamefont {Parrinello}, \citenamefont {Markland},\ and\ \citenamefont {Manolopoulos}}]{CerParMar10}%
  \BibitemOpen
  \bibfield  {author} {\bibinfo {author} {\bibfnamefont {M.}~\bibnamefont {Ceriotti}}, \bibinfo {author} {\bibfnamefont {M.}~\bibnamefont {Parrinello}}, \bibinfo {author} {\bibfnamefont {T.~E.}\ \bibnamefont {Markland}},\ and\ \bibinfo {author} {\bibfnamefont {D.~E.}\ \bibnamefont {Manolopoulos}},\ }\bibfield  {title} {\bibinfo {title} {Efficient stochastic thermostatting of path integral molecular dynamics},\ }\href {https://doi.org/10.1063/1.3489925} {\bibfield  {journal} {\bibinfo  {journal} {The Journal of Chemical Physics}\ }\textbf {\bibinfo {volume} {133}},\ \bibinfo {pages} {124104} (\bibinfo {year} {2010})}\BibitemShut {NoStop}%
\bibitem [{\citenamefont {Ying}\ \emph {et~al.}(2025)\citenamefont {Ying}, \citenamefont {Zhou}, \citenamefont {Svensson}, \citenamefont {Berger}, \citenamefont {Fransson}, \citenamefont {Eriksson}, \citenamefont {Xu}, \citenamefont {Liang}, \citenamefont {Xu}, \citenamefont {Song}, \citenamefont {Chen}, \citenamefont {Erhart},\ and\ \citenamefont {Fan}}]{YinZhoSve25}%
  \BibitemOpen
  \bibfield  {author} {\bibinfo {author} {\bibfnamefont {P.}~\bibnamefont {Ying}}, \bibinfo {author} {\bibfnamefont {W.}~\bibnamefont {Zhou}}, \bibinfo {author} {\bibfnamefont {L.}~\bibnamefont {Svensson}}, \bibinfo {author} {\bibfnamefont {E.}~\bibnamefont {Berger}}, \bibinfo {author} {\bibfnamefont {E.}~\bibnamefont {Fransson}}, \bibinfo {author} {\bibfnamefont {F.}~\bibnamefont {Eriksson}}, \bibinfo {author} {\bibfnamefont {K.}~\bibnamefont {Xu}}, \bibinfo {author} {\bibfnamefont {T.}~\bibnamefont {Liang}}, \bibinfo {author} {\bibfnamefont {J.}~\bibnamefont {Xu}}, \bibinfo {author} {\bibfnamefont {B.}~\bibnamefont {Song}}, \bibinfo {author} {\bibfnamefont {S.}~\bibnamefont {Chen}}, \bibinfo {author} {\bibfnamefont {P.}~\bibnamefont {Erhart}},\ and\ \bibinfo {author} {\bibfnamefont {Z.}~\bibnamefont {Fan}},\ }\bibfield  {title} {\bibinfo {title} {Highly efficient path-integral molecular dynamics simulations with {{GPUMD}} using neuroevolution potentials: {{Case}} studies on thermal properties of
  materials},\ }\href {https://doi.org/10.1063/5.0241006} {\bibfield  {journal} {\bibinfo  {journal} {The Journal of Chemical Physics}\ }\textbf {\bibinfo {volume} {162}},\ \bibinfo {pages} {064109} (\bibinfo {year} {2025})}\BibitemShut {NoStop}%
\bibitem [{\citenamefont {Fransson}\ \emph {et~al.}(2021)\citenamefont {Fransson}, \citenamefont {Slabanja}, \citenamefont {Erhart},\ and\ \citenamefont {Wahnstr{\"o}m}}]{FraSlaErh21}%
  \BibitemOpen
  \bibfield  {author} {\bibinfo {author} {\bibfnamefont {E.}~\bibnamefont {Fransson}}, \bibinfo {author} {\bibfnamefont {M.}~\bibnamefont {Slabanja}}, \bibinfo {author} {\bibfnamefont {P.}~\bibnamefont {Erhart}},\ and\ \bibinfo {author} {\bibfnamefont {G.}~\bibnamefont {Wahnstr{\"o}m}},\ }\bibfield  {title} {\bibinfo {title} {Dynasor---{{A Tool}} for {{Extracting Dynamical Structure Factors}} and {{Current Correlation Functions}} from {{Molecular Dynamics Simulations}}},\ }\href {https://doi.org/10.1002/adts.202000240} {\bibfield  {journal} {\bibinfo  {journal} {Advanced Theory and Simulations}\ }\textbf {\bibinfo {volume} {4}},\ \bibinfo {pages} {2000240} (\bibinfo {year} {2021})}\BibitemShut {NoStop}%
\bibitem [{\citenamefont {Berger}\ \emph {et~al.}(2025)\citenamefont {Berger}, \citenamefont {Fransson}, \citenamefont {Eriksson}, \citenamefont {Lindgren}, \citenamefont {Wahnstr{\"o}m}, \citenamefont {Rod},\ and\ \citenamefont {Erhart}}]{BerFraEri25}%
  \BibitemOpen
  \bibfield  {author} {\bibinfo {author} {\bibfnamefont {E.}~\bibnamefont {Berger}}, \bibinfo {author} {\bibfnamefont {E.}~\bibnamefont {Fransson}}, \bibinfo {author} {\bibfnamefont {F.}~\bibnamefont {Eriksson}}, \bibinfo {author} {\bibfnamefont {E.}~\bibnamefont {Lindgren}}, \bibinfo {author} {\bibfnamefont {G.}~\bibnamefont {Wahnstr{\"o}m}}, \bibinfo {author} {\bibfnamefont {T.~H.}\ \bibnamefont {Rod}},\ and\ \bibinfo {author} {\bibfnamefont {P.}~\bibnamefont {Erhart}},\ }\bibfield  {title} {\bibinfo {title} {Dynasor 2: {{From}} simulation to experiment through correlation functions},\ }\href {https://doi.org/10.1016/j.cpc.2025.109759} {\bibfield  {journal} {\bibinfo  {journal} {Computer Physics Communications}\ }\textbf {\bibinfo {volume} {316}},\ \bibinfo {pages} {109759} (\bibinfo {year} {2025})}\BibitemShut {NoStop}%
\bibitem [{\citenamefont {Fair}\ \emph {et~al.}(2022)\citenamefont {Fair}, \citenamefont {Jackson}, \citenamefont {Voneshen}, \citenamefont {Jochym}, \citenamefont {Le}, \citenamefont {Refson},\ and\ \citenamefont {Perring}}]{FaiJacVon22}%
  \BibitemOpen
  \bibfield  {author} {\bibinfo {author} {\bibfnamefont {R.}~\bibnamefont {Fair}}, \bibinfo {author} {\bibfnamefont {A.}~\bibnamefont {Jackson}}, \bibinfo {author} {\bibfnamefont {D.}~\bibnamefont {Voneshen}}, \bibinfo {author} {\bibfnamefont {D.}~\bibnamefont {Jochym}}, \bibinfo {author} {\bibfnamefont {D.}~\bibnamefont {Le}}, \bibinfo {author} {\bibfnamefont {K.}~\bibnamefont {Refson}},\ and\ \bibinfo {author} {\bibfnamefont {T.}~\bibnamefont {Perring}},\ }\bibfield  {title} {\bibinfo {title} {Euphonic: Inelastic neutron scattering simulations from force constants and visualization tools for phonon properties},\ }\href {https://doi.org/10.1107/S1600576722009256} {\bibfield  {journal} {\bibinfo  {journal} {Journal of Applied Crystallography}\ }\textbf {\bibinfo {volume} {55}},\ \bibinfo {pages} {1689} (\bibinfo {year} {2022})}\BibitemShut {NoStop}%
\bibitem [{\citenamefont {Turanyi}\ \emph {et~al.}(2025)\citenamefont {Turanyi}, \citenamefont {Jackson},\ and\ \citenamefont {Wilkins}}]{TurJacWil25}%
  \BibitemOpen
  \bibfield  {author} {\bibinfo {author} {\bibfnamefont {R.}~\bibnamefont {Turanyi}}, \bibinfo {author} {\bibfnamefont {A.}~\bibnamefont {Jackson}},\ and\ \bibinfo {author} {\bibfnamefont {J.}~\bibnamefont {Wilkins}},\ }\href {https://github.com/pace-neutrons/resins} {\bibinfo {title} {Pace-neutrons/resins: {{Python}} library for resolution functions of inelastic neutron scattering instruments}} (\bibinfo {year} {2025}),\ \bibinfo {note} {(accessed 2025-04-10)}\BibitemShut {NoStop}%
\bibitem [{\citenamefont {Arnold}\ \emph {et~al.}(2014)\citenamefont {Arnold}, \citenamefont {Bilheux}, \citenamefont {Borreguero}, \citenamefont {Buts}, \citenamefont {Campbell}, \citenamefont {Chapon}, \citenamefont {Doucet}, \citenamefont {Draper}, \citenamefont {{Ferraz Leal}}, \citenamefont {Gigg}, \citenamefont {Lynch}, \citenamefont {Markvardsen}, \citenamefont {Mikkelson}, \citenamefont {Mikkelson}, \citenamefont {Miller}, \citenamefont {Palmen}, \citenamefont {Parker}, \citenamefont {Passos}, \citenamefont {Perring}, \citenamefont {Peterson}, \citenamefont {Ren}, \citenamefont {Reuter}, \citenamefont {Savici}, \citenamefont {Taylor}, \citenamefont {Taylor}, \citenamefont {Tolchenov}, \citenamefont {Zhou},\ and\ \citenamefont {Zikovsky}}]{ArnBilBor14}%
  \BibitemOpen
  \bibfield  {author} {\bibinfo {author} {\bibfnamefont {O.}~\bibnamefont {Arnold}}, \bibinfo {author} {\bibfnamefont {J.}~\bibnamefont {Bilheux}}, \bibinfo {author} {\bibfnamefont {J.}~\bibnamefont {Borreguero}}, \bibinfo {author} {\bibfnamefont {A.}~\bibnamefont {Buts}}, \bibinfo {author} {\bibfnamefont {S.}~\bibnamefont {Campbell}}, \bibinfo {author} {\bibfnamefont {L.}~\bibnamefont {Chapon}}, \bibinfo {author} {\bibfnamefont {M.}~\bibnamefont {Doucet}}, \bibinfo {author} {\bibfnamefont {N.}~\bibnamefont {Draper}}, \bibinfo {author} {\bibfnamefont {R.}~\bibnamefont {{Ferraz Leal}}}, \bibinfo {author} {\bibfnamefont {M.}~\bibnamefont {Gigg}}, \bibinfo {author} {\bibfnamefont {V.}~\bibnamefont {Lynch}}, \bibinfo {author} {\bibfnamefont {A.}~\bibnamefont {Markvardsen}}, \bibinfo {author} {\bibfnamefont {D.}~\bibnamefont {Mikkelson}}, \bibinfo {author} {\bibfnamefont {R.}~\bibnamefont {Mikkelson}}, \bibinfo {author} {\bibfnamefont {R.}~\bibnamefont {Miller}}, \bibinfo {author} {\bibfnamefont {K.}~\bibnamefont
  {Palmen}}, \bibinfo {author} {\bibfnamefont {P.}~\bibnamefont {Parker}}, \bibinfo {author} {\bibfnamefont {G.}~\bibnamefont {Passos}}, \bibinfo {author} {\bibfnamefont {T.}~\bibnamefont {Perring}}, \bibinfo {author} {\bibfnamefont {P.}~\bibnamefont {Peterson}}, \bibinfo {author} {\bibfnamefont {S.}~\bibnamefont {Ren}}, \bibinfo {author} {\bibfnamefont {M.}~\bibnamefont {Reuter}}, \bibinfo {author} {\bibfnamefont {A.}~\bibnamefont {Savici}}, \bibinfo {author} {\bibfnamefont {J.}~\bibnamefont {Taylor}}, \bibinfo {author} {\bibfnamefont {R.}~\bibnamefont {Taylor}}, \bibinfo {author} {\bibfnamefont {R.}~\bibnamefont {Tolchenov}}, \bibinfo {author} {\bibfnamefont {W.}~\bibnamefont {Zhou}},\ and\ \bibinfo {author} {\bibfnamefont {J.}~\bibnamefont {Zikovsky}},\ }\bibfield  {title} {\bibinfo {title} {Mantid—data analysis and visualization package for neutron scattering and \textmu {SR} experiments},\ }\href {https://doi.org/10.1016/j.nima.2014.07.029} {\bibfield  {journal} {\bibinfo  {journal} {Nuclear
  Instruments and Methods in Physics Research Section A: Accelerators, Spectrometers, Detectors and Associated Equipment}\ }\textbf {\bibinfo {volume} {764}},\ \bibinfo {pages} {156} (\bibinfo {year} {2014})}\BibitemShut {NoStop}%
\bibitem [{\citenamefont {Liang}\ \emph {et~al.}(2026)\citenamefont {Liang}, \citenamefont {Xu}, \citenamefont {Lindgren}, \citenamefont {Chen}, \citenamefont {Zhao}, \citenamefont {Liu}, \citenamefont {Berger}, \citenamefont {Tang}, \citenamefont {Zhang}, \citenamefont {Wang}, \citenamefont {Song}, \citenamefont {Ying}, \citenamefont {Xu}, \citenamefont {Dong}, \citenamefont {Chen}, \citenamefont {Erhart}, \citenamefont {Fan}, \citenamefont {Ala-Nissila},\ and\ \citenamefont {Xu}}]{liang2026NEP89Dataset}%
  \BibitemOpen
  \bibfield  {author} {\bibinfo {author} {\bibfnamefont {T.}~\bibnamefont {Liang}}, \bibinfo {author} {\bibfnamefont {K.}~\bibnamefont {Xu}}, \bibinfo {author} {\bibfnamefont {E.}~\bibnamefont {Lindgren}}, \bibinfo {author} {\bibfnamefont {Z.}~\bibnamefont {Chen}}, \bibinfo {author} {\bibfnamefont {R.}~\bibnamefont {Zhao}}, \bibinfo {author} {\bibfnamefont {J.}~\bibnamefont {Liu}}, \bibinfo {author} {\bibfnamefont {E.}~\bibnamefont {Berger}}, \bibinfo {author} {\bibfnamefont {B.}~\bibnamefont {Tang}}, \bibinfo {author} {\bibfnamefont {B.}~\bibnamefont {Zhang}}, \bibinfo {author} {\bibfnamefont {Y.}~\bibnamefont {Wang}}, \bibinfo {author} {\bibfnamefont {K.}~\bibnamefont {Song}}, \bibinfo {author} {\bibfnamefont {P.}~\bibnamefont {Ying}}, \bibinfo {author} {\bibfnamefont {N.}~\bibnamefont {Xu}}, \bibinfo {author} {\bibfnamefont {H.}~\bibnamefont {Dong}}, \bibinfo {author} {\bibfnamefont {S.}~\bibnamefont {Chen}}, \bibinfo {author} {\bibfnamefont {P.}~\bibnamefont {Erhart}}, \bibinfo {author} {\bibfnamefont
  {Z.}~\bibnamefont {Fan}}, \bibinfo {author} {\bibfnamefont {T.}~\bibnamefont {Ala-Nissila}},\ and\ \bibinfo {author} {\bibfnamefont {J.}~\bibnamefont {Xu}},\ }\href {https://doi.org/10.5281/zenodo.19440423} {\bibinfo {title} {{Dataset and models supporting ``NEP89: Universal neuroevolution potential for inorganic and organic materials across 89 elements''. {\textit{Zenodo}} \url{https://doi.org/10.5281/zenodo.19440423}}}} (\bibinfo {year} {2026})\BibitemShut {NoStop}%
\bibitem [{\citenamefont {Fan}(2026)}]{gpumd40}%
  \BibitemOpen
  \bibfield  {author} {\bibinfo {author} {\bibfnamefont {Z.}~\bibnamefont {Fan}},\ }\href {https://doi.org/10.5281/zenodo.18977569} {\bibinfo {title} {{brucefan1983/GPUMD: GPUMD-v5.0. {\textit{Zenodo}} \url{https://doi.org/10.5281/zenodo.18977569}}}} (\bibinfo {year} {2026})\BibitemShut {NoStop}%
\end{thebibliography}%
\end{document}